\documentclass[12pt,epsf]{article}
\usepackage{graphicx,epsf}
\usepackage{lscape}
\usepackage{citesort}

\begin{document}

\def\ds{\displaystyle}
\def\beq{\begin{equation}}
\def\eeq{\end{equation}}
\def\bea{\begin{eqnarray}}
\def\eea{\end{eqnarray}}
\def\beeq{\begin{eqnarray}}
\def\eeeq{\end{eqnarray}}
\def\ve{\vert}
\def\vel{\left|}
\def\ver{\right|}
\def\nnb{\nonumber}
\def\ga{\left(}
\def\dr{\right)}
\def\aga{\left\{}
\def\adr{\right\}}
\def\lla{\left<}
\def\rra{\right>}
\def\rar{\rightarrow}
\def\nnb{\nonumber}
\def\la{\langle}
\def\ra{\rangle}
\def\ba{\begin{array}}
\def\ea{\end{array}}
\def\tr{\mbox{Tr}}
\def\ssp{{\Sigma^{*+}}}
\def\sso{{\Sigma^{*0}}}
\def\ssm{{\Sigma^{*-}}}
\def\xis0{{\Xi^{*0}}}
\def\xism{{\Xi^{*-}}}
\def\qs{\la \bar s s \ra}
\def\qu{\la \bar u u \ra}
\def\qd{\la \bar d d \ra}
\def\qq{\la \bar q q \ra}
\def\gGgG{\la g^2 G^2 \ra}
\def\q{\gamma_5 \not\!q}
\def\x{\gamma_5 \not\!x}
\def\g5{\gamma_5}
\def\sb{S_Q^{cf}}
\def\sd{S_d^{be}}
\def\su{S_u^{ad}}
\def\ss{S_s^{??}}
\def\sbp{{S}_Q^{'cf}}
\def\sdp{{S}_d^{'be}}
\def\sup{{S}_u^{'ad}}
\def\ssp{{S}_s^{'??}}
\def\sig{\sigma_{\mu \nu} \gamma_5 p^\mu q^\nu}
\def\fo{f_0(\frac{s_0}{M^2})}
\def\ffi{f_1(\frac{s_0}{M^2})}
\def\fii{f_2(\frac{s_0}{M^2})}
\def\O{{\cal O}}
\def\sl{{\Sigma^0 \Lambda}}
\def\es{\!\!\! &=& \!\!\!}
\def\ap{\!\!\! &\approx& \!\!\!}
\def\ar{&+& \!\!\!}
\def\ek{&-& \!\!\!}
\def\kek{\!\!\!&-& \!\!\!}
\def\cp{&\times& \!\!\!}
\def\se{\!\!\! &\simeq& \!\!\!}
\def\eqv{&\equiv& \!\!\!}
\def\kpm{&\pm& \!\!\!}
\def\kmp{&\mp& \!\!\!}


\def\simlt{\stackrel{<}{{}_\sim}}
\def\simgt{\stackrel{>}{{}_\sim}}


\title{
         {\Large
                 {\bf
Investigation of the $D_{s1}$ structure via $B_c$ to $D_{s1} l^+l^-/\nu\bar\nu$ transitions in QCD.
                 }
         }
      }

\author{\vspace{1cm}\\
{\small R. Khosravi$^1$\thanks {e-mail: khosravi.reza @
gmail.com}~\,}, {\small K. Azizi $^2$\thanks {e-mail: e146342 @
metu.edu.tr}~\,}, {\small M. Ghanaatian$^3$\thanks {e-mail:
m$_{_-}$ghanatian57 @ yahoo.com}~\,},
{\small F. Falahati$^1$\thanks {e-mail: fatemehfalahati58 @ gmail.com}~\,}\\
 {\small $^1$ Physics Department , Shiraz University, Shiraz 71454,
Iran}\\
{\small  $^2$ Department of Physics, Middle East Technical
University, 06531 Ankara, Turkey}
\\
{\small  $^3$ Physics Department, Payame Noor University, Iran}\\}
\date{}

\begin{titlepage}
\maketitle \thispagestyle{empty}

\begin{abstract}
We investigate the structure of the $D_{s1}(2460,2536) (J^P=1^+)$
mesons via analyzing the semileptonic $B_{c}\to D_{s1}l^+l^-$,
$l=\tau, \mu, e$ and $B_{c}\to D_{s1}\nu\bar{\nu}$ transitions  in
the framework of the three--point QCD sum rules.   We consider the
$D_{s1}$  meson in two ways,  the pure $|c\bar{s}\rangle$ state and
then as a mixture of two $|^3P_1\rangle$ and $|^1P_1\rangle$ states.
Such type rare transitions take place at loop level by electroweak
penguin and weak box diagrams in the standard model via the flavor
changing neutral current transition of $b \to  s$. The relevant form
factors are calculated taking into account the gluon condensate
contributions. These form factors are numerically obtained for
$|c\bar s\rangle$ case and plotted in terms of the unknown mixing
angle $\theta_s$, when  the $D_{s1}$ meson are considered as mixture
of two $|^3P_1\rangle$ and $|^1P_1\rangle$ states. The obtained
results for the form factors are used to evaluate the decay rates
and  branching ratios.  Any future  experimental measurement on
these form factors as well as decay rates and branching fractions
and their comparison with the obtained results in the present work
can give considerable information about the structure of this meson
and the mixing angle $\theta_s$.
\end{abstract}

\end{titlepage}

\section{Introduction}
The structure of  the even--parity charmed  $D_{sJ}$ mesons has
not known exactly yet and  has been debated in the quark model.
The observation of two narrow resonances with charm and
strangeness, $D_{s0}(2317)$ in the invariant mass distribution of
$D_{s}\pi^{0}$ \cite{1,2,3,4,5,6} and $D_{s1}(2460)$ in the
$D_{s}^{\ast}\pi^{0}$ and $D_{s}\gamma$ mass distributions
\cite{2,3,4,6,7,8}, has raised discussions about the structure  of
these states and their quark contents \cite{9,10}.
  Analysis of the $D_{s_{0}}(2317)\rightarrow
D_{s}^{\ast}\gamma$, $D_{s1}(2460)\rightarrow
    D_{s}^{\ast}\gamma$ and $ D_{s_{1}}(2460)\rightarrow D_{s_{0}}(2317)\gamma$ shows that the quark
content of these mesons are probably $c\bar s$ \cite{11}. Among
these mesons, the axial vector charm--strange meson $D_{s1}$ is
more attractive ones, because the discovery of the $D_{s1}(2460)
(J^P=1^{+})$ meson \cite{1,2,3,4} and its  measured mass indicated
a lower mass than expected in potential model (PM) \cite{Godfrey1}
and quark model (QM) \cite{Godfrey2,Pierro} predictions. In other
words, the $D_{s1}(2460)$ does not fit easily in to the $c\bar{s}$
spectroscopy \cite{Close}. However, some physicists presumed that
this discovered state is conventional $c\bar{s}$ meson
\cite{Godfrey3,Bardeen,Nowak,Deandrea,Cahn,Dai,Lucha,Hofmann,Sadzikowski,Becirevic,Lee}.
Many different theoretical efforts have been dedicated to the
understanding of this unexpected and surprising disparity between
theory and experiment
\cite{Kolomeitsev,Barnes,Cheng1,Terasaki,Browder,Dmitrasinovic,Bracco,Kim,Szczepaniak}.
As a result of the above discussion, we will consider the $D_{s1}$
meson  in two ways,  the pure
$|c\bar{s}\rangle$ state and also as a mixture of two
$|^3P_1\rangle$ and $|^1P_1\rangle$ states.

Heavy--light mesons are not charge conjugation eigen states and so
mixing can occur among states with the same $J^P$ and different
mass that are forbidden for neutral states. These occur between
states with $J=L$ and $S=1$ or 0 \cite{Close}. Hence, the mixing of
the physical $D_{s1}$ and $D'_{s1}$ states can be parameterized in
terms of a mixing angle $\theta_{s}$, as follow:
\begin{eqnarray}\label{eq1}
|D_{s1}\rangle &=& sin\theta_{s} |^3P_{1}\rangle ~+
cos\theta_{s} |^1P_{1}\rangle, \nonumber\\
|D'_{s1}\rangle &=& cos\theta_{s} |^3P_{1}\rangle ~-sin\theta_{s}
|^1P_{1}\rangle.
\end{eqnarray}
where,  the spectroscopic notation $^{2S+1}L_{J}$  has been used to introduce
 the mixing states. Considering
$|^3P_1\rangle\equiv |D_{s1}1\rangle$ and $|^1P_1\rangle\equiv
|D_{s1}2\rangle$ with different masses and decay constants
\cite{Thomas}, we can apply these relations for axial vectors
$D_{s1}(2460)$ and $D_{s1}(2536)$ mesons with two different
masses. i.e.,
\begin{eqnarray}\label{eq2}
|D_{s1}(2460)~\rangle &=& sin\theta_{s} |D_{s1}1\rangle ~+
cos\theta_{s} |D_{s1}2\rangle, \nonumber\\
|D_{s1}(2536)~\rangle &=& cos\theta_{s} |D_{s1}1\rangle
~-sin\theta_{s} |D_{s1}2\rangle.
\end{eqnarray}
The masses of $D_{s1}1$ and $D_{s1}2$ states are presented in
Table \ref{T1}. These values have been obtained in QM approach.
\begin{table}[th]
\centering
\begin{tabular}{cccc}
\cline{1-4}Ref&\cite{Godfrey1}&\cite{Godfrey2}&\cite{Pierro}\\
\cline{1-4}\lower0.35cm \hbox{{\vrule width 0pt height 1.0cm }}
$D_{s1}1(^3P_{1})$ &2.57&2.55&2.535\\
\cline{1-4}\lower0.35cm \hbox{{\vrule width 0pt height 1.0cm }}
$D_{s1}2(^1P_{1})$ &2.53&2.55&2.605\\
\hline
\end{tabular}\label{T1}
\vspace{0.10cm} \caption{Masses of $1^1P_1$ and $1^3P_1$
heavy-light mesons in quark models.}
\end{table}

Note that, in the heavy quark limit the physical eigen states
$D_{s1}$ and $D'_{s1}$ can be identified with $P_1^{1/2}$ and
$P_1^{3/2}$ with notation $L^j_J$, where $j$ is the total angular
momentum of the light quark \cite{Cheng2}, corresponding to
$\theta_{s}=-54.7^\circ$ \cite{Thomas}.

In this work, taking into account the gluon condensate
corrections, we analyze the rare semileptonic $B_{c}\to
D_{s1}~l^+l^-$, $l=\tau, \mu, e$ and $B_{c}\to D_{s1}\nu\bar{\nu}$
transitions in three--point QCD sum rules (3PSR) approach. Note
that, the $B_c\to (D^*,D_s^*,D_{s1}(2460)) \nu\bar{\nu}$
transitions have been studied  in Ref. \cite{Azizi3}, but
assuming the  $D_{s1}$ only as $c\bar s$. The $B_c\to
D_q~l^+l^-/\nu \bar {\nu}$ \cite{Azizi1}, $B_c\to D^*_q ~l^+l^-$,
$(q=d, s)$ \cite{Azizi2} transitions have also been analyzed in
the same framework.

The heavy  $B_c$ meson contains two heavy quarks $b$ and $c$ with
different charges. This meson is similar to the charmonium and
bottomonium in the spectroscopy, but in contrast to the charmonium
and bottomonium, the $B_c$ decays only via weak interaction and
has a long lifetime. The study of the $B_c$ transitions are useful
for  more precise determination of the Cabibbo, Kabayashi, Maskawa
(CKM) matrix elements in the weak decays.

The rare semileptonic $B_c\to D_{s1} l^+l^-/\nu\bar{\nu}$ decays
occur at loop level by electroweak penguin and weak box diagrams
in the standard model (SM) via the flavor changing neutral current
(FCNC) transition of $b\to s l^+l^-$. The FCNC decays of $B_c$
meson are sensitive to new physics (NP) contributions to penguin
operators. Therefore, the  study of such FCNC transitions can
improve the information about:
\begin{itemize}
\item
The CP violation, T violation and polarization asymmetries in
$b\to  s$ penguin channels, that occur in weak interactions,
\item
New operators or operators that are subdominant in the SM,
\item
Establishing NP and flavor physics beyond the SM.
\end{itemize}

To obtain the form factors of the semileptonic $B_c\to
D_{s1}(2460[2536])$ transitions, first, we will suppose the
$D_{s1}(2460)$ and $D_{s1}(2536)$ axial vector mesons as the pure
$|c\bar{s}\rangle$ state and calculate the related form factors.
Second, we will consider the $D_{s1}$ meson as a mixture of two
components $|D_{s1}1\rangle$ and $|D_{s1}2 \rangle$ states and
calculate the form factors of the $B_c\to D_{s1}1$ and $B_c\to
D_{s1}2$ transitions. With the help of  Eq. (\ref{eq2}) and the
definition of the form factors which will be presented in the next
section, we will derive the transition form factors of $B_c\to
D_{s1}(2460[2536])$ decays as a function of the mixing angle
$\theta_s$.  The future experimental study of such rare decays and
comparison of the  results with the predictions of theoretical
calculations can improve the information about the structure of
$D_{s1}$ meson and the mixing angle $\theta_s$.

This paper is organized as follow. In section 2, we calculate the
form factors for $B_c\to D_{s1}$ transition in the 3PSR.  In
section 3, the two-gluon condensate contributions as
non-perturbative corrections are calculated. The calculation of
the decay rates  for $B_c\to D_{s1}l^{+} l^{-}$ and $B_c\to D_{s1}
\nu \bar{\nu}$ transitions are presented in section 4. Finally,
section 5  is devoted to the numeric results and discussions.

\section{The form factors of $B_{c}\rightarrow D_{s1} $
transition in 3PSR}

In the standard model, the effective Hamiltonian responsible for the rare semileptonic $B_{c}
\rightarrow D_{s1} l^{+}l^{-}$ and $B_{c} \rightarrow
D_{s1}\nu\bar{\nu}$ decays, which are described  via
$b \rightarrow s~l^{+}l^{-}$ loop transitions (see Fig. 1) at quark-level, can be
written as:
\begin{eqnarray}\label{eq3}
\mathcal{H}_{eff} &=& \frac{G_{F}\alpha}{2\pi\sqrt{2} }~ V_{tb}V_{ts}^\ast %
\Bigg[ C_9^{eff} \, \overline{s}~ \gamma_\mu (1-\gamma_5) b \,
\overline \ell \gamma_\mu \ell + C_{10}~ \overline {s} \gamma_\mu
(1-\gamma_5) b \, \overline \ell \gamma_\mu \gamma_5
\ell  \nonumber \\
&-& 2 C_7^{eff} \frac{m_b}{q^2}~ \overline {s} ~i\sigma_{\mu\nu}
q^\nu (1+\gamma_5) b \, \overline \ell \gamma_\mu \ell \Bigg]~.
\end{eqnarray}
where $C_7^{eff}$, $C_9^{eff}$ and $C_{10}$ are the Wilson
coefficients, $G_{F}$ is the Fermi constant, $\alpha$ is the fine
structure constant at the $Z$ mass scale and $V_{ij}$ are the
elements of the CKM matrix.
\begin{figure}[th]
\vspace*{0.4cm}
\begin{center}
\begin{picture}(160,50)
\centerline{ \epsfxsize=14cm \epsfbox{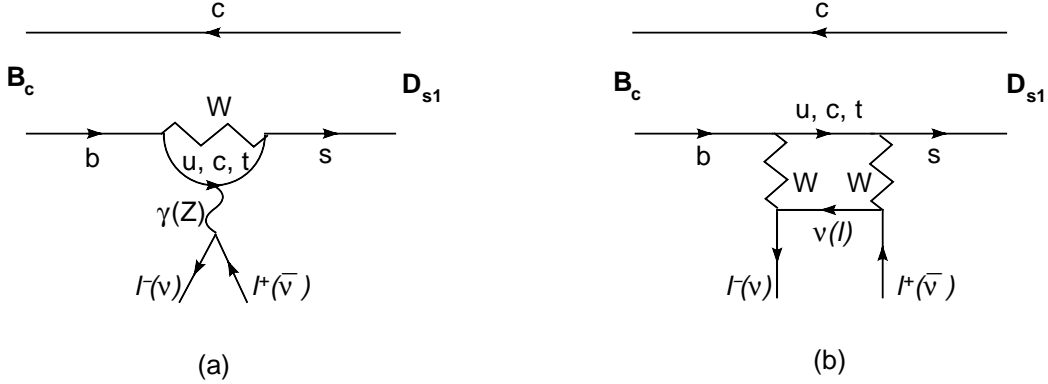}}
\end{picture}
\end{center}
\vspace*{-0.1cm} \caption{The loop diagrams of the semileptonic
decay of $B_c$ to $D_{s1}$. The electroweak penguin and box diagrams
are shown in parts (a) and (b), respectively.} \label{F1}
\end{figure}
\normalsize

These loop transitions occur via the intermediate $u, c, t$
quarks. In the SM, the measurement of the forward-backward
asymmetry and invariant dilepton mass distribution in $b\to
q^{\prime}l^{+}l^{-}$, $(q^{\prime}=d, s)$ transitions provide
information on the short distance contributions dominated by the
top quark loops \cite{Deshpande}. The electroweak penguin
involving the contributions of photon and $Z$ bosons is shown in
Fig. \ref{F1}(a) and Fig. \ref{F1}(b) presents the contribution of
the $W$ box diagram. It is reminded that the $b\to s~\nu\bar{\nu}$
transition receives contributions only from $Z$-penguin and box
diagrams.

The transition amplitude of $B_{c} \rightarrow
D_{s1}l^{+}l^{-}/\nu\bar{\nu}$ decays is obtained  sandwiching
Eq. (\ref{eq3}) between the initial and final states, i.e.,
\begin{eqnarray}\label{eq4}
\mathcal{M}&=&\frac{G_{F}\alpha }{2\pi\sqrt{2}
}~V_{tb}V_{ts}^{\ast } \Bigg[C_{9}^{eff}\,\langle D_{s1}(p')\mid
~\overline{s}~\gamma _{\mu }(1-\gamma _{5})b\mid
B_{c}(p)\rangle\overline{ \ell}\gamma _{\mu }\ell\nonumber
\\ &&+ ~C_{10}~\langle D_{s1}(p')\mid \overline{s}~\gamma _{\mu }(1-\gamma _{5})b\mid
B_{c}(p)\rangle\overline{\ell} \gamma _{\mu }\gamma
_{5}\ell\nonumber
\\ &&-2~C_{7}^{eff}~\frac{m_{b}}{q^{2}}\langle D_{s1}(p')\mid \overline{s}~i\sigma _{\mu \nu
}q^{\nu }(1+\gamma _{5})b\mid B_{c}(p)\rangle\overline{\ell}\gamma
_{\mu }\ell \Bigg]~,
\end{eqnarray}
where, $p$ and $p^{\prime}$ are the momentum of initial and final meson states,
respectively, and $\varepsilon$ is the polarization vector
of the $D_{s1}$ meson. Our aim is to parameterized the matrix elements
appearing in Eq. (\ref{eq4}) in terms of the transition form
factors considering  the Lorentz invariance and parity considerations.
\begin{eqnarray}\label{eq5}
\langle D_{s1}(p',\varepsilon)\mid \overline
{s}\gamma_{\mu}\gamma_{5} b\mid B_c(p)\rangle&=&\frac{2A^{B_c\to
D_{s1}}_{V}(q^2)}{m_{B_{c}}+m_{D_{s1}}}~\varepsilon_{\mu\nu\alpha\beta}
\varepsilon^{\ast\nu}p^\alpha
p'^\beta,\nonumber\\
\langle D_{s1}(p',\varepsilon)\mid\overline {s}\gamma_{\mu}b\mid
B_c(p)\rangle&=& -~iA^{B_c\to D_{s1}}_{0}(q^2)(m_{B_{c}}
+m_{D_{s1}})\varepsilon_{\mu}^{\ast} +i\frac{A^{B_c\to
D_{s1}}_{1}(q^2)}{m_{B_{c}}+m_{D_{s1}}}(\varepsilon^{*}p)P_{\mu}\nonumber \\
&+&i\frac{A^{B_c\to
D_{s1}}_{2}(q^2)}{m_{B_{c}}+m_{D_{s1}}}(\varepsilon^{*}p)q_{\mu}~,\nonumber\\ \nonumber\\
\langle D_{s1}(p',\varepsilon)\mid\overline {s}\sigma_{\mu\nu}
q^\nu \gamma_5 b\mid B_{c}(p)\rangle&=&2~T^{B_c\to
D_{s1}}_{V}(q^2)~i\varepsilon_{\mu\nu\alpha\beta}
\varepsilon^{\ast\nu}p^\alpha p'^\beta,\nonumber\\ \nonumber\\
\langle D_{s1}(p',\varepsilon)\mid\overline {s}\sigma_{\mu\nu}
q^\nu b\mid B_{c}(p)\rangle&=& T^{B_c\to
D_{s1}}_{0}(q^2)[\varepsilon_{\mu}^{\ast}(m_{B_{c}}^2-m_{D_{s1}}^2)
-(\varepsilon^{*}p)P_{\mu}]\nonumber\\
&+&T^{B_c\to
D_{s1}}_{1}(q^2)~(\varepsilon^{*}p)[q_{\mu}-\frac{q^2}{m_{B_{c}}^2-m_{D_{s1}}^2}P_{\mu}],
\end{eqnarray}
where $A^{B_c\to D_{s1}}_{i}(q^2)$, $i=V, 0, 1, 2$ and $T^{B_c\to
D_{s1}}_{j}(q^2)$, $j=V, 0, 1$ are the transition form factors,
$P_{\mu}=(p+p')_{\mu}$ and $q_{\mu}=(p-p')_{\mu}$. Here, $q^2$ is
the momentum transfer squared of the $Z$ boson (photon). In order to our calculations be simple, the following
redefinitions of the transition form factors are considered :
\begin{eqnarray}\label{eq6}
A^{'B_c\to D_{s1}}_{V}(q^2)&=&\frac{2A^{B_c\to
D_{s1}}_{V}(q^2)}{m_{B_{c}}+m_{D_{s1}}}~,~~~~~~~~~A^{'B_c\to
D_{s1}}_{0}(q^2)~=~A^{B_c\to D_{s1}}_{0}(q^2)(m_{B_{c}}
+m_{D_{s1}}),
\nonumber\\
A^{'B_c\to D_{s1}}_{1}(q^2)&=&-\frac{A^{B_c\to
D_{s1}}_{1}(q^2)}{m_{B_{c}}+m_{D_{s1}}}~,~~~~~~~~ A^{'B_c\to
D_{s1}}_{2}(q^2)~=~-\frac{A^{B_c\to D_{s1}}_{2
}(q^2)}{m_{B_{c}}+m_{D_{s1}}},\nonumber\\ \nonumber\\
T^{'B_c\to D_{s1}}_{V}(q^2)&=&-2T^{B_c\to
D_{s1}}_{V}(q^2)~,~~~~~~~T^{'B_c\to D_{s1}}_{0}(q^2)~=~-~T^{B_c\to
D_{s1}}_{0}(q^2)(m_{B_{c}}^2-m_{D_{s1}}^2),\nonumber\\ \nonumber\\
T^{'B_c\to D_{s1}}_{1}(q^2)&=&-~T^{B_c\to D_{s1}}_{1}(q^2).
\end{eqnarray}

To calculate the form factors within three-point QCD
sum rules method,  the following three-point
correlation functions are used:
\begin{eqnarray}\label{eq7}
\Pi _{\mu\nu}^{V-A}(p^2,p'^2,q^2)&=&i^2\int
d^{4}xd^4ye^{-ipx}e^{ip'y}\langle0\mid T[J _{\nu}^ {D_{s1}}(y)
J_{\mu}^{V-A}(0) {J^{B_{c}}}^{\dag}(x)]\mid  0\rangle,\nonumber\\
\Pi _{\mu\nu}^{T-PT}(p^2,p'^2,q^2)&=&i^2\int
d^{4}xd^4ye^{-ipx}e^{ip'y}\langle0\mid T[J _{\nu}^{D_{s1}}(y)
J_{\mu}^{T-PT}(0) {J^{B_{c}}}^{\dag}(x)]\mid 0\rangle,
\end{eqnarray}
where $J _{\nu}^{ D_{s1}}(y)=\overline{c}\gamma_{\nu}\gamma_{5}s$
and $J^{B_{c}}(x)=\overline{c}\gamma_{5}b$~are the interpolating
currents of the initial and final meson states, respectively.
$J_{\mu}^{V-A}=~\overline {s}\gamma_{\mu}(1-\gamma_{5})b $ ~and~
$J_{\mu}^{T-PT}=~\overline {s}{\sigma_{\mu\nu}
q^\nu(1+\gamma_5)}b$ are the vector-axial vector and  tensor-pseudo tensor parts of the transition currents. In QCD sum
rules approach, we can obtain the correlation function of Eq. (\ref{eq7})
in two sides. The phenomenological or physical part is calculated saturating the
correlator by a tower of hadrons with the same quantum numbers as interpolating currents. The QCD or
theoretical part, on the other side, is obtained in terms of
the quarks and gluons interacting in the QCD vacuum. To drive the phenomenological
part of the correlators given in Eq. (\ref{eq7}), two complete sets
of intermediate states with the same quantum numbers as the
currents $J_{D_{s1}}$ and $J_{B_{c}}$ are inserted. This procedure
leads to  the following representations of the above-mentioned
correlators:
\begin{eqnarray} \label{eq8}
\Pi _{\mu\nu}^{V-A}(p^2,p'^2,q^2)&=&\frac{\langle0\mid J_{\nu}^{
D_{s1}} \mid D_{s1}(p',{\varepsilon})\rangle\langle
D_{s1}(p',{\varepsilon})\mid J_{\mu}^{V-A}\mid
B_{c}(p)\rangle\langle B_{c}(p)\mid {J^{B_c}}^{\dag}\mid
0\rangle}{(p'^2-m_{D_{s1}}^2)(p^2-m_{B_c}^2)}
\nonumber\\&+&\mbox{higher resonances and continuum states}~,
\nonumber \\ \nonumber \\
\Pi _{\mu\nu}^{T-PT}(p^2,p'^2,q^2)&=&\frac{\langle0\mid J_{\nu}^
{D_{s1}} \mid D_{s1}(p',{\varepsilon})\rangle\langle
D_{s1}(p',{\varepsilon})\mid J_{\mu}^{T-PT}\mid
B_{c}(p)\rangle\langle B_{c}(p)\mid {J^{B_c}}^{\dag}\mid
0\rangle}{(p'^2-m_{D_{s1}}^2)(p^2-m_{B_c}^2)}
\nonumber\\&+&\mbox{higher resonances and continuum states}~.
\end{eqnarray}
The following matrix elements are defined in the standard way in
terms of the leptonic decay constants of the $D_{s1}$ and $B_c$
mesons as:
\begin{equation}\label{eq9}
 \langle0\mid J^{\nu}_{D_{s1}} \mid
D_{s1}(p',{\varepsilon})\rangle=f_{D_{s1}}m_{D_{s1}}\varepsilon^{\nu}~,~~\langle0\mid
J_{B_c}\mid
B_{c}(p)\rangle=i\frac{f_{B_{c}}m_{B_{c}}^2}{m_{b}+m_{c}}.
\end{equation}
Using Eq. (\ref{eq5}), Eq. (\ref{eq6}) and Eq. (\ref{eq9}) in
Eq. (\ref{eq8}) and performing summation over the polarization of
$D_{s1}$ meson we obtain:
\begin{eqnarray}\label{eq10}
\Pi_{\mu\nu}^{V-A}(p^2,p'^2,q^2)&=&-\frac{f_{B_{c}}m_{B_{c}}^2}{(m_{b}+m_{c})}\frac{f_{D_{s1}}m_{D_{s1}}}
{(p'^2-m_{D_{s1}}^2)(p^2-m_{B_c}^2)} \times \left[iA^{'B_c\to
D_{s1}}_{V}(q^2)\varepsilon_{\mu\nu\alpha\beta}p^{\alpha}p'^{\beta}\right.\nonumber\\
&+&\left.A^{'B_c\to D_{s1}}_{0}(q^2)g_{\mu\nu} + A^{'B_c\to
D_{s1}}_{1}(q^2)P_{\mu}p_{\nu} +A^{'B_c\to
D_{s1}}_{2}(q^2)q_{\mu}p_{\nu}\right]
+ \mbox{excited states,}\nonumber\\
\Pi_{\mu\nu}^{T-PT}(p^2,p'^2,q^2)&=&-\frac{f_{B_{c}}m_{B_{c}}^2}{(m_{b}+m_{c})}\frac{f_{D_{s1}}m_{D_{s1}}}
{(p'^2-m_{D_{s1}}^2)(p^2-m_{B_c}^2)} \times \left[T^{'B_c\to
D_{s1}}_{V}(q^2)\varepsilon_{\mu\nu\alpha\beta}p^{\alpha}p'^{\beta}\right.\nonumber\\
&-&\left.i~T^{'B_c\to D_{s1}}_{0}(q^2)g_{\mu\nu}-i~T^{'B_c\to
D_{s1}}_{1}(q^2)q_{\mu}p_{\nu} \right]+ \mbox{excited states.}
\end{eqnarray}

To calculate the form factors, $A^{'}_V$, $A^{'}_0$, $A^{'}_1$,
$A^{'}_2$, $T^{'}_V$, $T^{'}_0$ and $T^{'}_1$, we will choose the
structures,
$i\varepsilon_{\mu\nu\alpha\beta}p^{\alpha}p'^{\beta}$,
$g_{\mu\nu}$, $P_{\mu}p_{\nu}$, $q_{\mu}p_{\nu}$, from
$\Pi_{\mu\nu}^{V-A}$ and
$\varepsilon_{\mu\nu\alpha\beta}p^{\alpha}p'^{\beta}$,
$ig_{\mu\nu}$ and $iq_{\mu}p_{\nu}$ from $\Pi_{\mu\nu}^{T-PT}$,
respectively.

On the QCD  side,  using the operator product expansion (OPE),
we can obtain the correlation function in quark-gluon language in the deep Euclidean region where $p^2\ll
(m_b+m_c)^2$ and  ${p^{'}}^2\ll (m_c^2+m_{s}^2)$. For
this aim, the correlators are written as:
\begin{eqnarray}\label{eq11}
\Pi_{\mu\nu}^{V-A}(p^2,p'^2,q^2)&=&i~\Pi^{V-A}_{V}\varepsilon_{\mu\nu\alpha\beta}p^{\alpha}p'^{\beta}+\Pi^{V-A}_{0}g_{\mu\nu}+\Pi^{V-A}_{1}P_{\mu}p_{\nu}+
\Pi^{V-A}_{2}q_{\mu}p_{\nu},
\nonumber\\
\Pi_{\mu\nu}^{T-PT}(p^2,p'^2,q^2)&=&\Pi^{T-PT}_{V}\varepsilon_{\mu\nu\alpha\beta}p^{\alpha}p'^{\beta}
-i~\Pi^{T-PT}_{0}g_{\mu\nu}-i~\Pi^{T-PT}_{1}q_{\mu}p_{\nu},
\end{eqnarray}
where, each $\Pi_{i}$ function is defined in terms of the perturbative and
non-perturbative parts as:
\begin{eqnarray} \label{eq12}
\Pi_{i}(p^2,p'^2,q^2) = \Pi_{i}^{per}(p^2,p'^2,q^2)
+\Pi_{i}^{nonper}(p^2,p'^2,q^2)~.
\end{eqnarray}

To obtain the perturbative part of the correlation function, we
should study the bare loop diagrams in Fig. \ref{F1}. In
calculating the bare loop contributions, we first write the double
dispersion representation for the coefficients of the
corresponding Lorentz structures appearing in each correlation
function, as:
\begin{equation}\label{eq13}
\Pi_i^{per}=-\frac{1}{(2\pi)^2}\int ds'\int
ds\frac{\rho_{i}(s,s',q^2)}{(s-p^2)(s'-p'^2)}+\textrm{ subtraction
terms}.
\end{equation}
The spectral densities $\rho_{i}^{per} (s,s^\prime,q^2)$ are
calculated by  the help of the Gutkosky rules, i.e., the
propagators are replaced by Dirac--delta functions
\begin{eqnarray} \label{eq14}
\frac{1}{p^2-m^2} \rar -2i\pi \delta(p^2-m^2)~,
\end{eqnarray}
expressing  that all quarks are real. Note that, there are two
main vertexes related to the bare loop diagrams that describe
$b\to s~l^+ l^-$ transition in Fig. \ref{F1}, i.e.,
$\gamma_{\mu}(1-\gamma_{5})$ and
$\sigma_{\mu\nu}q^{\nu}(1+\gamma_{5})$. First, we calculate the
spectral densities related to $\gamma_{\mu}(1-\gamma_{5})$ vertex.
Straightforward calculations end up in  the following results:
\begin{eqnarray}\label{eq17}
\rho^{V-A}_V &=&4 N_c~I_0(s,s^{\prime },q^2)~\left\{B_1(m_b-m_c)-B_2(m%
_{s}+m_c)-m_c\right\} ~,\nonumber \\ \nonumber\\
\rho^{V-A}_0 &=&-2 N_c~I_0(s,s^{\prime },q^2)~\{\Delta (m_c+m_{s})
-\Delta ^{\prime }(m_b-m_c)-4 A_1(m_b-%
m_c)
\nonumber \\
&&+2 m_c^2(m_b-m_c-m_{s})+m_c(2m_b m_{s}-u)%
\}~, \nonumber \\ \nonumber \\
\rho^{V-A}_1&=&2N_c~I_0(s,s^{\prime },q^2)\{B_1(m_b-%
3m_c)-B_2(m_c+m_{s})+2A_2(m_b-m_c)
\nonumber \\
&&+2A_3(m_b-m_c)-m_c\}~,\nonumber \\ \nonumber \\
\rho^{V-A}_2&=&2N_c~I_0(s,s^{\prime },q^2)\{2A_2(m_b-%
m_c)-2A_3(m_b-m_c)-B_1(m_b+m_c)
\nonumber \\
&&+B_2(m_c+m_{s})+m_c\}~.
\end{eqnarray}
Then, the spectral densities related to
$\sigma_{\mu\nu}q^{\nu}(1+\gamma_{5})$ vertex are presented as:
\begin{eqnarray}\label{eq18}
\rho_{V}^{T-PT}(s,s',q^2)&=&2N_c~I_0(s,s',q^2)~\{m_c(m_b-m_s)+B_1~[\Delta-m_c m_s-m_c^2+m_b(m_c+m_s)-s]\nonumber\\
&+&B_2~[\Delta'-m_c m_s-m_c^2+m_b(m_c+m_s)-s']~\},\nonumber\\ \nonumber \\
\rho_{0}^{T-PT}(s,s',q^2)&=&2N_c~I_0(s,s',q^2)~\{2A_1(2s-u)-\Delta~[m_b(m_c+m_s)-m_c(m_c+m_s)+s']\nonumber\\
&+&\Delta'~[m_b(m_c+m_s)-m_c(m_c+m_s)+s]
-2m_c(m_{b}-m_{c})s'\nonumber\\&-&2m_c(m_c+m_s)s +2
m_c(m_{b}+m_s)u\},\nonumber \\ \nonumber \\
\rho_{1}^{T-PT}(s,s',q^2)&=&2N_c~I_0(s,s',q^2)~\{m_c(m_b+m_s)
+B_2~[(m_b-m_c)(m_c+m_s)+s']\nonumber\\
&+&B_1~[\Delta-\Delta'
-m_b(m_c+m_{s})+m_c(m_c+m_s)-s]\nonumber \\
&+&(A_2-A_3)(2s-u)\},
\end{eqnarray}
where
\begin{eqnarray*}\label{19}
I_{0}(s,s',q^2)&=&\frac{1}{4\lambda^{1/2}(s,s',q^2)},\nonumber\\
\lambda(a,b,c)&=&a^2+b^2+c^2-2ac-2bc-2ab,\nonumber \\
\Delta'&=&(s'+m_{c}^2 - m_{s}^2),\nonumber\\
\Delta&= &(s+m_{c}^2 -
m_{b}^2),\nonumber\\
u &=& s + s' - q^2,\nonumber\\
B_{1}&=&\frac{1}{\lambda(s,s',q^2)}[2s'\Delta-\Delta'u],\nonumber\\
B_{2}&=&\frac{1}{\lambda(s,s',q^2)}[2s\Delta'-\Delta u],\nonumber\\
A_{1}&=&-\frac{1}{2\lambda(s,s',q^2)}[(4ss'm_c^2-s\Delta'^2
-s'\Delta^2-u^2m_c^2+u\Delta\Delta')]
,\nonumber\\
A_{2}&=&-\frac{1}{\lambda^2(s,s',q^2)}[8ss'^2m_c^2-2ss'\Delta'^2-6s'^2\Delta^2
-2u^2s'm_c^2\nonumber\\
&+&6s'u\Delta\Delta'-u^2\Delta'^2]
,\nonumber\\%
A_{3}&=&\frac{1}{\lambda^{2}(s,s',q^2)}[4ss'um_c^2+4ss'\Delta\Delta'
-3su\Delta'^2
-3u\Delta^2s'-u^3m_c^2+2u^2\Delta\Delta'].\nonumber\\
\end{eqnarray*}
and $N_c=3$ is the color factor.

The integration region in Eq. (\ref{eq13}) is obtained
requiring that the argument of three delta vanish, simultaneously.
The physical region in the $s$ and $s^{\prime}$ plane is described
by the following inequalities:
\begin{equation}\label{eq15}
 -1\leq\frac{2ss'+(s+s'-q^2)(m_{b}^2-s-m_{c}^2)+(m_{c}^2-m_{s}^2)2s}
 {\lambda^{1/2}(m_{b}^2,s,m_{c}^2)\lambda^{1/2}(s,s',q^2)}\leq+1.
\end{equation}
 From this
inequality, to use in the lower limit of the integration over $s$
in continuum subtractions, it is easy to express $s$ in terms of $s'$, i.e.,
$s_{L}$ is as follow:
\begin{eqnarray}\label{eq16}
s_{L}=\frac{(m_c^2+q^2-m_b^2-s^{\prime})(m_b^2 s^{\prime}-q^2
m_c^2)}{(m_b^2-q^2)(m_c^2-s^{\prime})}~.
\end{eqnarray}

\section{Gluon condensate contribution}

\begin{figure}[th]
\vspace*{0.0cm}
\begin{center}
\begin{picture}(160,100)
\centerline{ \epsfxsize=14cm \epsfbox{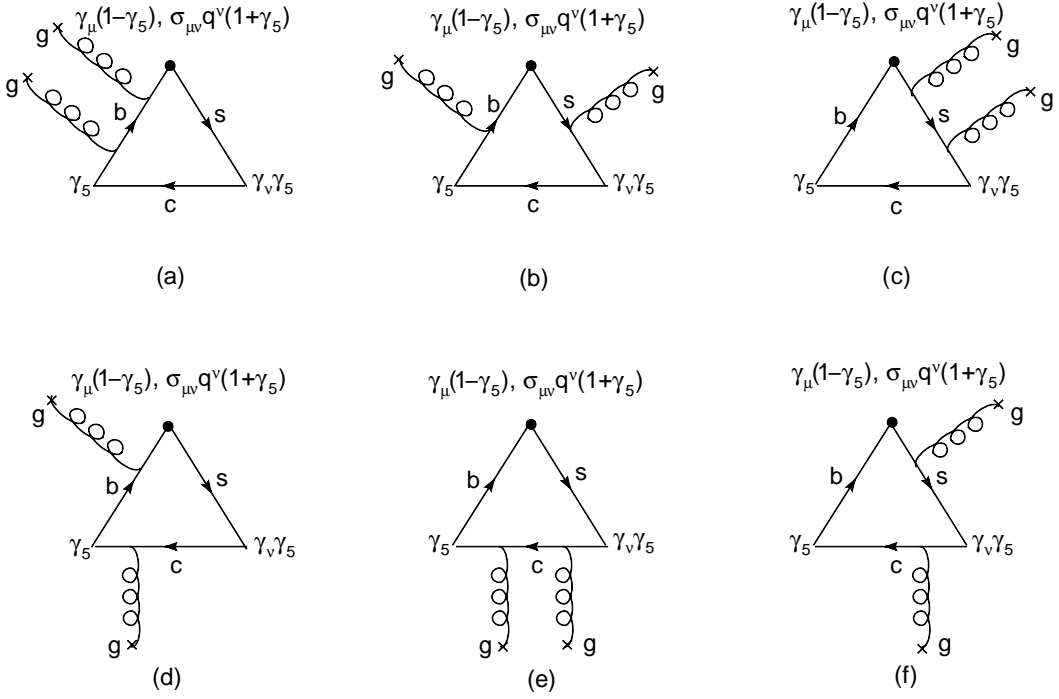}}
\end{picture}
\end{center}
\vspace*{-0.1cm} \caption{Contribution of gluon condensates for
$B_c \to D_{s1}$ transition.} \label{F2}
\end{figure}

In this section, the non-perturbative part contributions to the
correlation function are discussed. Here, we will follow the same procedure as stated in \cite{Azizi3,Azizi1,Azizi2,Azizi4}.   The non-perturbative part contains the quark and gluon condensate diagrams. For this aim, we
consider the condensate terms of dimension $3, 4$ and $ 5$.  It's found that
the heavy quark condensate contributions are suppressed by inverse
of the heavy quark mass and can be safely omitted. The light $s$
quark condensate contribution is zero after applying the double
Borel transformation with respect to the both variables $p^2$ and
${p^{'}}^2$, because only one variable appears in the denominator.
 Therefore in this case, we consider the two gluon condensate diagrams with mass dimension $4$ as non-perturbative corrections. The diagrams for contribution of the gluon condensates are
depicted in Fig. \ref{F2}.
\normalsize To obtain the contributions of these diagrams, the
Fock-Schwinger fixed-point gauge, $x^\mu A_\mu^a=0$, are used,
where $A_\mu^a$ is the gluon field. In the evaluation of
diagrams in Fig. \ref{F2}, integrals of the following types are encountered.
\begin{eqnarray}\label{eq23}
I_0(a,b,c) \es \int \frac{d^4k}{(2 \pi)^4} \frac{1}{\left[
k^2-m_c^2 \right]^a \left[ (p+k)^2-m_b^2 \right]^b \left[
(p^\prime+k)^2-m_{s}^2\right]^c}~,
\nnb \\ \nnb \\
I_\mu(a,b,c) \es \int \frac{d^4k}{(2 \pi)^4} \frac{k_\mu}{\left[
k^2-m_c^2 \right]^a \left[ (p+k)^2-m_b^2 \right]^b \left[
(p^\prime+k)^2-m_{s}^2\right]^c}~,
\nnb \\ \nnb \\
I_{\mu\nu}(a,b,c) \es \int \frac{d^4k}{(2 \pi)^4} \frac{k_\mu
k_\nu}{\left[ k^2-m_c^2 \right]^a \left[ (p+k)^2-m_b^2 \right]^b
\left[ (p^\prime+k)^2-m_{s}^2\right]^c}~.
\end{eqnarray}
These integrals can be calculated  using the Schwinger representation for the Euclidean
propagator
\begin{eqnarray}\label{eq24}
\frac{1}{(k^2+m^2)^n} = \frac{1}{\Gamma(n)} \int_0^\infty d\alpha
\, \alpha^{n-1} e^{-\alpha(k^2+m^2)}~,
\end{eqnarray}
After Borel transformation using:
\begin{eqnarray}\label{eq25}
{\cal B}_{\hat{p}^2} (M^2) e^{-\alpha p^2} = \delta
(1/M^2-\alpha)~,
\end{eqnarray}
we obtain
\begin{eqnarray}\label{eq26}
\hat{I}_{0}(a,b,c)\!\!\! &=&\!\!\!\frac{(-1)^{a+b+c}}{16\pi
^{2}\,\Gamma
(a)\Gamma (b)\Gamma (c)}(M_{1}^{2})^{2-a-b}(M_{2}^{2})^{2-a-c}\,\mathcal{U}%
_{0}(a+b+c-4,1-c-b)~, \nnb \\ \nnb \\
\hat{I}_{\mu }(a,b,c) &=&\hat{I}_{1}(a,b,c)p_{\mu
}+\hat{I}_{2}(a,b,c)p_{\mu
}^{\prime }~, \nnb \\ \nnb \\
\hat{I}_{\mu \nu }(a,b,c) &=&\hat{I}_{6}(a,b,c)g_{\mu \nu }+\hat{I}%
_{3}(a,b,c)p_{\mu }p_{\nu }+\hat{I}_{4}(a,b,c)p_{\mu }p_{\nu }^{\prime }+\hat{I}%
_{4}(a,b,c)p_{\mu }^{\prime }p_{\nu }+\hat{I}_{5}(a,b,c)p_{\mu
}^{\prime }p_{\nu }^{\prime }~.\nnb \\
\end{eqnarray}
$\hat{I}$ in Eq. (\ref{eq26}) stands for double Borel transformed
form of Eq. (\ref{eq23}). In Schwinger representation:
\begin{eqnarray}\label{eq27}
\hat{I}_k(a,b,c) \es i \frac{(-1)^{a+b+c+1}}{16 \pi^2\,\Gamma(a)
\Gamma(b) \Gamma(c)}
(M_1^2)^{1-a-b+k} (M_2^2)^{4-a-c-k} \, {\cal U}_0(a+b+c-5,1-c-b)~, \nnb \\ \nnb \\
\hat{I}_m(a,b,c) \es i \frac{(-1)^{a+b+c+1}}{16 \pi^2\,\Gamma(a)
\Gamma(b) \Gamma(c)}
(M_1^2)^{-a-b-1+m} (M_2^2)^{7-a-c-m} \, {\cal U}_0(a+b+c-5,1-c-b)~, \nnb \\ \nnb \\
\hat{I}_6(a,b,c) \es i \frac{(-1)^{a+b+c+1}}{32 \pi^2\,\Gamma(a)
\Gamma(b) \Gamma(c)} (M_1^2)^{3-a-b} (M_2^2)^{3-a-c} \, {\cal
U}_0(a+b+c-6,2-c-b)~.
\end{eqnarray}
where $k=1,2$, $m=3,4,5$, $M_1^2$ and $M_2^2$ are the Borel
parameters in the $s$ and $s^\prime$ channel, respectively, and
the function ${\cal U}_0(a,b)$ is defined as
\begin{eqnarray}\label{eq28}
{\cal U}_0(a,b) = \int_0^\infty dy (y+M_1^2+M_2^2)^a y^b
\,exp\left[ -\frac{B_{-1}}{y} - B_0 - B_{1}y \right]~, \nnb
\end{eqnarray}
where
\begin{eqnarray}\label{eq29}
B_{-1} \es \frac{1}{M_1^2M_2^2} \left[m_{s}^2M_1^4+m_b^2 M_2^4 +
M_2^2M_1^2 (m_b^2+m_{s}^2
-q^2) \right] ~, \nnb \\
B_0 \es \frac{1}{M_1^2 M_2^2} \left[ (m_{s}^2+m_c^2) M_1^2 + M_2^2
(m_b^2+m_c^2)
\right] ~, \nnb \\
B_{1} \es \frac{m_c^2}{M_1^2 M_2^2}~.
\end{eqnarray}
Performing the double Borel transformation over the variables
$p^2$ and $p^{'2}$ on the physical as well as perturbative parts of the correlation
functions and equating the coefficients of the selected structures from both sides, the sum rules for
the form factors $A_{i}^{'B_c\to D_{s1}}$ are obtained:
\begin{eqnarray}\label{eq30}
A_{i}^{'B_c\to D_{s1}}&=&-\frac{(m_b+m_c)}{f_{B_c} m_{B_c}^2
f_{D_{s1}} m_{D_{s1}}} e^\frac{m_{B_c}^2}{M_1^2}
e^\frac{m_{D_{s1}}^2}{M_2^2}\Bigg\{-\frac{1}{4 \pi^2}
\int_{m_c^2}^{s_0^\prime} ds^\prime \int_{s_L}^{s_0}
\rho_{i}^{V-A} (s,s^\prime,q^2) e^\frac{-s}{M_1^2}
e^\frac{-s^{'}}{M_2^2} \nnb \\
&-& iM^{2}_{1}M^{2}_{2} \lla \frac{\alpha_s}{\pi} G^2 \rra
\frac{C_{i}^{V-A}}{6} \Bigg\}~,
\end{eqnarray}
where $i=V, 0, 1, 2$ and for the
form factors $T_{j}^{'B_c\to D_{s1}}$, we get
\begin{eqnarray}\label{eq31}
T_{j}^{'B_c\to D_{s1}}&=&-\frac{(m_b+m_c)}{f_{B_c} m_{B_c}^2
f_{D_{s1}} m_{D_{s1}}} e^\frac{m_{B_c}^2}{M_1^2}
e^\frac{m_{D_{s1}}^2}{M_2^2}\Bigg\{-\frac{1}{4 \pi^2}
\int_{m_c^2}^{s_0^\prime} ds^\prime \int_{s_L}^{s_0}
\rho_{j}^{T-PT} (s,s^\prime,q^2) e^\frac{-s}{M_1^2}
e^\frac{-s^{'}}{M_2^2} \nnb \\
&-& iM^{2}_{1}M^{2}_{2} \lla \frac{\alpha_s}{\pi} G^2 \rra
\frac{C_{j}^{T-PT}}{6} \Bigg\}~.
\end{eqnarray}
where $j=V, 0, 1$. The  $s_0$ and $s_0^{'}$ are the continuum
thresholds in $B_c$ and  $D_{s1}$
channels, respectively and lower bound
$s_{L}$ in the integrals is given in Eq. (\ref{eq16}). We present the explicit
expressions of the coefficients  $C_{i(j)}^{V-A(T-PT)}$ correspond to gluon condensates in Appendix--A.

Now,  the
$A_{i}^{'B_{c}\to D_{s1}1(2)}$ and $T_{j}^{'B_{c}\to D_{s1}1(2)}$
form factors are obtained from the above equations  replacing the
$f_{D_{s1}}$ by decay constant $f_{D_{s1}1(2)}$, and $m_{D_{s1}}$
with $m_{D_{s1}1(2)}$, i.e.,
\begin{eqnarray}\label{eq32}
A_{i}^{'B_c\to D_{s1}1(2)}&=&-\frac{(m_b+m_c)}{f_{B_c} m_{B_c}^2
f_{D_{s1}1(2)} m_{D_{s1}1(2)}} e^\frac{m_{B_c}^2}{M_1^2}
e^\frac{m_{D_{s1}1(2)}^2}{M_2^2}\Bigg\{-\frac{1}{4 \pi^2}
\int_{m_c^2}^{s_0^\prime} ds^\prime \int_{s_L}^{s_0}
\rho_{i}^{V-A} (s,s^\prime,q^2) e^\frac{-s}{M_1^2}
e^\frac{-s^{'}}{M_2^2} \nnb \\
&-& iM^{2}_{1}M^{2}_{2} \lla \frac{\alpha_s}{\pi} G^2 \rra
\frac{C_{i}^{V-A}}{6} \Bigg\}~,
\end{eqnarray}
and
\begin{eqnarray}\label{eq33}
T_{j}^{'B_c\to D_{s1}1(2)}&=&-\frac{(m_b+m_c)}{f_{B_c} m_{B_c}^2
f_{D_{s1}1(2)} m_{D_{s1}1(2)}} e^\frac{m_{B_c}^2}{M_1^2}
e^\frac{m_{D_{s1}1(2)}^2}{M_2^2}\Bigg\{-\frac{1}{4 \pi^2}
\int_{m_c^2}^{s_0^\prime} ds^\prime \int_{s_L}^{s_0}
\rho_{i}^{T-PT} (s,s^\prime,q^2) e^\frac{-s}{M_1^2}
e^\frac{-s^{'}}{M_2^2} \nnb \\
&-& iM^{2}_{1}M^{2}_{2} \lla \frac{\alpha_s}{\pi} G^2 \rra
\frac{C_{i}^{T-PT}}{6} \Bigg\}~.
\end{eqnarray}
Using the straight forward calculations, the form factors of the
$A_{i}^{B_{c}\to D_{s1}(2460)}$ and $T_{j}^{B_{c}\to
D_{s1}(2460)}$ are found as follows:
\begin{eqnarray}\label{eq34}
A_0^{B_{c}\to
D_{s1}(2460)}&=&\left(\frac{m_{B_c}+m_{D_{s1}1}}{m_{B_c}+m_{D_{s1}}}\right)A_0^{B_{c}\to
D_{s1}1}~sin\theta_{s}+\left(\frac{m_{B_c}+m_{D_{s1}2}}{m_{B_c}+m_{D_{s1}}}\right)A_0^{B_{c}\to
D_{s1}2}~cos\theta_{s}~,\nonumber\\
A_{i'}^{B_{c}\to
D_{s1}(2460)}&=&\left(\frac{m_{B_c}+m_{D_{s1}}}{m_{B_c}+m_{D_{s1}1}}\right)~A_{i'}^{B_{c}\to
D_{s1}1}~sin\theta_{s}+\left(\frac{m_{B_c}+m_{D_{s1}}}{m_{B_c}+m_{D_{s1}2}}\right)~A_{i'}^{B_{c}\to
D_{s1}2}~cos\theta_{s}~,\nonumber\\
T_0^{B_{c}\to
D_{s1}(2460)}&=&\left(\frac{m_{B_c}+m_{D_{s1}1}}{m_{B_c}+m_{D_{s1}}}\right)~T_0^{B_{c}\to
D_{s1}1}~sin\theta_{s}+\left(\frac{m_{B_c}+m_{D_{s1}2}}{m_{B_c}+m_{D_{s1}}}\right)~T_0^{B_{c}\to
D_{s1}2}~cos\theta_{s}~,\nonumber\\
T_{j^{\prime}}^{B_{c}\to D_{s1}(2460)}&=&T_{j'}^{B_{c}\to
D_{s1}1}~sin\theta_{s}+T_{j'}^{B_{c}\to D_{s1}2}~cos\theta_{s}~.
\end{eqnarray}
where $i'=V, 1, 2$ and $j'=V, 1$. Note that, the $A_{i}^{B_{c}\to
D_{s1}(2536)}$ and $T_{j}^{B_{c}\to D_{s1}(2536)}$ form factors
are obtained from the above equations by replacing the
$sin{\theta_s}\rightarrow cos{\theta_s}$ and $cos{\theta_s}\rightarrow -sin{\theta_s}$.

\section{Decay widths }

Now, we present the dilepton invariant mass distribution for the
$B_c \rar D_{s1}\nu\bar{\nu}$ and $B_c \rar D_{s1}l\bar{l}$
decays. Using the parameterization of the $B_c \to D_{s1}$
transition in terms of form factors and also  Eq. (\ref{eq3}), the
dilepton invariant mass distribution of the $B_c \rar
D_{s1}\nu\bar{\nu}$ decay can be written as \cite{Geng1}:
\begin{eqnarray}\label{eq35}
\frac{d\Gamma \left( B_c\to D_{s1} \nu \bar{\nu }\right) }{dq^2}
&=&\frac{3G_F^2m_{B_c}^3|V_{tb}V_{ts}^\ast|^2\alpha^2|D\left(
x_t\right) |^2} {2^8\pi ^5sin^4\theta_W} \phi_{_{D_{s1}}}^{1\over
2}\left[ s\alpha _1+ {\phi_{_{D_{s1}}}\over 3} \beta _1\right] ,
\end{eqnarray}
where $s=q^2/m_{B_c}^2$, $x_t=m_t^2/m_W^2$. The parameters
$D(x_t)$, $\phi_{_{D_{s1}}}$, $\alpha_1$ and  $\beta_1$  are defined by

\begin{eqnarray}\label{eq36}
D(x_t)&=&\frac{x_t}{8}\left(\frac{2+x_t}{x_t-1}+\frac{3x_t-6}{(x_t-1)^2}~\ln
x_t \right) ,
\nonumber \\
\phi_{_{D_{s1}}} &=&\left( 1-r_{{D_{s1}}}\right) ^{2}-2s\left(
1+r_{{D_{s1}}}\right) +s^{2},
\nonumber \\
\alpha _1 &=& (1-\sqrt{r_{D_{s1}}}) ^{2} \left| A^{B_c\to
D_{s1}}_{0}\right| ^{2} + \frac{\phi_{_{D_{s1}}}
}{(1+\sqrt{r_{D_{s1}}})^{2}}\left| A^{B_c\to D_{s1}}_V\right| ^{2}
\ ,
\nonumber \\
\beta _1 &=& \frac{(1-\sqrt{r_{D_{s1}}})^2}{4r_{D_{s1}}} \left|
A^{B_c\to D_{s1}}_{0} \right|^2
-\frac{s}{(1+\sqrt{r_{D_{s1}}})^2}\left| A^{B_c\to
D_{s1}}_V\right| ^2 +\frac{\phi _{_{D_{s1}}} \left| A^{B_c\to
D_{s1}}_1\right| ^2}{4r_{D_{s1}}(1+\sqrt{r_{D_{s1}}})^2}
\nonumber \\
&&+\frac{1}{2}\left( \frac{1-s}{r_{D_{s1}}}-1\right)
\frac{1-\sqrt{r_{D_{s1}}}}{1+\sqrt{r_{D_{s1}}}}Re(A^{B_c\to
D_{s1}}_{0}A_{1}^{*B_c\to D_{s1}})\,,
\end{eqnarray}
where $r_{{D_{s1}}}=m_{{D_{s1}}}^2/m_{B_c}^2$. The differential
decay rates for $B_c\to D_{s1} l \bar{l }$ are found to be
\cite{Breub,Geng2}:
\begin{eqnarray}\label{eq37}
\frac{d\Gamma \left( B_c^+\rightarrow D_{s1}l^{+}l^{-}\right)
}{dq^2}
&=&\frac{G_F^2m_{B_c}^3|V_{tb}V_{ts}^\ast|^2\alpha^2}{2^9\pi ^5}
v~\phi_{_{D_{s1}}} ^{1\over 2} \left[ \Bigg(
1+\frac{2m_l^2}{q^2}\Bigg) \Bigg( s~\alpha _3+\frac{\phi
_{_{D_{s1}}}}{3}\beta _3 \Bigg) +4t~\delta \right],\nnb\\
\end{eqnarray}
where $t= m_l^2/ m_{B_c}^2$, $v=\sqrt{1-4m_l^2/q^2}$ and the
expressions of $\alpha_3$, $\beta_3$ and $\delta$ are given as:
\begin{eqnarray}\label{eq38}
\alpha _3 &=& (1-\sqrt{r_{{D_{s1}}}})^2\left[ \Bigg|
C_9^{eff}A^{B_c\to D_{s1}}_0
-\frac{2\hat{m}_bC_7^{eff}(1+\sqrt{r_{{D_{s1}}}})T^{B_c\to
D_{s1}}_0}{s}\Bigg|^2 +\left| C_{10}A^{B_c\to D_{s1}}_0\right|
^2\right]
\nonumber \\
&&+\frac{\phi _{_{D_{s1}}}}{(1+\sqrt{r_{{D_{s1}}}})^2}\left[ \
\Bigg| C_9^{eff}A^{B_c\to D_{s1}}_V
-\frac{2\hat{m}_bC_7^{eff}(1+\sqrt{r_{{D_{s1}}}})T^{B_c\to
D_{s1}}_V}{s} \Bigg| ^2 +\left| C_{10}A^{B_c\to D_{s1}}_V\right|
^2\right] \,, \nonumber\\ \\
\beta _3 &=&
\frac{(1-\sqrt{r_{{D_{s1}}}})^2}{4r_{{D_{s1}}}}\left[ \Bigg|
C_9^{eff}A^{B_c\to D_{s1}}_0
-\frac{2\hat{m}_bC_7^{eff}(1+\sqrt{r_{{D_{s1}}}})T^{B_c\to
D_{s1}}_0}{s} \Bigg| ^2 +\left| C_{10}A^{B_c\to D_{s1}}_0\right|
^2\right]
\nonumber \\
&&-\frac{s}{(1+\sqrt{r_{{D_{s1}}}})^2}\left[ \Bigg|
C_9^{eff}A^{B_c\to D_{s1}}_V
-\frac{2\hat{m}_bC_7^{eff}(1+\sqrt{r_{{D_{s1}}}})T^{B_c\to
D_{s1}}_V}{s} \Bigg| ^2 +\left| C_{10}A^{B_c\to D_{s1}}_V\right|
^2\right]
\nonumber \\
&&+\frac{\phi
_{_{D_{s1}}}}{4r_{{D_{s1}}}(1+\sqrt{r_{{D_{s1}}}})^2}\left[ \Bigg|
C_9^{eff}A^{B_c\to
D_{s1}}_1-\frac{2\hat{m}_bC_7^{eff}(1+\sqrt{r_{{D_{s1}}}})T^{B_c\to
D_{s1}}_1}{s} \Bigg| ^2 +\left| C_{10}A^{B_c\to D_{s1}}_1\right|
^2\right]
\nonumber \\
&&+\frac{1}{2}\left(\frac{1-s}{r_{{D_{s1}}}}-1\right)\frac{1-\sqrt{r_{{D_{s1}}}}}{1+\sqrt{r_{{D_{s1}}}}}
Re\left\{ \Bigg[ C_9^{eff}A^{B_c\to
D_{s1}}_0-\frac{2\hat{m}_bC_7^{eff}(1+\sqrt{r_{{D_{s1}}}})T^{B_c\to
D_{s1}}_0}{s} \Bigg] \right.
\nonumber \\
&&\left. \times  \Bigg[ C_9^{eff}A^{B_c\to
D_{s1}}_1-\frac{2\hat{m}_bC_7^{eff}(1+\sqrt{r_{{D_{s1}}}})T^{B_c\to
D_{s1}}_1}{s} \Bigg] +|C_{10}|^2Re(A^{B_c\to D_{s1}}_0A^{*B_c\to
D_{s1}}_1)\right\}.
\end{eqnarray}
and
\begin{eqnarray}\label{eq39}
\delta &=& \frac{|C_{10}|^2}{2(1+\sqrt{r_{{D_{s1}}}})^2}\Bigg\{
-2\phi _{_{D_{s1}}}|A^{B_c\to
D_{s1}}_V|^2-3(1-r_{{D_{s1}}})^2|A^{B_c\to
D_{s1}}_0|^2\nonumber \\
&&+\frac{\phi _{_{D_{s1}}}}{4r_{{D_{s1}}}}\left[
2(1+r_{{D_{s1}}})-s\right]|A^{B_c\to D_{s1}}_1|^2 +\frac{\phi
_{_{D_{s1}}}s}{4r_{{D_{s1}}}}|A^{B_c\to
D_{s1}}_2|^2\nonumber \\
&&+\frac{\phi _{_{D_{s1}}}(1-r_{{D_{s1}}})}{2r_{{D_{s1}}}}
Re\left( A^{B_c\to D_{s1}}_0A^{*B_c\to D_{s1}}_1+A^{B_c\to
D_{s1}}_0A^{*B_c\to D_{s1}}_2+A^{B_c\to D_{s1}}_1{A^{*B_c\to
D_{s1}}_2}\right) \Bigg\},\nonumber \\
\end{eqnarray}
where $\hat{m}_b=m_b/m_{B_c}$.

\section{Numerical analysis}
In this section, we present our numerical analysis of the form factors $%
A_{i}~,(i=V, 0, 1, 2)$ and $T_{j}~,(j=V, 0, 1)$. From the sum
rules expressions of the form factors, it is clear that the main
input parameters entering the expressions are gluon condensates,
elements of the CKM matrix $V_{tb}$ and $V_{ts}$, leptonic decay
constants $f_{B_c}$, $f_{D_{s1}}$, $f_{D_{s1}1}$ and
$f_{D_{s1}2}$, Borel parameters $M_{1}^2$ and $M_{2}^2$ as well as
the continuum thresholds $s_{0}$ and $s'_{0}$. We choose the
values of the condensates (at a fixed renormalization scale of
about $1~GeV$), leptonic decay constants , CKM matrix elements,
quark and meson masses as:
$\langle\frac{\alpha_s}{\pi}G^2\rangle=0.012~GeV^4$
\cite{Shifman}, $\mid V_{tb}\mid=0.77^{+0.18}_{-0.24}$, $\mid
V_{ts}\mid=(40.6\pm2.7)\times10^{-3}$ \cite{Ceccucci},
$C_7^{eff}=-0.313$, $C_9^{eff}=4.344$, $C_{10}=-4.669$
\cite{Buras,Bashiry}, $f_{D_{s1}}=(225\pm25)~MeV$,
$f_{D_{s1}1}=(240\pm25)~MeV$, $f_{D_{s1}2}=(63\pm7)~MeV$
\cite{Thomas,Veseli}, $f_{B_c}=(350\pm25)~MeV$
\cite{Colangelo1,Kiselev2,Aliev1},
$m_{s}(1~GeV)=(104^{+26}_{-34})~MeV$,
$m_{c}=\left(1.27^{+0.07}_{-0.11}\right)~GeV$, $m_{b}=(4.7\pm
0.07)~GeV$, $m_{D_{s1}(2460)}=(2459.6\pm 0.6)~MeV$,
$m_{D_{s1}(2536)}=(2535.35\pm 0.34\pm 0.5)~MeV$ and
$m_{B_c}=(6.276\pm0.004)~GeV$ \cite{Yao}.

The sum rules for the form factors contain also four auxiliary
parameters: Borel mass squares $M_{1}^2$ and $M_{2}^2$ and
continuum thresholds $s_{0}$ and $s'_{0}$. These are not physical
quantities, so the the form factors as physical quantities should
be independent of them. The parameters $s_0$ and $s_0^\prime$,
which are the continuum thresholds of $B_c$ and $D_{s1}$ mesons,
respectively, are determined from the condition that guarantees
the sum rules to practically be stable in the allowed regions for
$M_1^2$ and $M_2^2$. The values of the continuum thresholds
calculated from the two--point QCD sum rules are taken to be
$s_0=(45-50)~GeV^2$ and $s_0^\prime=(6-8)~GeV^2$
\cite{Aliev2,Shifman,Colangelo2}. The working regions for $M_1^2$
and $M_2^2$ are determined requiring that not only the
contributions of the higher states and continuum are small, but
the contributions of the operators with higher dimensions are also
small. Both conditions are satisfied in the regions $10~GeV^2 \le
M_1^2 \le 25~GeV^2$ and $4~GeV^2 \le M_2^2 \le 10~GeV^2$. First,
we would like to consider the $D_{s1}$ meson as the pure
$|c\bar{s}\rangle$ state. The values of the form factors  at
$q^2=0$ are presented in Table \ref{T22}.
\begin{table}[th]
\centering
\begin{tabular}{|c|c||c|c|}
\cline{1-4}\lower0.35cm \hbox{{\vrule width 0pt height 1.0cm }}
$A_{V}^{B_c\to D_{s1}(2460)}(0)$ &$-0.23\pm0.07$&$A_{V}^{B_c\to D_{s1}(2536)}(0)$&$-0.22\pm0.06$\\
\cline{1-4}\lower0.35cm \hbox{{\vrule width 0pt height 1.0cm }}
$A_{0}^{B_c\to D_{s1}(2460)}(0)$ &$0.09\pm0.02$&$A_{0}^{B_c\to D_{s1}(2536)}(0)$&$0.07\pm0.02$\\
\cline{1-4}\lower0.35cm \hbox{{\vrule width 0pt height 1.0cm }}
$A_{1}^{B_c\to D_{s1}(2460)}(0)$ &$0.16\pm0.05$&$A_{1}^{B_c\to D_{s1}(2536)}(0)$&$0.17\pm0.05$\\
\cline{1-4}\lower0.35cm \hbox{{\vrule width 0pt height 1.0cm }}
$A_{2}^{B_c\to D_{s1}(2460)}(0)$ &$-0.26\pm0.08$&$A_{2}^{B_c\to D_{s1}(2536)}(0)$&$-0.28\pm0.09$\\
\cline{1-4}\lower0.35cm \hbox{{\vrule width 0pt height 1.0cm }}
$T_{V}^{B_c\to D_{s1}(2460)}(02)$ &$0.12\pm0.03$&$T_{V}^{B_c\to D_{s1}(2536)}(0)$&$0.14\pm0.04$\\
\cline{1-4}\lower0.35cm \hbox{{\vrule width 0pt height 1.0cm }}
$T_{0}^{B_c\to D_{s1}(2460)}(0)$ &$0.11\pm0.03$&$T_{0}^{B_c\to D_{s1}(2536)}(0)$&$0.14\pm0.04$\\
\cline{1-4}\lower0.35cm \hbox{{\vrule width 0pt height 1.0cm }}
$T_{1}^{B_c\to D_{s1}(2460)}(0)$ &$-0.14\pm0.04$&$T_{1}^{B_c\to D_{s1}(2536)}(0)$&$-0.16\pm0.05$\\
\cline{1-4}
\end{tabular}\label{T22}
\vspace{0.01cm} \caption{The value of the form factors of the  $B_{c}\to D_{s1}(2460)$ and $B_{c}\to
D_{s1}(2536)$ transitions at $q^2=0$, $M_1^2=15~GeV^2$ and $M_2^2=8~GeV^2$, when the  $D_{s1}$ meson are considered as the pure
$|c\bar{s}\rangle$ state
.}\label{T22}
\end{table}
The sum rules for the form factors are truncated at about $10
~GeV^2$, so to extend our results to the full physical region, $0 \leq q^2 \leq (m_{B_c}-m_{D_{s1}})^2~ GeV^2$, we
look for a parameterization of the form factors in such a way that
in the region $0 \leq q^2 \leq 10~ GeV^2$,
this parameterization coincides with the sum rules predictions. Our
numerical calculations show that the sufficient parameterization
of the form factors with respect to $q^2$ is as follows:
\begin{equation}\label{eq40}
f_{i}(q^2)=\frac{a}{(1- \frac{q^2}{m^{2}_{fit}})}+ \frac{b}{(1-
\frac{q^2}{m^{2}_{fit}})^2}.
\end{equation}
The values of the parameters $a, b$ and  $m_{fit}$ are
given in the Table \ref{T2}.
\begin{table}[th]
\centering
\begin{tabular}{c|ccc||c|ccc}
\cline{2-4}\cline{6-8}&$a$&$b$&$m_{fit}$&&$a$&$b$&$m_{fit}$\\
\cline{1-8}\lower0.35cm \hbox{{\vrule width 0pt height 1.0cm }}
$A_{V}^{B_c\to D_{s1}(2460)}(q^2)$ &-0.13&-0.10&5.30&$A_{V}^{B_c\to D_{s1}(2536)}(q^2)$&-0.12&-0.10&5.22\\
\cline{1-8}\lower0.35cm \hbox{{\vrule width 0pt height 1.0cm }}
$A_{0}^{B_c\to D_{s1}(2460)}(q^2)$ &0.05&0.04&5.98&$A_{0}^{B_c\to D_{s1}(2536)}(q^2)$&0.04&0.03&5.99\\
\cline{1-8}\lower0.35cm \hbox{{\vrule width 0pt height 1.0cm }}
$A_{1}^{B_c\to D_{s1}(2460)}(q^2)$ &0.09&0.07&5.95&$A_{1}^{B_c\to D_{s1}(2536)}(q^2)$&0.09&0.08&5.98\\
\cline{1-8}\lower0.35cm \hbox{{\vrule width 0pt height 1.0cm }}
$A_{2}^{B_c\to D_{s1}(2460)}(q^2)$ &-0.16&-0.10&5.15&$A_{2}^{B_c\to D_{s1}(2536)}(q^2)$&-0.17&-0.11&5.17\\
\cline{1-8}\lower0.35cm \hbox{{\vrule width 0pt height 1.0cm }}
$T_{V}^{B_c\to D_{s1}(2460)}(q^2)$ &0.08&0.04&5.22&$T_{V}^{B_c\to D_{s1}(2536)}(q^2)$&0.09&0.05&5.05\\
\cline{1-8}\lower0.35cm \hbox{{\vrule width 0pt height 1.0cm }}
$T_{0}^{B_c\to D_{s1}(2460)}(q^2)$ &0.10&0.01&5.85&$T_{0}^{B_c\to D_{s1}(2536)}(q^2)$&0.11&0.03&5.96\\
\cline{1-8}\lower0.35cm \hbox{{\vrule width 0pt height 1.0cm }}
$T_{1}^{B_c\to D_{s1}(2460)}(q^2)$ &-0.09&-0.05&5.22&$T_{1}^{B_c\to D_{s1}(2536)}(q^2)$&-0.08&-0.07&5.14\\
\cline{1-8}
\end{tabular}\label{T2}
\vspace{0.01cm} \caption{Parameters appearing in the fit function
for the form factors of the $B_{c}\to D_{s1}(2460)$ and $B_{c}\to
D_{s1}(2536)$ transitions at $M_1^2=15~GeV^2$ and $M_2^2=8~GeV^2$
.}\label{T2}
\end{table}
To calculate the branching ratios of the $B_c\to
D_{s1}(2460[2536])l^+l^-/\nu\bar{\nu}$ decays, we integrate Eqs.
(\ref{eq35}, \ref{eq37}) over $q^2$ in the whole physical region
and use the total mean life time $\tau_{B_c}=(0.46\pm0.07)~ps$
\cite{Yao}. Our numerical analysis shows that the contribution of
the non-perturbative part (the gluon condensate diagrams ) is
about $12\%$ of the total and the main contribution comes from the
perturbative part of the form factors. The values for the
branching ratio of these decays are obtained as presented in Table
\ref{T3}, when only the short distance (SD) effects are considered.
\begin{table}[th]
\centering
\begin{tabular}{c|c||c|c}
MODS& BR & MODS & BR \\
\cline{1-4}\lower0.35cm \hbox{{\vrule width 0pt height 1.0cm }}
$B_c\to D_{s1}(2460)\nu\bar{\nu}$ &$(3.26\pm 1.10) \times 10^{-7}$&$B_c\to D_{s1}(2536)\nu\bar{\nu}$&$(2.76\pm 0.88) \times 10^{-7}$\\
\cline{1-4}\lower0.35cm \hbox{{\vrule width 0pt height 1.0cm }}
$B_c\to D_{s1}(2460)e^+e^-$ &$(5.40\pm 1.70) \times 10^{-6}$&$B_c\to D_{s1}(2536)e^+e^-$&$(2.91\pm 0.93) \times 10^{-6}$\\
\cline{1-4}\lower0.35cm \hbox{{\vrule width 0pt height 1.0cm }}
$B_c\to D_{s1}(2460)\mu^+\mu^-$ &$(2.27\pm 0.95) \times 10^{-6}$&$B_c\to D_{s1}(2536)\mu^+\mu^-$&$(1.96\pm 0.63) \times 10^{-6}$\\
\cline{1-4}\lower0.35cm \hbox{{\vrule width 0pt height 1.0cm }}
$B_c\to D_{s1}(2460)\tau^+\tau^-$ &$(1.42\pm 0.45) \times 10^{-8}$&$B_c\to D_{s1}(2536)\tau^+\tau^-$&$(0.68\pm 0.21) \times 10^{-8}$\\
\cline{1-4}
\end{tabular}
\vspace{0.01cm} \caption{The branching ratios of the semileptonic
$B_c\to D_{s1}(2460)l^+l^-/\nu\bar{\nu}$ and $B_c\to
D_{s1}(2536)l^+l^-/\nu\bar{\nu}$ decays with SD effects.}\label{T3}
\end{table}
It should be noted that, the long distance (LD) effects for the charged
lepton modes are not included in the values of Table \ref{T3}.
With the LD effects, we introduce some cuts close to $q^2=0$ and
around the resonances of $J/\psi$ and $\psi^{\prime }$ and study
the three regions as follows:
\begin{eqnarray}
I: &&\ \ \ \ \ \ \ \ \sqrt{q^2_{min}} \;\leq\; \sqrt{q^2} \;\leq\;
M_{J/\psi }-0.20,
\nonumber\\
II: && M_{J/\psi}+0.04 \;\leq\; \sqrt{q^2} \;\leq\;
M_{\psi^{\prime}}-0.10,
\nonumber \\
III: &&\ \ M_{\psi^{\prime}}+0.02 \;\leq\; \sqrt{q^2} \;\leq\;
m_{B_c}-m_{D_{s1}}. \label{eq50}
\end{eqnarray}
where $\sqrt{q^2_{min}}=2m_l$. In Table \ref{T4}, we present the
 branching ratios in terms of the regions shown in Eq.
(\ref{eq50}).
\begin{table}[th]
\centering
\begin{tabular}{c|c|c|c}
\cline{1-4}\lower0.35cm \hbox{{\vrule width 0pt height 1.0cm }}
MODS&$I$&$II$&$III$\\
\hline\hline\lower0.35cm \hbox{{\vrule width 0pt height 1.0cm }}
$BR(B_c\to D_{s1}(2460)e^+e^-)$ &$(4.40\pm 1.35)\times 10^{-6}$&$(1.62\pm 0.52)\times 10^{-7}$&$(1.01\pm 0.35)\times 10^{-7}$\\
\cline{1-4}\lower0.35cm \hbox{{\vrule width 0pt height 1.0cm }}
$BR(B_c\to D_{s1}(2536)e^+e^-)$ &$(2.65\pm 0.82)\times 10^{-6}$&$(0.90\pm 0.28)\times 10^{-7}$&$(0.81\pm 0.25)\times 10^{-7}$\\
\hline\hline\lower0.35cm \hbox{{\vrule width 0pt height 1.0cm }}
$BR(B_c\to D_{s1}(2460)\mu^+\mu^-)$ &$(1.76\pm 0.58)\times 10^{-6}$&$(1.61\pm 0.53)\times 10^{-7}$&$(1.01\pm 0.35)\times 10^{-7}$\\
\cline{1-4}\lower0.35cm \hbox{{\vrule width 0pt height 1.0cm }}
$BR(B_c\to D_{s1}(2536)\mu^+\mu^-)$ &$(1.00\pm 0.31)\times 10^{-6}$&$(0.96\pm 0.29)\times 10^{-7}$&$(0.68\pm 0.21)\times 10^{-7}$\\
\hline\hline\lower0.35cm \hbox{{\vrule width 0pt height 1.0cm }}
$BR(B_c\to D_{s1}(2460)\tau^+\tau^-)$ & undefined & $(5.20\pm 1.61)\times 10^{-9}$ &$(8.12\pm 2.75)\times 10^{-9}$\\
\cline{1-4}\lower0.35cm \hbox{{\vrule width 0pt height 1.0cm }}
$BR(B_c\to D_{s1}(2536)\tau^+\tau^-)$ & undefined & $(4.16\pm 1.29)\times 10^{-9}$ &$(6.56\pm 2.16)\times 10^{-9}$\\
\cline{1-4}
\end{tabular}
\vspace{0.01cm} \caption{The branching ratios of the semileptonic
$B_c\to D_{s1}(2460)l^+l^-/\nu\bar{\nu}$ and $B_c\to
D_{s1}(2536)l^+l^-/\nu\bar{\nu}$ decays including  LD effects.}\label{T4}
\end{table}
The errors are estimated by the variation of the Borel parameters
$M_1^2$ and $M_2^2$, the variation of the continuum thresholds
$s_0$ and $s_0^\prime$, the variation of $b$ and $c$ quark masses
and leptonic decay constants $f_{B_c}$ and $f_{D_{s1}}$.

Now, we would like to analyze the form factors obtained when we
considered the $D_{s1}$ meson as a mixture of two $|^3P_1\rangle$
and $|^1P_1\rangle$ states (see Eq. \ref{eq34}). The transition form
factors of $B_c\to D_{s1}(2460[2536])l^+l^-/\nu\bar{\nu}$ at $q^2=0$
in the interval $-180^\circ\leq\theta_{s}\leq180^\circ$ are shown in
Figs. \ref{F3} and  \ref{F4}. From these figures, we see that all
form factors have  the following common behaviors: 1) they have
extrema at the same mixing angles and 2) they come across at two
points. The dependence of the form factors $B_c\to D_{s1}(2460)$ and
$B_c\to D_{s1}(2536)$ transitions on the mixing angle, $\theta_s $,
and the transferred momentum square, $q^2$, are plotted in Figs.
\ref{F5}, \ref{F6} in the regions $0 \leq q^2 \leq
(m_{B_c}-m_{D_{s1}})^2~ GeV^2$ and $-180^\circ\leq \theta_s \leq
180^\circ$ for $q^2$ and mixing angle, respectively. Using Eqs.
\ref{eq35} and \ref{eq37} we analyze the decay widths and the
branching ratios related to considered decays. For this aim, we
denote the variations of the decay widths  with respect to $q^2$ and
$\theta_s$ in the regions $4m_l^2 \leq q^2 \leq
(m_{B_c}-m_{D_{s1}})^2~ GeV^2$, $(l=\tau, \mu, e)$ and
$-180^\circ\leq \theta_s \leq 180^\circ$  and branching ratios only
in terms of mixing angle $\theta_s$ in Figs. \ref{F7}-\ref{F10}. The
results for electron and muon are approximately the same, so we
consider $l=\mu, \tau$.  In Fig.  \ref{F12}, as an example, we only
depict the variation of the branching ratio of $B_c\to D_{s1}(2460)
\nu \bar{\nu}$ decay in terms of the mixing angle.  The figures
\ref{F9} and \ref{F10} depict a regular variation of the branching
ratios for $l^+l^-$ case with respect to the mixing angle, while we
see an irregular variation of the branching ratio of the $B_c\to
D_{s1}(2460)\nu\bar{\nu}$ transition with respect to the $\theta_s$.

In summary, We analyzed the semileptonic $B_{c}\to
D_{s1}(2460[2536])l^+l^-$, $l=e, \mu, \tau$ and $B_{c}\to
D_{s1}(2460[2536])\nu \bar{\nu}$ decays in the framework of the
three--point QCD sum rules. First, we assumed the $D_{s1}(2460)$
and $D_{s1}(2536)$ axial vector mesons as the
pure $|c\bar{s}\rangle$ states. In this case, the  related form
factors were computed. The branching ratios of these decays were
also estimated with both the short distance (SD) and long distance
(LD) effects, for the charged lepton modes. Second,
$D_{s1}(2460[2536])$ mesons were considered as a combinations of
two states $|^3P_1\rangle\equiv |D_{s1}1\rangle$ and
$|^1P_1\rangle\equiv |D_{s1}2\rangle$ with different masses and
decay constants. We evaluated the transitions form factors and the
decay widths of these decays with respect to the mixing angle
$\theta_s$ and the transferred momentum square $q^2$. The
dependence of the branching ratios on $\theta_s$ was also
presented.  Detection of these channels and their comparison with
the phenomenological models like QCD sum rules could give useful
information about the structure of the $D_{s1}$ meson and the
mixing angle $\theta_s$.

\section*{Acknowledgments}
Partial support of Shiraz university research council is
appreciated. K. A.  would like to thank T. M. Aliev, M. T. Zeyrek and A.
Ozpineci for their useful discussions and also TUBITAK, Turkish
Scientific and Technical Research Council, for their partial financial
support provided under the
project 108T502.

\newpage

\appendix
\begin{center}
{\Large \textbf{Appendix--A}}
\end{center}
\setcounter{equation}{0} \renewcommand{\theequation}

In this appendix, the explicit expressions of the coefficients of
the gluon condensate  entering  the sum rules of the form factors
$A^{'B_c\to D_{s1}}_i$, $(i=V, 0, 1, 2)$ and $T^{'B_c\to
D_{s1}}_j$, $(j=V, 0, 1)$ are given.

\begin{eqnarray*}
C_{V}^{V-A} &=&-10\,\hat{I}_{{1}}(3,2,2){m_{{b}}}^{3}{m_{{c}}}^{2}+10\,\hat{I}_{{1}}(3,2,2)%
{m_{{b}}}^{2}{m_{{c}}}^{3}+10\,\hat{I}_{{2}}(3,2,2){m_{{b}}}^{2}{m_{{c}}}%
^{3}+10\,\hat{I}_{{0}}(3,2,2){m_{{b}}}^{2}{m_{{c}}}^{3} \\
&&+10\,\hat{I}_{{1}}(3,2,2)m_{{b}}{m_{{c}}}^{4}+10\,\hat{I}_{0}^{[0,1]}(3,2,2){m_{{b}}}%
^{2}m_{{c}}-30\,\hat{I}_{{1}}(3,2,1){m_{{b}}}^{2}m_{{c}}+60\,\hat{I}_{{1}}(1,4,1){m_{{b}}%
}^{2}m_{{c}} \\
&&+60\,\hat{I}_{{2}}(1,4,1){m_{{b}}}^{2}m_{{c}}-20\,\hat{I}_{{2}}(3,2,1){m_{{b}}}^{2}m_{{%
c}}+10\,\hat{I}_{2}^{[0,1]}(3,2,2){m_{{b}}}^{2}m_{{c}}-20\,\hat{I}_{{0}}(3,2,1){m_{{b}}}%
^{2}m_{{c}} \\
&&+60\,\hat{I}_{{0}}(1,4,1){m_{{b}}}^{2}m_{{c}}+10\hat{I}_{1}^{[0,1]}(3,2,2){m_{{b}}}%
^{2}m_{{c}}+20\,\hat{I}_{{1}}(2,2,2)m_{{b}}{m_{{c}}}^{2}+10\,\hat{I}_{{2}}(3,2,1)m_{{b}}{%
m_{{c}}}^{2} \\
&&+10\,\hat{I}_{{1}}(3,2,1)m_{{b}}{m_{{c}}}^{2}+40\,\hat{I}_{{2}}(2,3,1)m_{{b}}{m_{{c}}}%
^{2}-10\,\hat{I}_{{0}}(3,2,1)m_{{b}}{m_{{c}}}^{2}+20\,\hat{I}_{{1}}(2,3,1)m_{{b}}{m_{{c}}%
}^{2} \\
&&-20\,\hat{I}_{1}^{[0,1]}(3,2,2)m_{{b}}{m_{{c}}}^{2}+30\,\hat{I}_{{1}}(4,1,1)m_{{b}}{m_{%
{c}}}^{2}-10\,\hat{I}_{{2}}(3,2,2){m_{{c}}}^{5}-10\,\hat{I}_{{1}}(3,2,2){m_{{c}}}^{5} \\
&&-10\,\hat{I}_{{0}}(3,2,2){m_{{c}}}^{5}+20\,\hat{I}_{{1}}(3,2,1){m_{{b}}}^{3}+10\,\hat{I}_{{1}%
}(2,2,2){m_{{b}}}^{3}-20\,\hat{I}_{{1}}(2,3,1){m_{{b}}}^{3} \\
&&-60\,\hat{I}_{{1}}(1,4,1){m_{{b}}}^{3}-10\,\hat{I}_{1}^{[0,1]}(3,2,2){m_{{b}}}%
^{3}-30\,\hat{I}_{{2}}(4,1,1){m_{{c}}}^{3}+20\,\hat{I}_{2}^{[0,1]}(3,2,2){m_{{c}}}^{3} \\
&&+10\,\hat{I}_{{0}}(3,2,1){m_{{c}}}^{3}-10\,\hat{I}_{{2}}(3,1,2){m_{{c}}}^{3}-20\,\hat{I}_{{0}%
}(2,2,2){m_{{c}}}^{3}-20\,\hat{I}_{{2}}(2,2,2){m_{{c}}}^{3} \\
&&-20\,\hat{I}_{{1}}(2,2,2){m_{{c}}}^{3}-30\,\hat{I}_{{0}}(4,1,1){m_{{c}}}^{3}-30\,\hat{I}_{{1}%
}(4,1,1){m_{{c}}}^{3}+20\,\hat{I}_{1}^{[0,1]}(3,2,2){m_{{c}}}^{3} \\
&&-10\,\hat{I}_{{0}}(3,1,2){m_{{c}}}^{3}+20\,\hat{I}_{0}^{[0,1]}(3,2,2){m_{{c}}}%
^{3}-50\,\hat{I}_{{1}}(2,2,1)m_{{b}}+20\,\hat{I}_{1}^{[0,1]}(2,3,1)m_{{b}} \\
&&-20\,\hat{I}_{1}^{[0,1]}(3,2,1)m_{{b}}+20\,\hat{I}_{{1}}(1,2,2)m_{{b}}+60\,\hat{I}_{{0}%
}(1,3,1)m_{{b}}-20\,\hat{I}_{{2}}(2,2,1)m_{{b}} \\
&&-20\,\hat{I}_{1}^{[0,1]}(3,1,2)m_{{b}}-20\,\hat{I}_{{0}}(2,2,1)m_{{b}}+30\,\hat{I}_{{1}%
}(2,1,2)m_{{b}}+100\,\hat{I}_{{2}}(1,3,1)m_{{b}} \\
&&+10\,\hat{I}_{1}^{[0,2]}(3,2,2)m_{{b}}-20\,\hat{I}_{1}^{[0,1]}(2,2,2)m_{{b}%
}+40\,\hat{I}_{0}^{[0,1]}(2,3,1)m_{{b}}+20\,\hat{I}_{{1}}(1,3,1)m_{{b}} \\
&&+30\,\hat{I}_{{0}}(2,2,1)m_{{c}}+30\,\hat{I}_{2}^{[0,1]}(3,1,2)m_{{c}%
}+20\,\hat{I}_{2}^{[0,1]}(3,2,1)m_{{c}}+10\,\hat{I}_{0}^{[0,1]}(3,2,1)m_{{c}} \\
&&+20\,\hat{I}_{1}^{[0,1]}(3,2,1)m_{{c}}-10\,\hat{I}_{0}^{[0,2]}(3,2,2)m_{{c}%
}+20\,\hat{I}_{1}^{[0,1]}(3,1,2)m_{{c}}+20\,\hat{I}_{1}^{[0,1]}(2,2,2)m_{{c}} \\
&&+20\,\hat{I}_{{2}}(2,2,1)m_{{c}}-30\,\hat{I}_{{2}}(2,1,2)m_{{c}}+10\,\hat{I}_{{0}}(3,1,1)m_{{%
c}}+20\,\hat{I}_{0}^{[0,1]}(2,2,2)m_{{c}} \\
&&-20\,\hat{I}_{{0}}(1,2,2)m_{{c}}-20\,\hat{I}_{{2}}(1,2,2)m_{{c}}+30%
\,\hat{I}_{0}^{[0,1]}(3,1,2)m_{{c}}-10\,\hat{I}_{{1}}(3,1,1)m_{{c}} \\
&&+20\,\hat{I}_{2}^{[0,1]}(2,2,2)m_{{c}}-10\,\hat{I}_{{2}}(3,1,1)m_{{c}}-20\,\hat{I}_{{1}%
}(2,1,2)m_{{c}}-30\,\hat{I}_{{0}}(2,1,2)m_{{c}} \\
&&-10\,\hat{I}_{2}^{[0,2]}(3,2,2)m_{{c}}+20\,\hat{I}_{{1}}(2,2,1)m_{{c}}-20\,\hat{I}_{{1}%
}(1,2,2)m_{{c}}-10\,\hat{I}_{1}^{[0,2]}(3,2,2)m_{{c}}
\end{eqnarray*}
\begin{eqnarray*}
C_{0}^{V-A} &=&-20\,\hat{I}_{{6}}(3,2,2){m_{{c}}}^{5}-40\,\hat{I}_{{6}}(3,2,1){m_{{c}}}%
^{3}-20\,\hat{I}_{{6}}(3,1,2){m_{{c}}}^{3}+40\,\hat{I}_{6}^{[0,6]}(3,2,2){m_{{c}}}^{3} \\
&&-40\,\hat{I}_{{6}}(2,2,2){m_{{c}}}^{3}-60\,\hat{I}_{{6}}(4,1,1){m_{{c}}}^{3}+5\,\hat{I}_{{0}%
}(3,1,1){m_{{c}}}^{3}-30\,\hat{I}_{0}^{[0,1]}(1,4,1){m_{{b}}}^{3} \\
&&+20\,\hat{I}_{{6}}(2,2,2){m_{{b}}}^{3}+5\,\hat{I}_{{0}}(2,2,1){m_{{b}}}^{3}-120\,\hat{I}_{{6}%
}(1,4,1){m_{{b}}}^{3}+40\,\hat{I}_{{6}}(2,3,1){m_{{b}}}^{3} \\
&&+15\,\hat{I}_{0}^{[0,1]}(3,2,1){m_{{b}}}^{3}-20\,\hat{I}_{{0}}(1,3,1){m_{{b}}}%
^{3}+10\,\hat{I}_{0}^{[0,1]}(2,3,1){m_{{b}}}^{3}-5\,\hat{I}_{0}^{[0,2]}(3,2,2){m_{{b}}}%
^{3} \\
&&+10\,\hat{I}_{0}^{[0,1]}(2,2,2){m_{{b}}}^{3}-5\,\hat{I}_{{0}}(1,2,2){m_{{b}}}%
^{3}-20\,\hat{I}_{6}^{[0,1]}(3,2,2){m_{{b}}}^{3}+20\,\hat{I}_{6}^{[0,1]}(3,1,2)m_{{c}} \\
&&+20\,\hat{I}_{{6}}(2,2,1)m_{{c}}+40\,\,\hat{I}_{6}^{[0,1]}(2,2,2)m_{{c}%
}-20\,\hat{I}_{6}^{[0,2]}(3,2,2)m_{{c}}+20\,\hat{I}_{{6}}(3,2,1){m_{{b}}}^{3} \\
&&+5\,\hat{I}_{0}^{[0,1]}(3,1,1)m_{{c}}+5\,\hat{I}_{{0}}(1,1,2)m_{{c}}+20\,\hat{I}_{{6}%
}(2,1,2)m_{{c}}+40\,\hat{I}_{{6}}(3,1,1)m_{{c}} \\
&&+5\,\hat{I}_{{0}}(1,2,1)m_{{c}}-5\,\hat{I}_{{0}}(2,1,1)m_{{c}}-10%
\,\hat{I}_{0}^{[0,2]}(2,3,1)m_{{b}}-10\,\hat{I}_{0}^{[0,2]}(3,1,2)m_{{b}} \\
&&-10\,\hat{I}_{0}^{[0,1]}(1,3,1)m_{{b}}-15\,\hat{I}_{{0}}(1,2,1)m_{{b}}-40\,\hat{I}_{{6}%
}(2,2,1)m_{{b}}+15\,\hat{I}_{0}^{[0,1]}(2,2,1)m_{{b}} \\
&&+20\,\hat{I}_{0}^{[0,1]}(1,2,2)m_{{b}}-40\,\hat{I}_{{6}}(1,3,1)m_{{b}}-40\,\hat{I}_{{6}%
}(3,1,1)m_{{b}}+40\,\hat{I}_{6}^{[0,1]}(3,2,1)m_{{c}} \\
&&-20\,\hat{I}_{6}^{[0,1]}(2,2,2)m_{{b}}+20\,\hat{I}_{6}^{[0,2]}(3,2,2)m_{{b}%
}-40\,\hat{I}_{6}^{[0,1]}(3,1,2)m_{{b}}-15\,\hat{I}_{{0}}(1,1,2)m_{{b}} \\
&&-15\,\hat{I}_{0}^{[0,2]}(2,2,2)m_{{b}}+15\,\hat{I}_{{0}}(2,1,1)m_{{b}}-20\,\hat{I}_{{6}%
}(2,1,2)m_{{b}}+25\,\hat{I}_{0}^{[0,1]}(2,1,2)m_{{b}} \\
&&+10\,\hat{I}_{0}^{[0,1]}(3,1,1)m_{{b}}-15\,\hat{I}_{0}^{[0,2]}(3,2,1)m_{{b}}-20\,\hat{I}_{{6}%
}(1,2,2)m_{{b}}-40\,\hat{I}_{6}^{[0,1]}(2,3,1)m_{{b}} \\
&&-20\,\hat{I}_{6}^{[0,1]}(3,2,1)m_{{b}}-5\,\hat{I}_{{0}}(3,2,2){m_{{c}}}^{6}m_{{b}%
}+10\,\hat{I}_{{0}}(2,2,2){m_{{c}}}^{3}{m_{{b}}}^{2}+20\,\hat{I}_{{6}}(3,2,2){m_{{c}}}%
^{3}{m_{{b}}}^{2} \\
&&-10\,\hat{I}_{{0}}(2,3,1){m_{{c}}}^{4}m_{{b}}+15\,\hat{I}_{0}^{[0,1]}(3,2,2){m_{{c}}}%
^{4}m_{{b}}+20\,\hat{I}_{{6}}(3,2,2){m_{{c}}}^{4}m_{{b}}-15\,\hat{I}_{{0}}(2,2,2){m_{{c}}%
}^{4}m_{{b}} \\
&&-5\,\hat{I}_{{0}}(3,2,1){m_{{c}}}^{4}m_{{b}}-15\,\hat{I}_{{0}}(4,1,1){m_{{c}}}^{4}m_{{b%
}}-5\,\hat{I}_{{0}}(3,2,2){m_{{c}}}^{3}{m_{{b}}}^{4}+5\,\hat{I}_{{0}}(3,2,2){m_{{c}}}^{4}%
{m_{{b}}}^{3} \\
&&+5\,\hat{I}_{{0}}(3,2,2){m_{{c}}}^{5}{m_{{b}}}^{2}-30\,\hat{I}_{{0}}(1,4,1)m_{{c}}{m_{{%
b}}}^{4}-5\,\hat{I}_{0}^{[0,1]}(3,2,2)m_{{c}}{m_{{b}}}^{4}+10\,\hat{I}_{{0}}(3,2,1)m_{{c}%
}{m_{{b}}}^{4} \\
&&+10\,\hat{I}_{{0}}(2,3,1){m_{{c}}}^{2}{m_{{b}}}^{3}-10\,\hat{I}_{{0}}(3,2,1){m_{{c}}}%
^{2}{m_{{b}}}^{3}-20\,\hat{I}_{{6}}(3,2,2){m_{{c}}}^{2}{m_{{b}}}^{3}+30\,\hat{I}_{{0}%
}(1,4,1){m_{{c}}}^{2}{m_{{b}}}^{3} \\
&&-10\,\hat{I}_{0}^{[0,1]}(3,2,2){m_{{c}}}^{3}{m_{{b}}}^{2}+5\,\hat{I}_{{0}}(3,2,1){m_{{c%
}}}^{3}{m_{{b}}}^{2}+15\,\hat{I}_{{0}}(4,1,1){m_{{c}}}^{3}{m_{{b}}}^{2}+20\,\hat{I}_{{6}%
}(2,2,2){m_{{c}}}^{2}m_{{b}} \\
&&+10\,\hat{I}_{0}^{[0,1]}(3,1,2){m_{{c}}}^{2}m_{{b}}-40\,\hat{I}_{6}^{[0,1]}(3,2,2){m_{{%
c}}}^{2}m_{{b}}-15\,\hat{I}_{0}^{[0,2]}(3,2,2){m_{{c}}}^{2}m_{{b}} \\
&&+10\,\hat{I}_{{0}}(1,3,1){m_{{c}}}^{2}m_{{b}}+20\,\hat{I}_{0}^{[0,1]}(3,2,1){m_{{c}}}%
^{2}m_{{b}}-20\,\hat{I}_{{0}}(1,2,2){m_{{c}}}^{2}m_{{b}}-15\,\hat{I}_{{0}}(2,1,2){m_{{c}}%
}^{2}m_{{b}} \\
&&-40\,\hat{I}_{{6}}(2,3,1){m_{{c}}}^{2}m_{{b}}+15\,\hat{I}_{0}^{[0,1]}(4,1,1){m_{{c}}}%
^{2}m_{{b}}+30\,\hat{I}_{0}^{[0,1]}(2,2,2){m_{{c}}}^{2}m_{{b}}+60\,\hat{I}_{{6}}(4,1,1){%
m_{{c}}}^{2}m_{{b}} \\
&&-10\,\hat{I}_{{0}}(3,1,1){m_{{c}}}^{2}m_{{b}}+20\,\hat{I}_{{6}}(3,1,2){m_{{c}}}^{2}m_{{%
b}}+15\,\hat{I}_{{0}}(2,2,1){m_{{c}}}^{2}m_{{b}}+20\,\hat{I}_{0}^{[0,1]}(2,3,1){m_{{c}}}%
^{2}m_{{b}} \\
&&+40\,\hat{I}_{{6}}(3,2,1){m_{{c}}}^{2}m_{{b}}-10\,\hat{I}_{0}^{[0,1]}(2,2,2)m_{{c}}{m_{%
{b}}}^{2}-20\,\hat{I}_{{6}}(3,2,1)m_{{c}}{m_{{b}}}^{2}+20\,\hat{I}_{6}^{[0,1]}(3,2,2)m_{{%
c}}{m_{{b}}}^{2} \\
&&+15\,\hat{I}_{{0}}(2,1,2)m_{{c}}{m_{{b}}}^{2}+5\,\hat{I}_{{0}}(3,1,1)m_{{c}}{m_{{b}}}%
^{2}-20\,\hat{I}_{0}^{[0,1]}(3,1,2)m_{{c}}{m_{{b}}}^{2}-20\,\hat{I}_{{6}}(2,2,2)m_{{c}}{%
m_{{b}}}^{2} \\
&&-30\,\hat{I}_{{0}}(1,3,1)m_{{c}}{m_{{b}}}^{2}+120\,\hat{I}_{{6}}(1,4,1)m_{{c}}{m_{{b}}}%
^{2}+10\,\hat{I}_{{0}}(1,2,2)m_{{c}}{m_{{b}}}^{2}-5\,\hat{I}_{0}^{[0,1]}(3,2,1)m_{{c}}{%
m_{{b}}}^{2} \\
&&-10\,\hat{I}_{{0}}(2,2,1)m_{{c}}{m_{{b}}}^{2}+5\,\hat{I}_{0}^{[0,2]}(3,2,2)m_{{c}}{m_{{%
b}}}^{2}
\end{eqnarray*}
\begin{eqnarray*}
C_{1}^{V-A} &=&-40\,\hat{I}_{4}^{[0,1]}(2,3,1)m_{{b}}+20\,\,\hat{I}_{4}^{[0,2]}(3,2,2)m_{{b%
}}-40\,\hat{I}_{{3}}(2,2,1)m_{{b}}-20\,\hat{I}_{{1}}(1,2,2)m_{{b}} \\
&&-20\,\hat{I}_{{1}}(2,1,2)m_{{b}}-40\,\hat{I}_{3}^{[0,1]}(2,3,1)m_{{b}%
}-20\,\,\hat{I}_{3}^{[0,1]}(3,2,1)m_{{b}}-20\,\hat{I}_{{3}}(2,1,2)m_{{b}} \\
&&-20\,\hat{I}_{3}^{[0,1]}(2,2,2)m_{{b}}-20\,\hat{I}_{{3}}(1,2,2)m_{{b}}-20\,\hat{I}_{{4}%
}(1,2,2)m_{{b}}-10\,\hat{I}_{1}^{[0,1]}(2,3,1)m_{{b}} \\
&&+5\,\,\hat{I}_{1}^{[0,2]}(3,2,2)m_{{b}}-5\,\hat{I}_{{2}}(3,2,2){m_{{3}}}^{5}-5\,\hat{I}_{{0}%
}(3,2,2){m_{{c}}}^{5}-20\,\hat{I}_{{3}}(3,2,2){m_{{c}}}^{5} \\
&&-15\,\hat{I}_{{1}}(3,2,2){m_{{c}}}^{5}-45\,\hat{I}_{{1}}(3,2,1){m_{{c}}}^{3}-20\,\hat{I}_{{4}%
}(3,1,2){m_{{c}}}^{3}-20\,\hat{I}_{{2}}(3,2,1){m_{{c}}}^{3} \\
&&-15\,\hat{I}_{{2}}(4,1,1){m_{{c}}}^{3}-25\,\hat{I}_{{0}}(3,2,1){m_{{c}}}^{3}-40\,\hat{I}_{{4}%
}(3,2,1){m_{{c}}}^{3}-10\,\hat{I}_{{2}}(2,2,2){m_{{c}}}^{3} \\
&&-45\,\hat{I}_{{1}}(4,1,1){m_{{c}}}^{3}-20\,\hat{I}_{{4}}(3,2,2){m_{{c}}}^{5}-5\,\hat{I}_{{0}%
}(3,1,2){m_{{c}}}^{3}-40\,\hat{I}_{{3}}(2,2,2){m_{{c}}}^{3} \\
&&+10\,\hat{I}_{0}^{[0,1]}(3,2,2){m_{{c}}}^{3}-30\,\hat{I}_{{1}}(2,2,2){m_{{c}}}%
^{3}+40\,\hat{I}_{3}^{[0,1]}(3,2,2){m_{{c}}}^{3}-60\,\hat{I}_{{3}}(4,1,1){m_{{c}}}^{3} \\
&&-15\,\hat{I}_{{0}}(4,1,1){m_{{c}}}^{3}-5\,\hat{I}_{{2}}(3,1,2){m_{{c}}}^{3}-10\,\hat{I}_{{0}%
}(2,2,2){m_{{c}}}^{3}-20\,\hat{I}_{{1}}(3,1,2){m_{{c}}}^{3} \\
&&-40\,\hat{I}_{{3}}(3,2,1){m_{{c}}}^{3}+30\,\hat{I}_{1}^{[0,1]}(3,2,2){m_{{c}}}%
^{3}+40\,\hat{I}_{4}^{[0,1]}(3,2,2){m_{{c}}}^{3}-60\,\hat{I}_{{4}}(4,1,1){m_{{c}}}^{3} \\
&&-20\,\hat{I}_{{3}}(3,2,2){m_{{c}}}^{2}{m_{{b}}}^{3}+20\,\hat{I}_{{4}}(2,2,2){m_{{c}}}%
^{2}m_{{b}}-20\,\hat{I}_{{0}}(2,3,1){m_{{c}}}^{2}m_{{b}}+40\,\hat{I}_{{4}}(3,2,1){m_{{c}}%
}^{2}m_{{b}} \\
&&-40\,\hat{I}_{4}^{[0,1]}(3,2,2){m_{{c}}}^{2}m_{{b}}-20\,\hat{I}_{{2}}(2,3,1){m_{{c}}}%
^{2}m_{{b}}+60\,\hat{I}_{{4}}(4,1,1){m_{{c}}}^{2}m_{{b}}+60\,\hat{I}_{{3}}(4,1,1){m_{{c}}%
}^{2}m_{{b}} \\
&&+20\,\hat{I}_{{3}}(3,1,2){m_{{c}}}^{2}m_{{b}}+20\,\hat{I}_{{3}}(2,2,2){m_{{c}}}^{2}m_{{%
b}}+5\,\hat{I}_{{1}}(3,2,2){m_{{c}}}^{4}m_{{b}}+20\,\hat{I}_{{3}}(3,2,2){m_{{c}}}^{4}m_{{%
b}} \\
&&+20\,\hat{I}_{{4}}(3,2,2){m_{{c}}}^{4}m_{{b}}+5\,\hat{I}_{{0}}(3,2,2){m_{{c}}}^{3}{m_{{%
b}}}^{2}+20\,\hat{I}_{{4}}(3,2,2){m_{{c}}}^{3}{m_{{b}}}^{2}+5\,\hat{I}_{{2}}(3,2,2){m_{{c%
}}}^{3}{m_{{b}}}^{2} \\
&&+15\,\hat{I}_{{1}}(3,2,2){m_{{c}}}^{3}{m_{{b}}}^{2}+20\,\hat{I}_{{3}}(3,2,2){m_{{c}}}%
^{3}{m_{{b}}}^{2}-20\,\hat{I}_{{4}}(3,2,2){m_{{c}}}^{2}{m_{{b}}}^{3}-5\,\hat{I}_{{1}%
}(3,2,2){m_{{c}}}^{2}{m_{{b}}}^{3} \\
&&+15\,\hat{I}_{{2}}(3,2,1){m_{{c}}}^{2}m_{{b}}+15\,\hat{I}_{{1}}(4,1,1){m_{{c}}}^{2}m_{{%
b}}+5\,\hat{I}_{{0}}(3,2,1){m_{{c}}}^{2}m_{{b}}+10\,\hat{I}_{{1}}(3,1,2){m_{{c}}}^{2}m_{{%
b}} \\
&&-50\,\hat{I}_{{1}}(2,3,1){m_{{c}}}^{2}m_{{b}}-10\,\hat{I}_{1}^{[0,1]}(3,2,2){m_{{c}}}%
^{2}m_{{b}}+35\,\hat{I}_{{1}}(3,2,1){m_{{c}}}^{2}m_{{b}}+20\,\hat{I}_{{4}}(3,1,2){m_{{3}}%
}^{2}m_{{b}} \\
&&-40\,\hat{I}_{3}^{[0,1]}(3,2,2){m_{{c}}}^{2}m_{{b}}-40\,\hat{I}_{{3}}(2,3,1){m_{{c}}}%
^{2}m_{{b}}-20\,\hat{I}_{{3}}(3,1,2){m_{{c}}}^{3}+10\,\hat{I}_{2}^{[0,1]}(3,2,2){m_{{c}}}%
^{3} \\
&&+40\,\hat{I}_{{3}}(2,3,1){m_{{b}}}^{3}+20\,\hat{I}_{{3}}(2,2,2){m_{{b}}}%
^{3}-5\,\hat{I}_{1}^{[0,1]}(3,2,2){m_{{b}}}^{3}+40\,\hat{I}_{{4}}(2,3,1){m_{{b}}}^{3} \\
&&-20\,\hat{I}_{4}^{[0,1]}(3,2,2){m_{{b}}}^{3}-120\,\hat{I}_{{3}}(1,4,1){m_{{b}}}%
^{3}+20\,\hat{I}_{{3}}(3,2,1){m_{{b}}}^{3}-120\,\hat{I}_{{4}}(1,4,1){m_{{b}}}^{3} \\
&&+20\,\hat{I}_{{4}}(3,2,1){m_{{b}}}^{3}+10\,\hat{I}_{{1}}(2,3,1){m_{{b}}}^{3}-30\,\hat{I}_{{1}%
}(1,4,1){m_{{b}}}^{3}-40\,\hat{I}_{{4}}(2,2,2){m_{{c}}}^{3} \\
&&+40\,\hat{I}_{{3}}(3,2,1){m_{{c}}}^{2}m_{{b}}-40\,\hat{I}_{{4}}(2,3,1){m_{{c}}}^{2}m_{{%
b}}+20\,\hat{I}_{4}^{[0,1]}(3,2,2)m_{{c}}{m_{{b}}}^{2}+5\,\hat{I}_{0}^{[0,1]}(3,2,2)m_{{c%
}}{m_{{b}}}^{2} \\
&&-20\,\hat{I}_{{4}}(2,2,2)m_{{c}}{m_{{b}}}^{2}-30\,\hat{I}_{{1}}(3,2,1)m_{{c}}{m_{{b}}}%
^{2}+90\,\hat{I}_{{1}}(1,4,1)m_{{c}}{m_{{b}}}^{2}+120\,\hat{I}_{{3}}(1,4,1)m_{{3}}{m_{{b}%
}}^{2} \\
&&+20\,\hat{I}_{3}^{[0,1]}(3,2,2)m_{{c}}{m_{{b}}}^{2}+120\,\hat{I}_{{4}}(1,4,1)m_{{c}}{%
m_{{b}}}^{2}+10\,\hat{I}_{{1}}(3,2,1){m_{{b}}}^{3}-20\,\hat{I}_{3}^{[0,1]}(3,2,2){m_{{b}}%
}^{3} \\
&&+40\,\hat{I}_{{3}}(3,1,1)m_{{c}}-5\,\hat{I}_{{0}}(2,2,1)m_{{c}}+10%
\,\hat{I}_{0}^{[0,1]}(2,2,2)m_{{c}}+20\,\hat{I}_{{4}}(2,1,2)m_{{c}}+40%
\,\hat{I}_{3}^{[0,1]}(3,2,1)m_{{c}} \\
&&+20\,\hat{I}_{4}^{[0,1]}(3,1,2)m_{{c}}+10\,\hat{I}_{2}^{[0,1]}(2,2,2)m_{{c}}+20\,\hat{I}_{{3}%
}(2,2,1)m_{{c}}-20\,\hat{I}_{4}^{[0,2]}(3,2,2)m_{{c}} \\
&&+40\,\hat{I}_{4}^{[0,1]}(2,2,2)m_{{c}}+5\,\hat{I}_{{2}}(3,1,1)m_{{c}}+20\,\hat{I}_{{4}%
}(2,2,2){m_{{b}}}^{3}+30\,\hat{I}_{{0}}(1,4,1)m_{{c}}{m_{{b}}}^{2} \\
&&+5\,\hat{I}_{{1}}(3,1,2)m_{{c}}{m_{{b}}}^{2}-15\,\hat{I}_{{0}}(3,2,1)m_{{c}}{m_{{b}}}%
^{2}+5\,\hat{I}_{2}^{[0,1]}(3,2,2)m_{{c}}{m_{{b}}}^{2}-10\,\hat{I}_{{2}}(3,2,1)m_{{c}}{%
m_{{b}}}^{2} \\
&&+30\,\hat{I}_{{2}}(1,4,1)m_{{c}}{m_{{b}}}^{2}-20\,\hat{I}_{{4}}(3,2,1)m_{{c}}{m_{{b}}}%
^{2}+15\,\hat{I}_{1}^{[0,1]}(3,2,2)m_{{c}}{m_{{b}}}^{2}-10\,\hat{I}_{{1}}(2,2,2)m_{{c}}{%
m_{{b}}}^{2} \\
&&-20\,\hat{I}_{{3}}(3,2,1)m_{{c}}{m_{{b}}}^{2}-20\,\hat{I}_{{3}}(2,2,2)m_{{c}}{m_{{b}}}%
^{2}-15\,\hat{I}_{1}^{[0,2]}(3,2,2)m_{{c}}-10\,\hat{I}_{{1}}(2,1,2)m_{{c}} \\
&&-5\,\hat{I}_{2}^{[0,2]}(3,2,2)m_{{c}}+5\,\hat{I}_{{1}}(2,2,1)m_{{c}}+40%
\,\hat{I}_{4}^{[0,1]}(3,2,1)m_{{c}}+10\,\hat{I}_{2}^{[0,1]}(3,2,1)m_{{c}} \\
&&+15\,\hat{I}_{0}^{[0,1]}(3,1,2)m_{{c}}+20\,\hat{I}_{3}^{[0,1]}(3,1,2)m_{{c}%
}+35\,\hat{I}_{1}^{[0,1]}(3,2,1)m_{{c}}+30\,\hat{I}_{1}^{[0,1]}(3,1,2)m_{{c}} \\
&&-5\,\hat{I}_{0}^{[0,2]}(3,2,2)m_{{c}}+40\,\hat{I}_{3}^{[0,1]}(2,2,2)m_{{c}}+20\,\hat{I}_{{3}%
}(2,1,2)m_{{c}}-15\,\hat{I}_{{0}}(2,1,2)m_{{c}} \\
&&-40\,\hat{I}_{3}^{[0,1]}(3,1,2)m_{{b}}-40\,\hat{I}_{{4}}(3,1,1)m_{{b}%
}-10\,\hat{I}_{1}^{[0,1]}(3,2,1)m_{{b}}-40\,\hat{I}_{{4}}(2,2,1)m_{{b}} \\
&&+20\,\hat{I}_{3}^{[0,2]}(3,2,2)m_{{b}}-40\,\hat{I}_{{3}}(1,3,1)m_{{b}%
}-40\,\hat{I}_{4}^{[0,1]}(3,1,2)m_{{b}}+10\,\hat{I}_{{1}}(1,3,1)m_{{b}} \\
&&-10\,\hat{I}_{{0}}(2,2,1)m_{{b}}-20\,\hat{I}_{4}^{[0,1]}(2,2,2)m_{{b}}-10\,\hat{I}_{{2}%
}(2,2,1)m_{{b}}-10\,\hat{I}_{1}^{[0,1]}(3,1,2)m_{{b}} \\
&&+10\,\hat{I}_{{0}}(1,3,1)m_{{b}}-20\,\hat{I}_{4}^{[0,1]}(3,2,1)m_{{b}%
}-20\,\hat{I}_{3}^{[0,2]}(3,2,2)m_{{c}}-10\,\hat{I}_{{0}}(3,1,1)m_{{c}} \\
&&+40\,\hat{I}_{{4}}(3,1,1)m_{{c}}-15\,\hat{I}_{{2}}(2,1,2)m_{{c}}+20\,\hat{I}_{{4}}(2,2,1)m_{{%
c}}+15\,\hat{I}_{{1}}(3,1,1)m_{{c}}+15\,\hat{I}_{2}^{[0,1]}(3,1,2)m_{{c}} \\
&&+15\,\hat{I}_{0}^{[0,1]}(3,2,1)m_{{c}}-40\,\hat{I}_{{3}}(3,1,1)m_{{b}}-20\,\hat{I}_{{4}%
}(2,1,2)m_{{b}}+10\,\hat{I}_{{2}}(1,3,1)m_{{b}}-30\,\hat{I}_{{1}}(2,2,1)m_{{b}} \\
&&-40\,\hat{I}_{{4}}(1,3,1)m_{{b}}+30\,\hat{I}_{1}^{[0,1]}(2,2,2)m_{{c}}
\end{eqnarray*}
\begin{eqnarray*}
C_{2}^{V-A} &=&15\,\hat{I}_{{2}}(4,1,1){m_{{c}}}^{2}m_{{b}}-40\,\hat{I}_{3}^{[0,1]}(3,2,2){%
m_{{c}}}^{2}m_{{b}}-40\,\hat{I}_{{4}}(3,2,1){m_{{c}}}^{2}m_{{b}}-10\,\hat{I}_{{2}}(2,3,1)%
{m_{{3}}}^{2}m_{{b}} \\
&&+20\,\hat{I}_{{3}}(3,1,2){m_{{c}}}^{2}m_{{b}}+60\,\hat{I}_{{3}}(4,1,1){m_{{c}}}^{2}m_{{%
b}}-20\,\hat{I}_{{4}}(2,2,2){m_{{c}}}^{2}m_{{b}}+40\,\hat{I}_{{3}}(3,2,1){m_{{c}}}^{2}m_{%
{b}} \\
&&+40\,\hat{I}_{4}^{[0,1]}(3,2,2){m_{{c}}}^{2}m_{{b}}-60\,\hat{I}_{{4}}(4,1,1){m_{{c}}}%
^{2}m_{{b}}+40\,\hat{I}_{{4}}(2,3,1){m_{{c}}}^{2}m_{{b}}+20\,\hat{I}_{{3}}(2,2,2){m_{{c}}%
}^{2}m_{{b}} \\
&&-5\,\hat{I}_{{0}}(3,2,1){m_{{c}}}^{2}m_{{b}}+15\,\hat{I}_{{1}}(3,2,1){m_{{c}}}^{2}m_{{b%
}}-20\,\hat{I}_{{1}}(2,3,1){m_{{c}}}^{2}m_{{b}}-40\,\hat{I}_{{3}}(2,3,1){m_{{c}}}^{2}m_{{%
b}} \\
&&-20\,\hat{I}_{{4}}(3,1,2){m_{{c}}}^{2}m_{{b}}+10\,\hat{I}_{2}^{[0,1]}(3,2,2){m_{{c}}}%
^{3}+60\,\hat{I}_{{4}}(4,1,1){m_{{c}}}^{3}-20\,\hat{I}_{{3}}(3,1,2){m_{{c}}}^{3} \\
&&+40\,\hat{I}_{{4}}(2,2,2){m_{{c}}}^{3}-15\,\hat{I}_{{1}}(4,1,1){m_{{c}}}^{3}-40\,\hat{I}_{{3}%
}(3,2,1){m_{{c}}}^{3}-40\,\hat{I}_{{3}}(2,2,2){m_{{c}}}^{3} \\
&&-15\,\hat{I}_{{2}}(4,1,1){m_{{c}}}^{3}-5\,\hat{I}_{{2}}(3,2,1){m_{{c}}}%
^{3}+10\,\hat{I}_{1}^{[0,1]}(3,2,2){m_{{c}}}^{3}+5\,\hat{I}_{{0}}(3,1,2){m_{{c}}}^{3} \\
&&-10\,\hat{I}_{0}^{[0,1]}(3,2,2){m_{{c}}}^{3}-10\,\hat{I}_{{1}}(2,2,2){m_{{c}}}%
^{3}+10\,\hat{I}_{{0}}(2,2,2){m_{{c}}}^{3}-10\,\hat{I}_{{2}}(3,1,2){m_{{c}}}^{3} \\
&&-5\,\hat{I}_{{1}}(3,1,2){m_{{c}}}^{3}+15\,\hat{I}_{{0}}(4,1,1){m_{{c}}}^{3}-20\,\hat{I}_{{1}%
}(3,2,1){m_{{c}}}^{3}+20\,\hat{I}_{4}^{[0,1]}(3,2,2){m_{{b}}}^{3} \\
&&-30\,\hat{I}_{{2}}(1,4,1){m_{{b}}}^{3}-20\,\hat{I}_{3}^{[0,1]}(3,2,2){m_{{b}}}%
^{3}+40\,\hat{I}_{{3}}(2,3,1){m_{{b}}}^{3}+5\,\hat{I}_{{2}}(3,2,1){m_{{c}}}^{2}m_{{b}} \\
&&-20\,\hat{I}_{4}^{[0,1]}(3,2,2)m_{{c}}{m_{{b}}}^{2}+5\,\hat{I}_{{2}}(3,1,2)m_{{c}}{m_{{%
b}}}^{2}-20\,\hat{I}_{{3}}(3,2,1)m_{{c}}{m_{{b}}}^{2}+20\,\hat{I}_{{4}}(2,2,2)m_{{c}}{m_{%
{b}}}^{2} \\
&&-5\,\hat{I}_{0}^{[0,1]}(3,2,2)m_{{c}}{m_{{b}}}^{2}-10\,\hat{I}_{{2}}(3,2,1)m_{{c}}{m_{{%
b}}}^{2}+15\,\hat{I}_{{0}}(3,2,1)m_{{c}}{m_{{b}}}^{2}+20\,\hat{I}_{3}^{[0,1]}(3,2,2)m_{{c%
}}{m_{{b}}}^{2} \\
&&-10\,\hat{I}_{{2}}(2,2,2)m_{{c}}{m_{{b}}}^{2}-30\,\hat{I}_{{0}}(1,4,1)m_{{c}}{m_{{b}}}%
^{2}+120\,\hat{I}_{{3}}(1,4,1)m_{{c}}{m_{{b}}}^{2}+5\,\hat{I}_{1}^{[0,1]}(3,2,2)m_{{c}}{%
m_{{b}}}^{2} \\
&&+30\,\hat{I}_{{1}}(1,4,1)m_{{c}}{m_{{b}}}^{2}-120\,\hat{I}_{{4}}(1,4,1)m_{{c}}{m_{{b}}}%
^{2}+5\,\hat{I}_{2}^{[0,1]}(3,2,2)m_{{c}}{m_{{b}}}^{2}-20\,\hat{I}_{{3}}(2,2,2)m_{{c}}{%
m_{{b}}}^{2} \\
&&+20\,\hat{I}_{{4}}(3,2,1)m_{{c}}{m_{{b}}}^{2}-10\,\hat{I}_{{1}}(3,2,1)m_{{c}}{m_{{b}}}%
^{2}+30\,\hat{I}_{{2}}(1,4,1)m_{{c}}{m_{{b}}}^{2}+20\,\hat{I}_{{3}}(2,2,2){m_{{b}}}^{3}
\\
&&-20\,\hat{I}_{{4}}(2,2,2){m_{{b}}}^{3}-5\,\hat{I}_{2}^{[0,1]}(3,2,2){m_{{b}}}%
^{3}-40\,\hat{I}_{{4}}(2,3,1){m_{{b}}}^{3}+120\,\hat{I}_{{4}}(1,4,1){m_{{b}}}^{3} \\
&&-20\,\hat{I}_{{4}}(3,2,1){m_{{b}}}^{3}+10\,\hat{I}_{{2}}(2,3,1){m_{{b}}}^{3}+10\,\hat{I}_{{2}%
}(3,2,1){m_{{b}}}^{3}-120\,\hat{I}_{{3}}(1,4,1){m_{{b}}}^{3} \\
&&+20\,\hat{I}_{{3}}(3,2,1){m_{{b}}}^{3}+20\,\hat{I}_{4}^{[0,2]}(3,2,2)m_{{c}%
}+10\,\hat{I}_{1}^{[0,1]}(3,2,1)m_{{c}}-40\,\hat{I}_{4}^{[0,1]}(2,2,2)m_{{c}} \\
&&+5\,\hat{I}_{{0}}(2,2,1)m_{{c}}+15\,\hat{I}_{1}^{[0,1]}(3,1,2)m_{{c}}-5\,{C1}_{{2}%
}(3,2,2)m_{{c}}-15\,\hat{I}_{0}^{[0,1]}(3,2,1)m_{{c}} \\
&&+5\,\hat{I}_{0}^{[0,2]}(3,2,2)m_{{c}}-40\,\hat{I}_{{4}}(3,1,1)m_{{c}}+15\,\hat{I}_{{0}%
}(2,1,2)m_{{c}}+5\,\hat{I}_{{2}}(3,1,1)m_{{c}} \\
&&+40\,\hat{I}_{{3}}(3,1,1)m_{{c}}+5\,\hat{I}_{{1}}(3,1,1)m_{{c}}-20%
\,\hat{I}_{4}^{[0,1]}(3,1,2)m_{{c}}+10\,\hat{I}_{{0}}(3,1,1)m_{{c}} \\
&&+20\,\hat{I}_{{3}}(2,2,1)m_{{c}}+15\,\hat{I}_{2}^{[0,1]}(3,2,1)m_{{c}}-20\,\hat{I}_{{4}%
}(2,2,1)m_{{c}}+20\,\hat{I}_{{2}}(2,1,2)m_{{c}} \\
&&+10\,\hat{I}_{1}^{[0,1]}(2,2,2)m_{{c}}+20\,\,\hat{I}_{3}^{[0,1]}(3,1,2)m_{{c}}-20\,\hat{I}_{{%
4}}(2,1,2)m_{{c}}+20\,\hat{I}_{{3}}(2,1,2)m_{{c}} \\
&&-40\,\hat{I}_{4}^{[0,1]}(3,2,1)m_{{c}}+40\,\hat{I}_{3}^{[0,1]}(3,2,1)m_{{c}%
}-5\,\,\hat{I}_{1}^{[0,2]}(3,2,2)m_{{c}}+40\,\hat{I}_{3}^{[0,1]}(2,2,2)m_{{c}} \\
&&+5\,\hat{I}_{{2}}(2,2,1)m_{{c}}+10\,\hat{I}_{2}^{[0,1]}(2,2,2)m_{{c}}-15\,\hat{I}_{{1}%
}(2,1,2)m_{{c}}-15\,\hat{I}_{0}^{[0,1]}(3,1,2)m_{{c}} \\
&&-10\,\hat{I}_{0}^{[0,1]}(2,2,2)m_{{c}}-20\,\,\hat{I}_{3}^{[0,2]}(3,2,2)m_{{c}}+10\,\hat{I}_{{%
1}}(1,3,1)m_{{b}}+20\,\hat{I}_{{4}}(1,2,2)m_{{b}} \\
&&-40\,\hat{I}_{3}^{[0,1]}(3,1,2)m_{{b}}+40\,\hat{I}_{4}^{[0,1]}(3,1,2)m_{{b}}-20\,\hat{I}_{{2}%
}(2,1,2)m_{{b}}+20\,\hat{I}_{4}^{[0,1]}(3,2,1)m_{{b}} \\
&&-10\,\hat{I}_{2}^{[0,1]}(3,2,1)m_{{b}}-20\,\hat{I}_{{3}}(1,2,2)m_{{b}%
}-10\,\hat{I}_{2}^{[0,1]}(2,3,1)m_{{b}}-40\,\hat{I}_{{3}}(1,3,1)m_{{b}} \\
&&+5\,\hat{I}_{2}^{[0,2]}(3,2,2)m_{{b}}+10\,\hat{I}_{{0}}(2,2,1)m_{{b}}-20\,\hat{I}_{{3}%
}(2,1,2)m_{{b}}-10\,\hat{I}_{{1}}(2,2,1)m_{{b}} \\
&&-10\,\hat{I}_{{2}}(2,2,1)m_{{b}}-10\,\hat{I}_{{0}}(1,3,1)m_{{b}}-20%
\,\hat{I}_{4}^{[0,2]}(3,2,2)m_{{b}}-40\,\hat{I}_{3}^{[0,1]}(2,3,1)m_{{b}} \\
&&-20\,\hat{I}_{{2}}(1,2,2)m_{{b}}+20\,\hat{I}_{{4}}(2,1,2)m_{{b}}-40\,\hat{I}_{{3}}(3,1,1)m_{{%
b}}-10\,\hat{I}_{{2}}(1,3,1)m_{{b}} \\
&&-10\,\hat{I}_{2}^{[0,1]}(3,1,2)m_{{b}}-20\,\hat{I}_{3}^{[0,1]}(2,2,2)m_{{\ b}}+40\,\hat{I}_{{%
4}}(1,3,1)m_{{b}}+40\,\hat{I}_{{4}}(3,1,1)m_{{b}} \\
&&+20\,\hat{I}_{3}^{[0,2]}(3,2,2)m_{{b}}-40\,\hat{I}_{{3}}(2,2,1)m_{{b}%
}+20\,\hat{I}_{4}^{[0,1]}(2,2,2)m_{{b}}+40\,\hat{I}_{4}^{[0,1]}(2,3,1)m_{{b}} \\
&&-20\,\hat{I}_{3}^{[0,1]}(3,2,1)m_{{b}}+40\,\hat{I}_{{4}}(2,2,1)m_{{b}}+25\,\hat{I}_{{0}%
}(3,2,1){m_{{c}}}^{3}-60\,\hat{I}_{{3}}(4,1,1){m_{{c}}}^{3} \\
&&+10\,\hat{I}_{{2}}(3,1,2){m_{{c}}}^{2}m_{{b}}+20\,\hat{I}_{{4}}(3,2,2){m_{{c}}}%
^{5}+5\,\hat{I}_{{0}}(3,2,2){m_{{c}}}^{5}-5\,\hat{I}_{{2}}(3,2,2){m_{{c}}}^{5} \\
&&-20\,\hat{I}_{{3}}(3,2,2){m_{{c}}}^{5}-5\,\hat{I}_{{1}}(3,2,2){m_{{\ c}}}%
^{5}-40\,\hat{I}_{4}^{[0,1]}(3,2,2){m_{{c}}}^{3}-10\,\hat{I}_{{2}}(2,2,2){m_{{c}}}^{3} \\
&&+40\,\hat{I}_{{4}}(3,2,1){m_{{c}}}^{3}+40\,\hat{I}_{3}^{[0,1]}(3,2,2){m_{{c}}}%
^{3}+20\,\hat{I}_{{4}}(3,1,2){m_{{c}}}^{3}+5\,\hat{I}_{{2}}(3,2,2){m_{{c}}}^{4}m_{{b}} \\
&&-20\,\hat{I}_{{4}}(3,2,2){m_{{c}}}^{4}m_{{b}}+20\,\hat{I}_{{3}}(3,2,2){m_{{c}}}^{4}m_{{%
\ b}}+5\,\hat{I}_{{1}}(3,2,2){m_{{c}}}^{3}{m_{{b}}}^{2}-5\,\hat{I}_{{0}}(3,2,2){m_{{c}}}%
^{3}{m_{{b}}}^{2} \\
&&+20\,\hat{I}_{{3}}(3,2,2){m_{{c}}}^{3}{m_{{b}}}^{2}-20\,\hat{I}_{{4}}(3,2,2){m_{{c}}}%
^{3}{m_{{b}}}^{2}+5\,\hat{I}_{{2}}(3,2,2){m_{{c}}}^{3}{m_{{b}}}^{2}-5\,\hat{I}_{{2}%
}(3,2,2){m_{{c}}}^{2}{m_{{b}}}^{3} \\
&&+20\,\hat{I}_{{4}}(3,2,2){m_{{c}}}^{2}{m_{{b}}}^{3}-20\,\hat{I}_{{3}}(3,2,2){m_{{c}}}%
^{2}{m_{{b}}}^{3}-10\,\hat{I}_{2}^{[0,1]}(3,2,2){m_{{\ c}}}^{2}m_{{b}}+20\,\hat{I}_{{0}%
}(2,3,1){m_{{c}}}^{2}m_{{b}}
\end{eqnarray*}
\begin{eqnarray*}
C_{V}^{T-PT} &=&10\,\hat{I}_{{1}}(3,2,2){m_{{b}}}^{4}{m_{{c}}}^{2}-20\,\hat{I}_{{1}%
}(1,1,2)+10\,\hat{I}_{0}^{[0,1]}(2,2,1)-10\,\hat{I}_{2}^{[0,2]}(3,2,1) \\
&&-10\,\hat{I}_{{0}}(3,2,2){m_{{b}}}^{3}{m_{{c}}}^{3}-10\,\hat{I}_{{1}}(3,2,2){m_{{b}}}%
^{3}{m_{{c}}}^{3}-10\,\hat{I}_{{1}}(1,2,1)-10\,\hat{I}_{{0}}(1,1,2) \\
&&-60\,\hat{I}_{{1}}(1,4,1){m_{{b}}}^{3}m_{{c}}-10\,\hat{I}_{0}^{[0,1]}(3,2,2){m_{{b}}}%
^{3}m_{{c}}-60\,\hat{I}_{{0}}(1,4,1){m_{{b}}}^{3}m_{{c}}+20\,\hat{I}_{{2}}(3,2,1){m_{{b}}%
}^{3}m_{{c}} \\
&&+20\,\hat{I}_{{1}}(3,2,1){m_{{b}}}^{3}m_{{c}}-10\,\hat{I}_{2}^{[0,1]}(3,2,2){m_{{b}}}%
^{3}m_{{c}}-10\,\hat{I}_{1}^{[0,1]}(3,2,2){m_{{b}}}^{3}m_{{c}}-60\,\hat{I}_{{2}}(1,4,1){%
m_{{b}}}^{3}m_{{c}} \\
&&+20\,\hat{I}_{{2}}(2,3,1){m_{{b}}}^{3}m_{{c}}-10\,\hat{I}_{{2}}(3,2,2){m_{{b}}}^{3}{m_{%
{c}}}^{3}+10\,\hat{I}_{{2}}(3,2,2)m_{{b}}{m_{{c}}}^{5}+10\,\hat{I}_{{1}}(3,2,2)m_{{b}}{%
m_{{c}}}^{5} \\
&&+10\,\hat{I}_{{0}}(3,2,2)m_{{b}}{m_{{c}}}^{5}+20\,\hat{I}_{{0}}(3,2,1){m_{{b}}}^{3}m_{{%
c}}-20\,\hat{I}_{{1}}(2,3,1){m_{{b}}}^{3}m_{{c}}+10\,\hat{I}_{{1}}(3,1,2){m_{{b}}}^{3}m_{%
{c}} \\
&&+20\,\hat{I}_{{0}}(2,3,1){m_{{b}}}^{3}m_{{c}}-10\,\hat{I}_{{1}}(3,2,2){m_{{b}}}^{2}{m_{%
{c}}}^{4}+10\,\hat{I}_{1}^{[0,1]}(3,2,2){m_{{b}}}^{4}-20\,\hat{I}_{{1}}(3,2,1){m_{{b}}}%
^{4} \\
&&+60\,\hat{I}_{{1}}(1,4,1){m_{{b}}}^{4}-10\,\hat{I}_{{0}}(3,1,2){m_{{c}}}^{4}-10\,\hat{I}_{{1}%
}(3,1,2){m_{{c}}}^{4}+10\,\hat{I}_{{1}}(3,2,1){m_{{c}}}^{4} \\
&&+10\,\hat{I}_{{0}}(3,2,1){m_{{c}}}^{4}+10\,\hat{I}_{2}^{[0,1]}(3,2,1){m_{{b}}}%
^{2}+60\,\hat{I}_{{1}}(1,3,1){m_{{b}}}^{2}+20\,\hat{I}_{1}^{[0,1]}(3,2,1){m_{{b}}}^{2} \\
&&-10\,\hat{I}_{{1}}(2,2,2){m_{{b}}}^{4}+20\,\hat{I}_{{0}}(2,3,1)m_{{b}}{m_{{c}}}%
^{3}-20\,\hat{I}_{2}^{[0,1]}(3,2,2)m_{{b}}{m_{{c}}}^{3}+20\,\hat{I}_{{1}}(2,2,2)m_{{b}}{%
m_{{c}}}^{3} \\
&&+10\,\hat{I}_{{2}}(3,2,1)m_{{b}}{m_{{c}}}^{3}+20\,\hat{I}_{{2}}(2,3,1)m_{{b}}{m_{{c}}}%
^{3}+10\,\hat{I}_{{1}}(3,2,1)m_{{b}}{m_{{c}}}^{3}+20\,\hat{I}_{{2}}(2,2,2)m_{{b}}{m_{{c}}%
}^{3} \\
&&+20\,\hat{I}_{{1}}(2,3,1)m_{{b}}{m_{{c}}}^{3}+30\,\hat{I}_{{0}}(4,1,1)m_{{b}}{m_{{c}}}%
^{3}+20\,\hat{I}_{1}^{[0,1]}(3,2,2){m_{{b}}}^{2}{m_{{c}}}^{2}+10\,\hat{I}_{{2}}(3,2,1){%
m_{{b}}}^{2}{m_{{c}}}^{2} \\
&&-20\,\hat{I}_{{1}}(3,2,1){m_{{b}}}^{2}{m_{{c}}}^{2}-30\,\hat{I}_{{1}}(4,1,1){m_{{b}}}%
^{2}{m_{{c}}}^{2}+20\,\hat{I}_{{0}}(3,2,1)m_{{b}}{m_{{c}}}^{3}+30\,\hat{I}_{{1}%
}(4,1,1)m_{{b}}{m_{{c}}}^{3} \\
&&+30\,\hat{I}_{{2}}(4,1,1)m_{{b}}{m_{{c}}}^{3}+20\,\hat{I}_{{0}}(2,2,2)m_{{b}}{m_{{c}}}%
^{3}-20\,\hat{I}_{{1}}(2,2,2){m_{{b}}}^{2}{m_{{c}}}^{2}-40\,\hat{I}_{{2}}(1,3,1)m_{{b}%
}m_{{c}} \\
&&-30\,\hat{I}_{{2}}(2,2,1)m_{{b}}m_{{c}}-70\,\hat{I}_{{1}}(2,2,1)m_{{b}}m_{{c}%
}-10\,\hat{I}_{2}^{[0,1]}(3,2,1)m_{{b}}m_{{c}}-20\,\hat{I}_{0}^{[0,1]}(2,2,2)m_{{b}}m_{{c%
}} \\
&&-20\,\hat{I}_{2}^{[0,1]}(2,2,2)m_{{b}}m_{{c}}+20\,\hat{I}_{{0}}(1,2,2)m_{{b}}m_{{c}%
}+10\,\hat{I}_{0}^{[0,2]}(3,2,2)m_{{b}}m_{{c}}+10\,\hat{I}_{1}^{[0,2]}(3,2,2)m_{{b}}m_{{c%
}} \\
&&-20\,\hat{I}_{0}^{[0,1]}(3,2,1)m_{{b}}m_{{c}}+20\,\hat{I}_{{2}}(1,2,2)m_{{b}}m_{{c}%
}-20\,\hat{I}_{1}^{[0,1]}(2,3,1)m_{{b}}m_{{c}}-20\,\hat{I}_{{0}}(2,2,1)m_{{b}}m_{{c}} \\
&&-20\,\hat{I}_{2}^{[0,1]}(3,1,2)m_{{b}}m_{{c}}+20\,\hat{I}_{{2}}(2,1,2)m_{{b}}m_{{c}%
}-40\,\hat{I}_{{0}}(1,3,1)m_{{b}}m_{{c}}-20\,\hat{I}_{0}^{[0,1]}(3,1,2)m_{{b}}m_{{c}} \\
&&-20\,\hat{I}_{2}^{[0,1]}(2,3,1)m_{{b}}m_{{c}}-10\,\hat{I}_{1}^{[0,1]}(3,2,1)m_{{b}}m_{{%
c}}+20\,\hat{I}_{{0}}(2,1,2)m_{{b}}m_{{c}}+10\,\hat{I}_{2}^{[0,2]}(3,2,2)m_{{b}}m_{{c}}
\\
&&-10\,\hat{I}_{{0}}(1,2,1)-20\,\hat{I}_{{1}}(1,2,2){m_{{b}}}^{2}+10\,\hat{I}_{{0}}(2,2,1){m_{{%
b}}}^{2}+20\,\hat{I}_{1}^{[0,1]}(3,1,2){m_{{b}}}^{2} \\
&&-30\,\hat{I}_{{1}}(2,1,2){m_{{b}}}^{2}+20\,\hat{I}_{1}^{[0,1]}(2,2,2){m_{{b}}}%
^{2}+30\,\hat{I}_{{2}}(2,2,1){m_{{b}}}^{2}-10\,\hat{I}_{1}^{[0,2]}(3,2,2){m_{{b}}}^{2} \\
&&+10\,\hat{I}_{0}^{[0,1]}(3,1,2){m_{{c}}}^{2}-20\,\hat{I}_{{0}}(2,1,2){m_{{c}}}%
^{2}-10\,\hat{I}_{0}^{[0,1]}(3,2,1){m_{{c}}}^{2}-20\,\hat{I}_{{1}}(2,1,2){m_{{c}}}^{2} \\
&&+60\,\hat{I}_{{1}}(2,2,1){m_{{b}}}^{2}-30\,\hat{I}_{{2}}(1,2,1)-20%
\,\hat{I}_{0}^{[0,1]}(3,2,2)m_{{b}}{m_{{c}}}^{3}+20\,\hat{I}_{{1}}(2,1,2)m_{{b}}m_{{c}}
\\
&&-20\,\hat{I}_{0}^{[0,1]}(2,3,1)m_{{b}}m_{{c}}+20\,\hat{I}_{{1}}(1,2,2)m_{{b}}m_{{c}%
}-10\,\hat{I}_{{1}}(3,1,1)m_{{b}}m_{{c}}-20\,\hat{I}_{1}^{[0,1]}(3,1,2)m_{{b}}m_{{c}} \\
&&-20\,\hat{I}_{1}^{[0,1]}(2,2,2)m_{{b}}m_{{c}}-40\,\hat{I}_{{1}}(1,3,1)m_{{b}}m_{{c}%
}-20\,\hat{I}_{1}^{[0,1]}(3,2,2)m_{{b}}{m_{{c}}}^{3}-10\,\hat{I}_{1}^{[0,1]}(3,2,1){m_{{c%
}}}^{2} \\
&&+10\,\hat{I}_{1}^{[0,1]}(3,1,2){m_{{c}}}^{2}-20\,\hat{I}_{{2}}(2,2,1){m_{{c}}}%
^{2}+10\,\hat{I}_{{0}}(3,1,1){m_{{c}}}^{2}+10\,\hat{I}_{2}^{[0,1]}(3,2,1){m_{{c}}}^{2} \\
&&-10\,\hat{I}_{2}^{[0,1]}(3,1,2){m_{{c}}}^{2}+10\,\hat{I}_{{2}}(3,1,1){m_{{c}}}^{2}
\end{eqnarray*}
\begin{eqnarray*}
C_{0}^{T-PT} &=&-15\,\hat{I}_{{0}}(2,2,2){m_{{c}}}^{4}{m_{{b}}}^{2}-15\,\hat{I}_{{0}}(4,1,1)%
{m_{{c}}}^{4}{m_{{b}}}^{2}+5\,\hat{I}_{{0}}(3,1,2){m_{{c}}}^{5}m_{{b}}-5\,\hat{I}_{{0}%
}(3,2,1){m_{{c}}}^{5}m_{{b}} \\
&&-5\,\hat{I}_{{0}}(3,2,2){m_{{c}}}^{3}{m_{{b}}}^{5}+5\,\hat{I}_{{0}}(3,2,2){m_{{c}}}^{4}%
{m_{{b}}}^{4}+5\,\hat{I}_{{0}}(3,2,2){m_{{c}}}^{5}{m_{{b}}}^{3}-5\,\hat{I}_{{0}}(3,2,2){%
m_{{c}}}^{6}{m_{{b}}}^{2} \\
&&+30\,\hat{I}_{0}^{[0,1]}(2,2,2){m_{{c}}}^{2}{m_{{b}}}^{2}-15\,\hat{I}_{{0}}(2,1,2){m_{{%
c}}}^{2}{m_{{b}}}^{2}-15\,\hat{I}_{0}^{[0,2]}(3,2,2){m_{{c}}}^{2}{m_{{b}}}%
^{2}\\
&&-20\,\hat{I}_{{0}}(2,2,1){m_{{c}}}^{3}m_{{b}}+10\,\hat{I}_{0}^{[0,1]}(3,2,1){m_{{c}}}%
^{3}m_{{b}}+10\,\hat{I}_{{0}}(2,1,2){m_{{c}}}^{3}m_{{b}}-30\,\hat{I}_{{0}}(1,4,1)m_{{c}}{%
m_{{b}}}^{5} \\
&&-5\,\hat{I}_{0}^{[0,1]}(3,2,2)m_{{c}}{m_{{b}}}^{5}+10\,\hat{I}_{{0}}(3,2,1)m_{{c}}{m_{{%
b}}}^{5}+10\,\hat{I}_{{0}}(2,3,1)m_{{c}}{m_{{b}}}^{5}-10\,\hat{I}_{{0}}(3,2,1){m_{{c}}}%
^{2}{m_{{b}}}^{4} \\
&&+30\,\hat{I}_{{0}}(1,4,1){m_{{c}}}^{2}{m_{{b}}}^{4}+15\,\hat{I}_{{0}}(4,1,1){m_{{c}}}%
^{3}{m_{{b}}}^{3}-10\,\hat{I}_{0}^{[0,1]}(3,2,2){m_{{c}}}^{3}{m_{{b}}}^{3} %
\\
&&-10\,\hat{I}_{{0}}(2,3,1){m_{{c}}}^{3}{m_{{b}}}^{3}+10\,\hat{I}_{{0}}(2,2,2){m_{{c}}}%
^{3}{m_{{b}}}^{3}+15\,\hat{I}_{0}^{[0,1]}(3,2,2){m_{{c}}}^{4}{m_{{b}}}^{2}%
 \\
&&+10\,\hat{I}_{{0}}(1,2,2)m_{{c}}{m_{{b}}}^{3}+15\,\hat{I}_{{0}}(2,1,2)m_{{c}}{m_{{b}}}%
^{3}-20\,\hat{I}_{0}^{[0,1]}(3,1,2)m_{{c}}{m_{{b}}}^{3}+10\,\hat{I}_{0}^{[0,1]}(2,3,1)m_{%
{c}}{m_{{b}}}^{3} \\
&&-5\,\hat{I}_{{0}}(2,2,1)m_{{c}}{m_{{b}}}^{3}-10\,\hat{I}_{0}^{[0,1]}(2,2,2)m_{{c}}{m_{{%
b}}}^{3}-20\,\hat{I}_{{0}}(1,2,2){m_{{c}}}^{2}{m_{{b}}}^{2}%
 \\
&&+15\,\hat{I}_{0}^{[0,1]}(3,2,1){m_{{c}}}^{2}{m_{{b}}}^{2}-5\,\hat{I}_{{0}}(3,1,1){m_{{c%
}}}^{2}{m_{{b}}}^{2}+30\,\hat{I}_{{0}}(1,3,1){m_{{c}}}^{2}{m_{{b}}}%
^{2} \\
&&+15\,\hat{I}_{0}^{[0,1]}(3,1,2){m_{{c}}}^{2}{m_{{b}}}^{2}-30\,\hat{I}_{{0}}(2,1,1)m_{{c%
}}m_{{b}}+10\,\hat{I}_{0}^{[0,1]}(3,1,1)m_{{c}}m_{{b}}-5\,\hat{I}_{0}^{[0,2]}(3,2,1)m_{{c%
}}m_{{b}} \\
&&-10\,\hat{I}_{0}^{[0,1]}(2,1,2)m_{{c}}m_{{b}}-50\,\hat{I}_{{0}}(1,2,1)m_{{c}}m_{{b}%
}+10\,\hat{I}_{{0}}(1,1,2)m_{{c}}m_{{b}}+5\,\hat{I}_{0}^{[0,2]}(3,1,2)m_{{c}}m_{{b}} \\
&&+20\,\hat{I}_{0}^{[0,1]}(2,2,1)m_{{c}}m_{{b}}-70\,\hat{I}_{{0}}(1,3,1)m_{{c}}{m_{{b}}}%
^{3}+5\,\hat{I}_{0}^{[0,2]}(3,2,2)m_{{c}}{m_{{b}}}^{3}+5\,\hat{I}_{{0}}(3,1,1)m_{{c}}{m_{%
{b}}}^{3} \\
&&-20\,\hat{I}_{0}^{[0,1]}(1,2,1)-5\,\hat{I}_{0}^{[0,2]}(3,2,1){m_{{b}}}^{2}-15\,\hat{I}_{{0}%
}(1,1,2){m_{{b}}}^{2}+30\,\hat{I}_{0}^{[0,1]}(2,1,2){m_{{b}}}^{2} \\
&&-15\,\hat{I}_{0}^{[0,2]}(3,1,2){m_{{b}}}^{2}-30\,\hat{I}_{0}^{[0,1]}(1,3,1){m_{{b}}}%
^{2}+10\,\hat{I}_{{0}}(1,2,1){m_{{b}}}^{2}-15\,\hat{I}_{0}^{[0,2]}(2,1,2) \\
&&+20\,\hat{I}_{0}^{[0,1]}(1,2,2){m_{{b}}}^{2}-15\,\hat{I}_{0}^{[0,2]}(2,2,2){m_{{b}}}%
^{2}+20\,\hat{I}_{{0}}(2,1,1){m_{{b}}}^{2}+30\,\hat{I}_{0}^{[0,1]}(2,1,2){m_{{c}}}^{2} \\
&&-30\,\hat{I}_{0}^{[0,1]}(2,2,1){m_{{c}}}^{2}+10\,\hat{I}_{0}^{[0,1]}(3,1,1){m_{{b}}}%
^{2}-5\,\hat{I}_{0}^{[0,1]}(2,2,1){m_{{b}}}^{2}+20\,\hat{I}_{{0}}(1,2,1){m_{{c}}}^{2} \\
&&-15\,\hat{I}_{0}^{[0,2]}(3,1,2){m_{{c}}}^{2}+15\,\hat{I}_{0}^{[0,2]}(3,2,1){m_{{c}}}%
^{2}-20\,\hat{I}_{{0}}(1,1,2){m_{{c}}}^{2}+15\,\hat{I}_{{0}}(2,1,1){m_{{c}}}^{2} \\
&&-30\,\hat{I}_{0}^{[0,1]}(1,4,1){m_{{b}}}^{4}+15\,\hat{I}_{{0}}(1,1,1)+10\,\hat{I}_{{0}%
}(2,2,1){m_{{b}}}^{4}+10\,\hat{I}_{0}^{[0,1]}(2,2,2){m_{{b}}}^{4} \\
&&+10\,\hat{I}_{0}^{[0,1]}(3,2,1){m_{{b}}}^{4}-5\,\hat{I}_{0}^{[0,2]}(3,2,2){m_{{b}}}%
^{4}+15\,\hat{I}_{0}^{[0,1]}(3,1,2){m_{{c}}}^{4}-5\,\hat{I}_{{0}}(1,2,2){m_{{b}}}^{4} \\
&&-15\,\hat{I}_{{0}}(2,1,2){m_{{c}}}^{4}-15\,\hat{I}_{0}^{[0,1]}(3,2,1){m_{{c}}}%
^{4}-5\,\hat{I}_{{0}}(3,1,2){m_{{c}}}^{6}+5\,\hat{I}_{{0}}(3,2,1){m_{{c}}}^{6} \\
&&+15\,\hat{I}_{{0}}(2,2,1){m_{{c}}}^{4}+20\,\hat{I}_{0}^{[0,1]}(1,1,2)+15%
\,\hat{I}_{0}^{[0,2]}(2,2,1)
\end{eqnarray*}
\begin{eqnarray*}
C_{1}^{T-PT} &=&+10\,\hat{I}_{{0}}(3,1,1)m_{{c}}m_{{b}}+10\,\hat{I}_{1}^{[0,1]}(2,2,2)m_{{c%
}}m_{{b}}-10\,\hat{I}_{2}^{[0,1]}(2,2,2)m_{{c}}m_{{b}} \\
&&-35\,\hat{I}_{{2}}(2,2,1)m_{{c}}m_{{b}}-10\,\hat{I}_{0}^{[0,1]}(2,2,2)m_{{c}}m_{{b}%
}-10\,\hat{I}_{{0}}(1,1,2)+20\,\hat{I}_{{1}}(1,2,2){m_{{c}}}^{2} \\
&&+15\,\hat{I}_{0}^{[0,1]}(4,1,1){m_{{c}}}^{2}+25\,\hat{I}_{2}^{[0,1]}(3,1,2){m_{{c}}}%
^{2}-15\,\hat{I}_{{0}}(2,1,2){m_{{c}}}^{2}-5\,\hat{I}_{{1}}(3,1,1){m_{{c}}}^{2} \\
&&-15\,\hat{I}_{1}^{[0,1]}(3,1,2){m_{{c}}}^{2}+30\,\hat{I}_{2}^{[0,1]}(2,2,2){m_{{c}}}%
^{2}+15\,\hat{I}_{{1}}(2,1,2){m_{{c}}}^{2}+30\,\hat{I}_{0}^{[0,1]}(2,2,2){m_{{c}}}^{2} \\
&&+5\,\hat{I}_{2}^{[0,1]}(3,2,1){m_{{c}}}^{2}-15\,\hat{I}_{1}^{[0,1]}(3,2,1){m_{{c}}}%
^{2}+15\,\hat{I}_{1}^{[0,2]}(3,2,2){m_{{c}}}^{2}-15\,\hat{I}_{0}^{[0,2]}(3,2,2){m_{{c}}}%
^{2} \\
&&+10\,\hat{I}_{{2}}(2,2,1){m_{{c}}}^{2}+10\,\hat{I}_{{2}}(3,1,1){m_{{c}}}^{2}-5\,\hat{I}_{{0}%
}(3,2,2){m_{{c}}}^{3}{m_{{b}}}^{3}-5\,\hat{I}_{{2}}(3,2,2){m_{{c}}}^{3}{m_{{b}}}%
^{3} \\
&&+5\,\hat{I}_{{2}}(3,2,2){m_{{c}}}^{2}{m_{{b}}}^{4}-15\,\hat{I}_{{1}}(4,1,1){m_{{c}}}%
^{3}m_{{b}}+10\,\hat{I}_{{2}}(2,2,2){m_{{c}}}^{3}m_{{b}}-15\,\hat{I}_{0}^{[0,2]}(2,2,2)
\\
&&-15\,\hat{I}_{{1}}(2,1,2)m_{{c}}m_{{b}}-5\,\hat{I}_{2}^{[0,1]}(3,2,1)m_{{c}}m_{{b}%
}+5\,\hat{I}_{0}^{[0,2]}(3,2,2)m_{{c}}m_{{b}}-10\,\hat{I}_{0}^{[0,1]}(3,2,1)m_{{c}}m_{{b}%
} \\
&&-15\,\hat{I}_{0}^{[0,1]}(3,1,2)m_{{c}}m_{{b}}+15\,\hat{I}_{1}^{[0,1]}(3,1,2)m_{{c}}m_{{%
b}}+10\,\hat{I}_{{2}}(1,2,2)m_{{c}}m_{{b}}-10\,\hat{I}_{1}^{[0,1]}(2,3,1)m_{{c}}m_{{b}}
\\
&&+5\,\hat{I}_{{1}}(3,1,1)m_{{c}}m_{{b}}+30\,\hat{I}_{2}^{[0,1]}(2,1,2)-20%
\,\hat{I}_{1}^{[0,1]}(2,1,2)+10\,\hat{I}_{1}^{[0,2]}(3,1,2) \\
&&-5\,\hat{I}_{2}^{[0,1]}(3,1,1)-10\,\hat{I}_{{1}}(1,2,2)m_{{c}}m_{{b}}+10\,\hat{I}_{{0}%
}(1,2,2)m_{{c}}m_{{b}}-15\,\hat{I}_{2}^{[0,1]}(3,1,2)m_{{c}}m_{{b}} \\
&&-15\,\hat{I}_{2}^{[0,2]}(3,1,2)+30\,\hat{I}_{{2}}(1,3,1){m_{{b}}}^{2}+20\,\hat{I}_{{2}%
}(2,2,1){m_{{b}}}^{2}+5\,\hat{I}_{{2}}(3,1,1){m_{{b}}}^{2} \\
&&-5\,\hat{I}_{0}^{[0,2]}(3,2,2){m_{{b}}}^{2}-10\,\hat{I}_{2}^{[0,2]}(3,2,2){m_{{b}}}%
^{2}+10\,\hat{I}_{0}^{[0,1]}(2,2,2){m_{{b}}}^{2}+5\,\hat{I}_{{0}}(2,2,1){m_{{b}}}^{2} \\
&&-30\,\hat{I}_{0}^{[0,1]}(1,4,1){m_{{b}}}^{2}-10\,\hat{I}_{{1}}(2,2,1){m_{{b}}}%
^{2}-5\,\hat{I}_{{0}}(1,2,2){m_{{b}}}^{2}+5\,\hat{I}_{1}^{[0,2]}(3,2,2){m_{{b}}}^{2} \\
&&+10\,\hat{I}_{2}^{[0,1]}(3,2,1){m_{{b}}}^{2}-10\,\hat{I}_{1}^{[0,1]}(2,2,2){m_{{b}}}%
^{2}-15\,\hat{I}_{{2}}(1,2,2){m_{{b}}}^{2}+30\,\hat{I}_{1}^{[0,1]}(1,4,1){m_{{b}}}^{2} \\
&&-5\,\hat{I}_{1}^{[0,1]}(2,2,1)-5\,\hat{I}_{2}^{[0,1]}(2,2,1)+20%
\,\hat{I}_{0}^{[0,1]}(1,2,2)-15\,\hat{I}_{{2}}(2,1,2){m_{{b}}}^{2}-30%
\,\hat{I}_{2}^{[0,1]}(1,4,1){m_{{b}}}^{2} \\
&&+15\,\hat{I}_{2}^{[0,1]}(3,1,2){m_{{b}}}^{2}+20\,\hat{I}_{2}^{[0,1]}(2,2,2){m_{{b}}}%
^{2}-10\,\hat{I}_{1}^{[0,1]}(3,2,1){m_{{b}}}^{2}+15\,\hat{I}_{0}^{[0,1]}(3,2,1){m_{{b}}}%
^{2} \\
&&+5\,\hat{I}_{{1}}(1,2,2){m_{{b}}}^{2}+15\,\hat{I}_{1}^{[0,2]}(2,2,2)-15%
\,\hat{I}_{2}^{[0,2]}(2,2,2)+10\,\hat{I}_{0}^{[0,1]}(3,1,1) \\
&&+15\,\hat{I}_{0}^{[0,1]}(2,2,1)-5\,\hat{I}_{1}^{[0,3]}(3,2,2)+5%
\,\hat{I}_{2}^{[0,3]}(3,2,2)-5\,\hat{I}_{{2}}(3,2,2){m_{{c}}}^{6}-5\,\hat{I}_{{0}}(3,2,2){m_{{c%
}}}^{6} \\
&&+5\,\hat{I}_{{1}}(3,2,1){m_{{c}}}^{4}-10\,\hat{I}_{{0}}(3,2,1){m_{{c}}}^{4}-15\,\hat{I}_{{2}%
}(4,1,1){m_{{c}}}^{4}-15\,\hat{I}_{{2}}(2,2,2){m_{{c}}}^{4} \\
&&+15\,\hat{I}_{{1}}(2,2,2){m_{{c}}}^{4}+15\,\hat{I}_{{1}}(4,1,1){m_{{c}}}%
^{4}+15\,\hat{I}_{2}^{[0,1]}(3,2,2){m_{{c}}}^{4}-15\,\hat{I}_{1}^{[0,1]}(3,2,2){m_{{c}}}%
^{4} \\
&&-15\,\hat{I}_{{0}}(2,2,2){m_{{c}}}^{4}+5\,\hat{I}_{{2}}(3,2,2){m_{{c}}}^{5}m_{{b}%
}+5\,\hat{I}_{{0}}(3,2,2){m_{{c}}}^{5}m_{{b}}-5\,\hat{I}_{{1}}(3,2,2){m_{{c}}}^{5}m_{{b}}
\\
&&+5\,\hat{I}_{{0}}(3,2,2){m_{{c}}}^{4}{m_{{b}}}^{2}-5\,\hat{I}_{{1}}(3,2,2){m_{{c}}}^{4}%
{m_{{b}}}^{2}+5\,\hat{I}_{{1}}(3,2,2){m_{{c}}}^{3}{m_{{b}}}^{3}+5\,\hat{I}_{{1}}(3,2,2){%
m_{{c}}}^{6} \\
&&-5\,\hat{I}_{{1}}(3,1,2){m_{{c}}}^{3}m_{{b}}+10\,\hat{I}_{1}^{[0,1]}(3,2,2){m_{{c}}}%
^{3}m_{{b}}+10\,\hat{I}_{{1}}(2,3,1){m_{{c}}}^{3}m_{{b}}-10\,\hat{I}_{0}^{[0,1]}(3,2,2){%
m_{{c}}}^{3}m_{{b}} \\
&&+5\,\hat{I}_{{0}}(3,1,2){m_{{c}}}^{3}m_{{b}}-5\,\hat{I}_{{1}}(3,2,1){m_{{c}}}^{3}m_{{b}%
}-10\,\hat{I}_{{1}}(2,2,2){m_{{c}}}^{3}m_{{b}}+5\,\hat{I}_{{2}}(3,1,2){m_{{c}}}^{3}m_{{b}%
} \\
&&+15\,\hat{I}_{{2}}(4,1,1){m_{{c}}}^{3}m_{{b}}+10\,\hat{I}_{{0}}(3,2,1){m_{{c}}}^{3}m_{{%
b}}-10\,\hat{I}_{2}^{[0,1]}(3,2,2){m_{{c}}}^{3}m_{{b}}+5\,\hat{I}_{2}^{[0,1]}(3,2,2){m_{{%
b}}}^{4} \\
&&-15\,\hat{I}_{{0}}(4,1,1){m_{{c}}}^{4}+5\,\hat{I}_{{1}}(3,1,2){m_{{c}}}^{4}-10\,\hat{I}_{{2}%
}(3,1,2){m_{{c}}}^{4}-5\,\hat{I}_{{0}}(3,1,2){m_{{c}}}^{4} \\
&&\hat{I}_{{0}}(2,2,2){m_{{c}}}^{3}m_{{b}}+5\,\hat{I}_{{2}}(3,2,1){m_{{c}}}^{3}m_{{b}%
}-10\,\hat{I}_{{2}}(2,3,1){m_{{c}}}^{3}m_{{b}}-10\,\hat{I}_{{2}}(2,2,2){m_{{c}}}^{2}{m_{{%
b}}}^{2} \\
&&-20\,\hat{I}_{{2}}(3,2,1){m_{{c}}}^{2}{m_{{b}}}^{2}+30\,\hat{I}_{{0}}(1,4,1){m_{{c}}}%
^{2}{m_{{b}}}^{2}+10\,\hat{I}_{{1}}(3,2,1){m_{{c}}}^{2}{m_{{b}}}^{2}-5\,\hat{I}_{{0}%
}(3,2,1){m_{{c}}}^{2}{m_{{b}}}^{2} \\
&&+10\,\hat{I}_{2}^{[0,1]}(3,2,2){m_{{c}}}^{2}{m_{{b}}}^{2}+30\,\hat{I}_{{2}}(1,4,1){m_{{%
b}}}^{4}-15\,\hat{I}_{{2}}(4,1,1){m_{{c}}}^{2}{m_{{b}}}^{2}+30\,\hat{I}_{{2}}(1,4,1){m_{{%
c}}}^{2}{m_{{b}}}^{2} \\
&&+5\,\hat{I}_{{2}}(3,2,1)m_{{c}}{m_{{b}}}^{3}+5\,\hat{I}_{1}^{[0,1]}(3,2,2)m_{{c}}{m_{{b%
}}}^{3}+5\,\hat{I}_{{2}}(3,1,2)m_{{c}}{m_{{b}}}^{3}-30\,\hat{I}_{{2}}(1,4,1)m_{{c}}{m_{{b%
}}}^{3} \\
&&-10\,\hat{I}_{{2}}(2,3,1)m_{{c}}{m_{{b}}}^{3}-5\,\hat{I}_{{2}}(3,2,1){m_{{b}}}%
^{4}-30\,\hat{I}_{{0}}(1,4,1)m_{{c}}{m_{{b}}}^{3}+10\,\hat{I}_{{0}}(2,3,1)m_{{c}}{m_{{b}}%
}^{3} \\
&&+10\,\hat{I}_{2}^{[0,1]}(2,3,1)m_{{c}}m_{{b}}-40\,\hat{I}_{{0}}(1,3,1)m_{{c}}m_{{b}%
}+15\,\hat{I}_{{0}}(2,1,2)m_{{c}}m_{{b}}-10\,\hat{I}_{{2}}(2,2,2){m_{{b}}}^{4}
\end{eqnarray*}
where
\begin{eqnarray}
\hat{I}_n^{[i,j]} (a,b,c) = \left( M_1^2 \right)^i \left( M_2^2
\right)^j
\frac{d^i}{d\left( M_1^2 \right)^i} \frac{d^j}{d\left( M_2^2 \right)^j} %
\left[\left( M_1^2 \right)^i \left( M_2^2 \right)^j \hat{I}_n(a,b,c) \right]%
~.  \nonumber
\end{eqnarray}

\newpage
\begin{figure}
\vspace*{0.4cm}
\begin{center}
\begin{picture}(160,150)
\centerline{ \epsfxsize=14cm \epsfbox{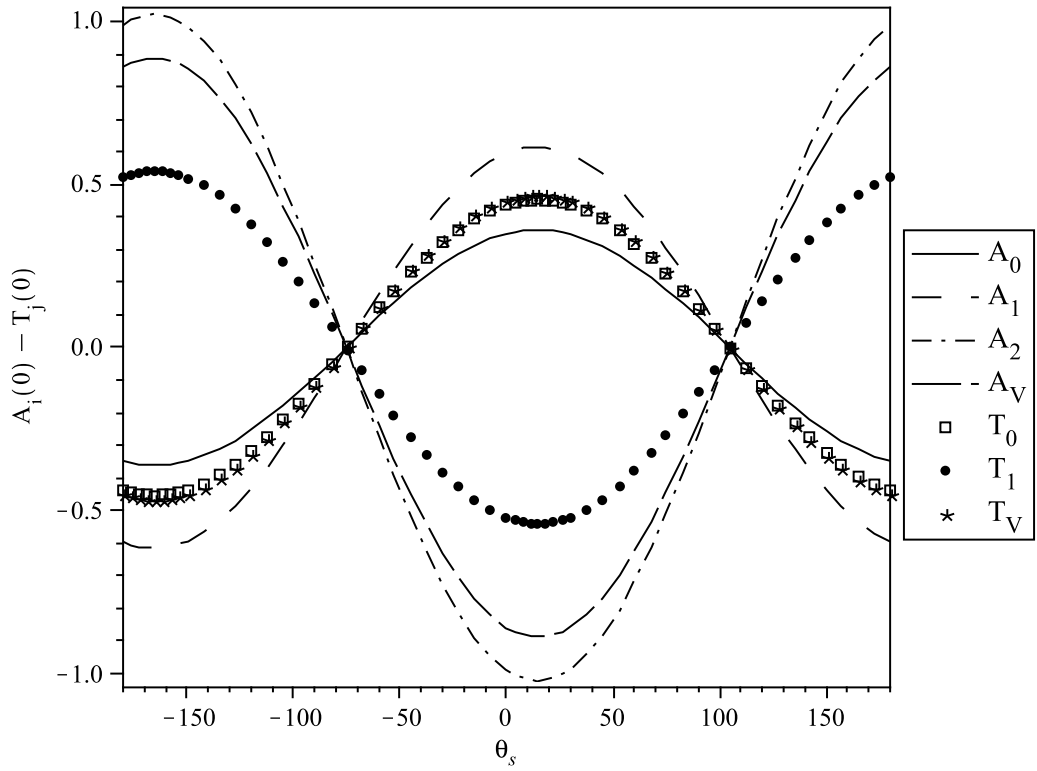}}
\end{picture}
\end{center}
\vspace*{-0.1cm} \caption{The dependence of the form factors on
$\theta_{s}$ at $q^2=0$ for $B_c\to D_{s1}(2460)$ transition.}
\label{F3}
\end{figure}
\normalsize
\newpage
\begin{figure}
\vspace*{0.4cm}
\begin{center}
\begin{picture}(160,150)
\centerline{ \epsfxsize=14cm \epsfbox{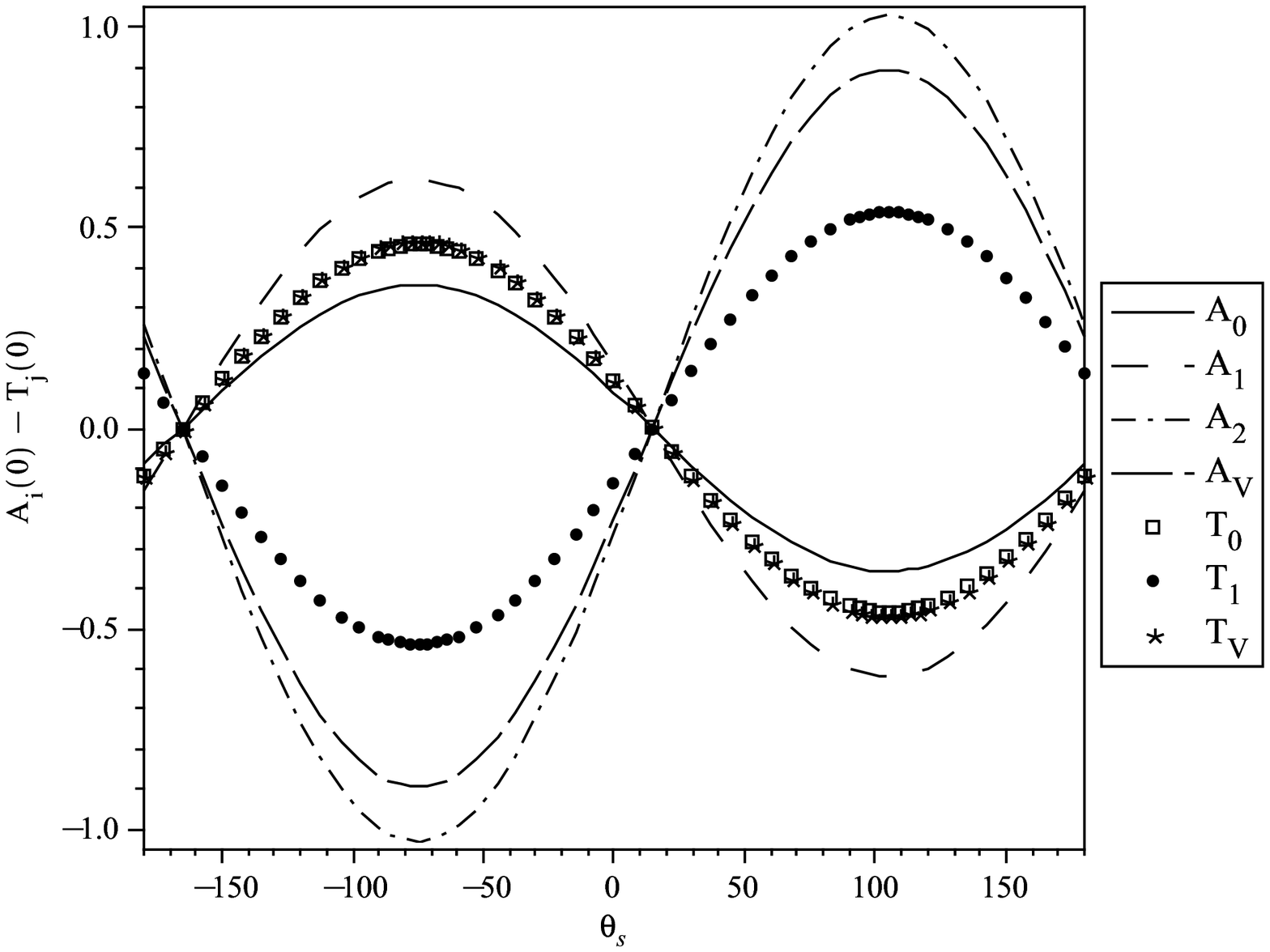}}
\end{picture}
\end{center}
\vspace*{-0.1cm} \caption{The dependence of the form factors on
$\theta_{s}$ at $q^2=0$ for $B_c\to D_{s1}(2536)$ transition.}
\label{F4}
\end{figure}
\normalsize
\begin{figure}[th]
\begin{center}
\begin{picture}(0,0)
\put(-110,-40){ \epsfxsize=7cm \epsfbox{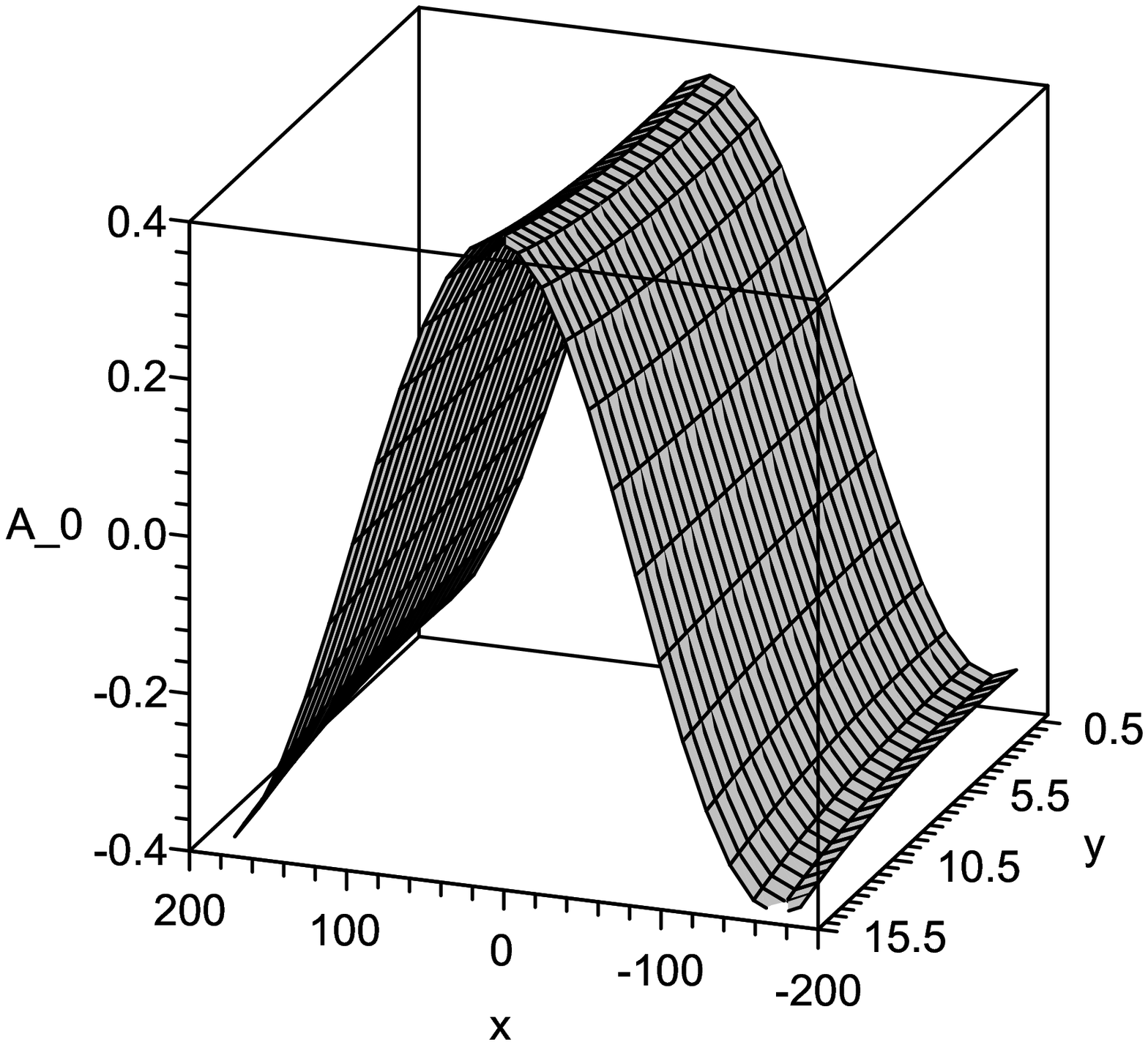}
\epsfxsize=7cm \epsfbox{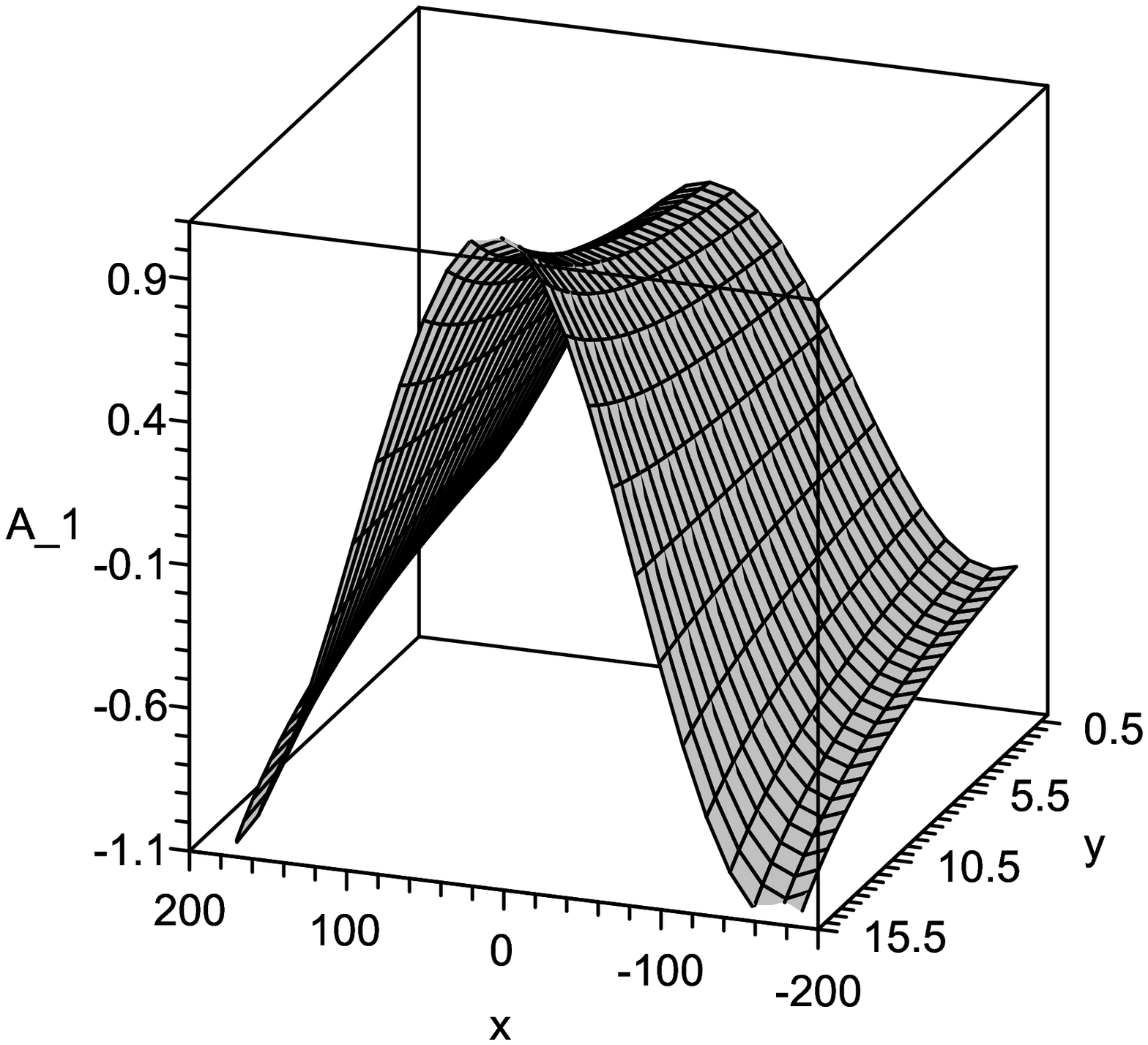} \epsfxsize=7cm
\epsfbox{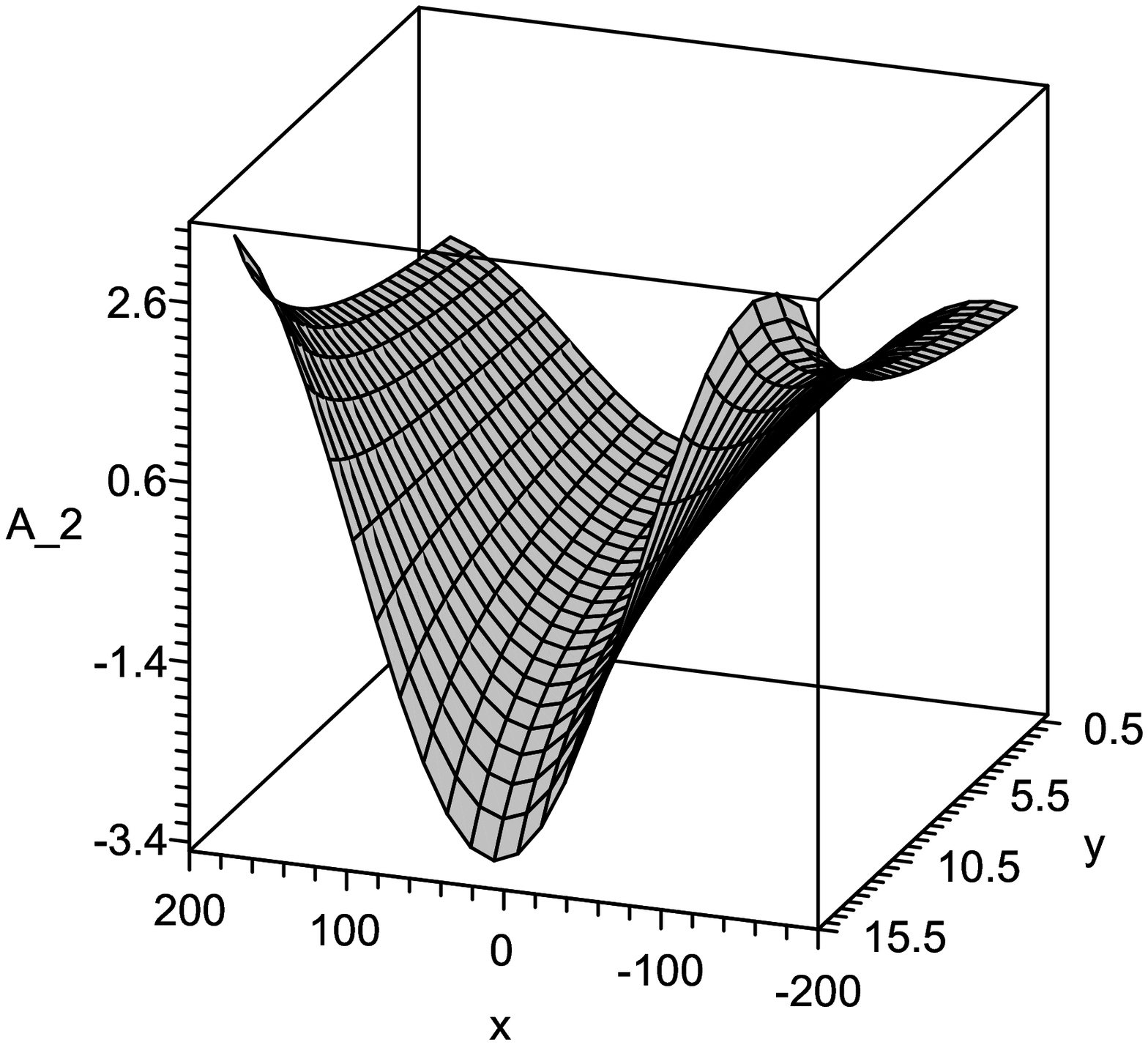}} \put(-110,-105){ \epsfxsize=7cm
\epsfbox{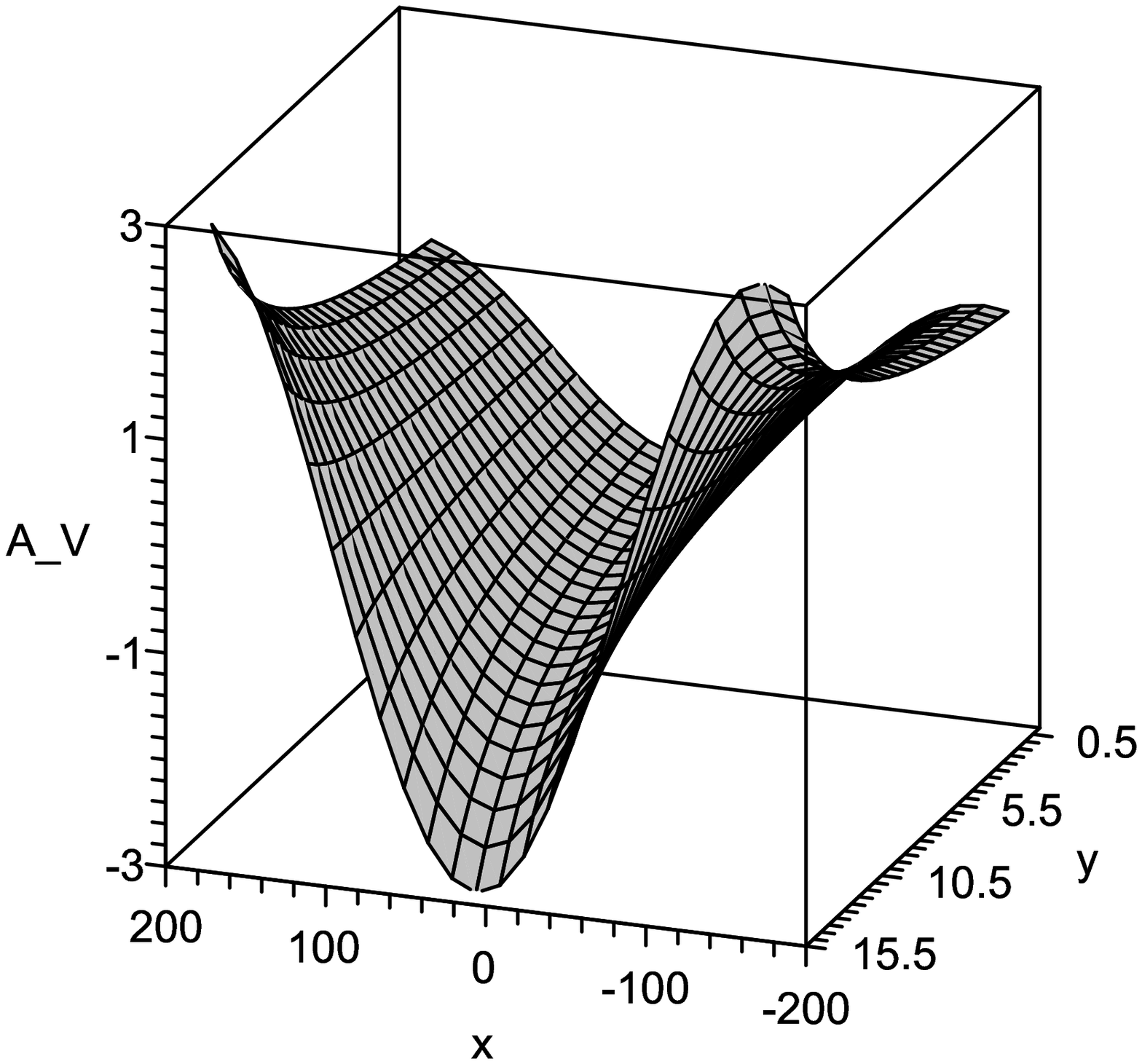} \epsfxsize=7cm
\epsfbox{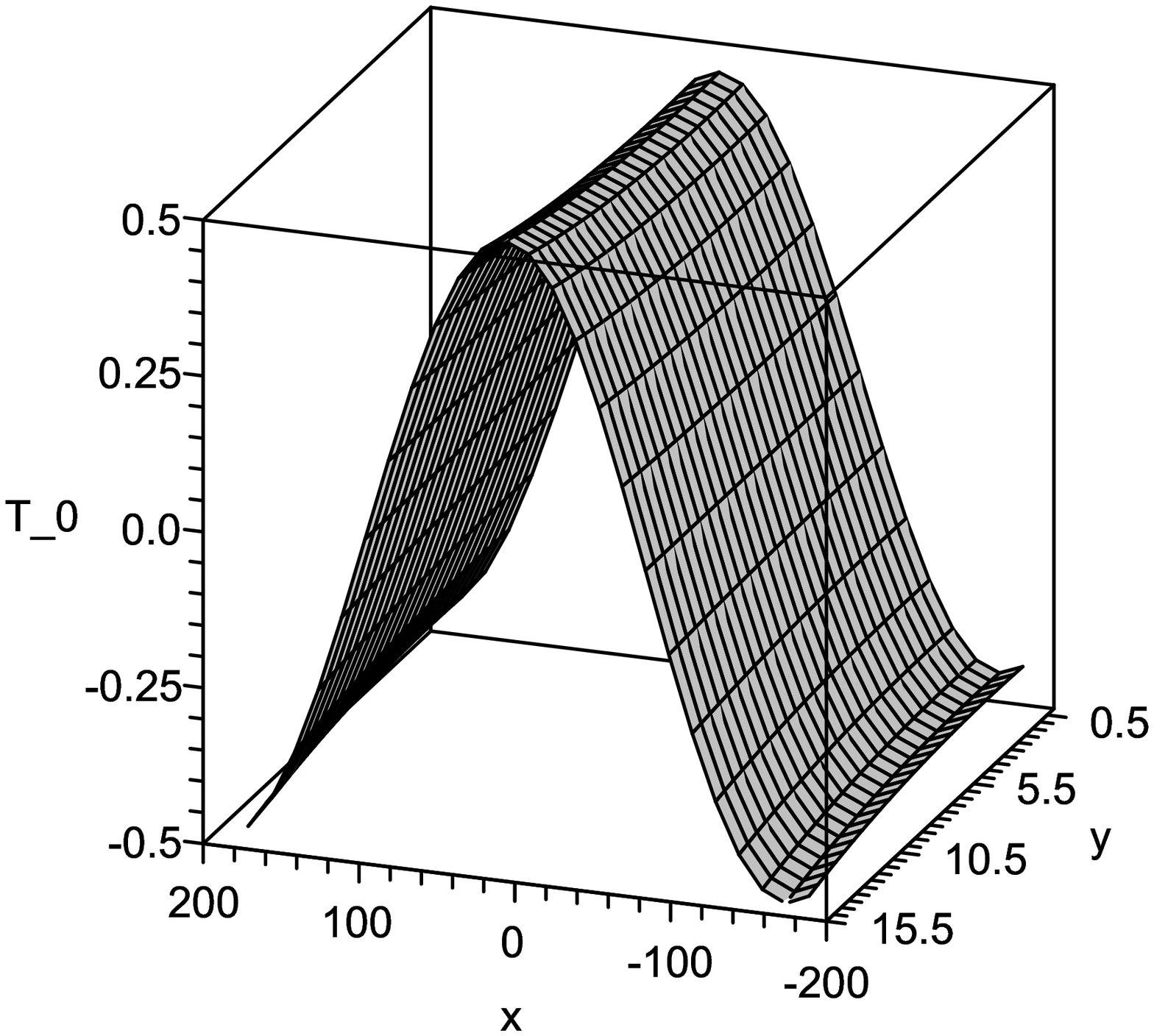}\epsfxsize=7cm \epsfbox{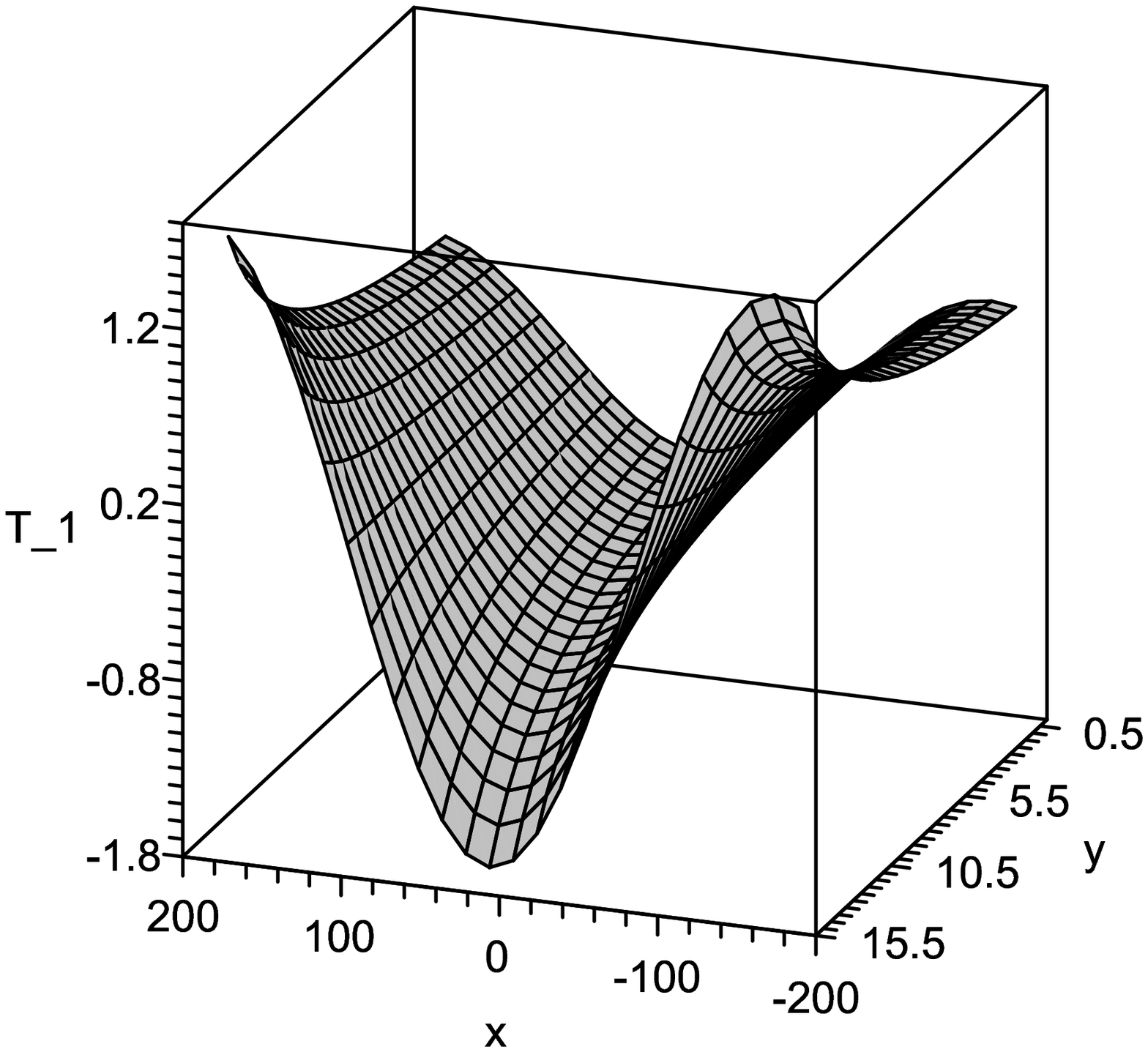}}
\put(-40,-170){ \epsfxsize=7cm \epsfbox{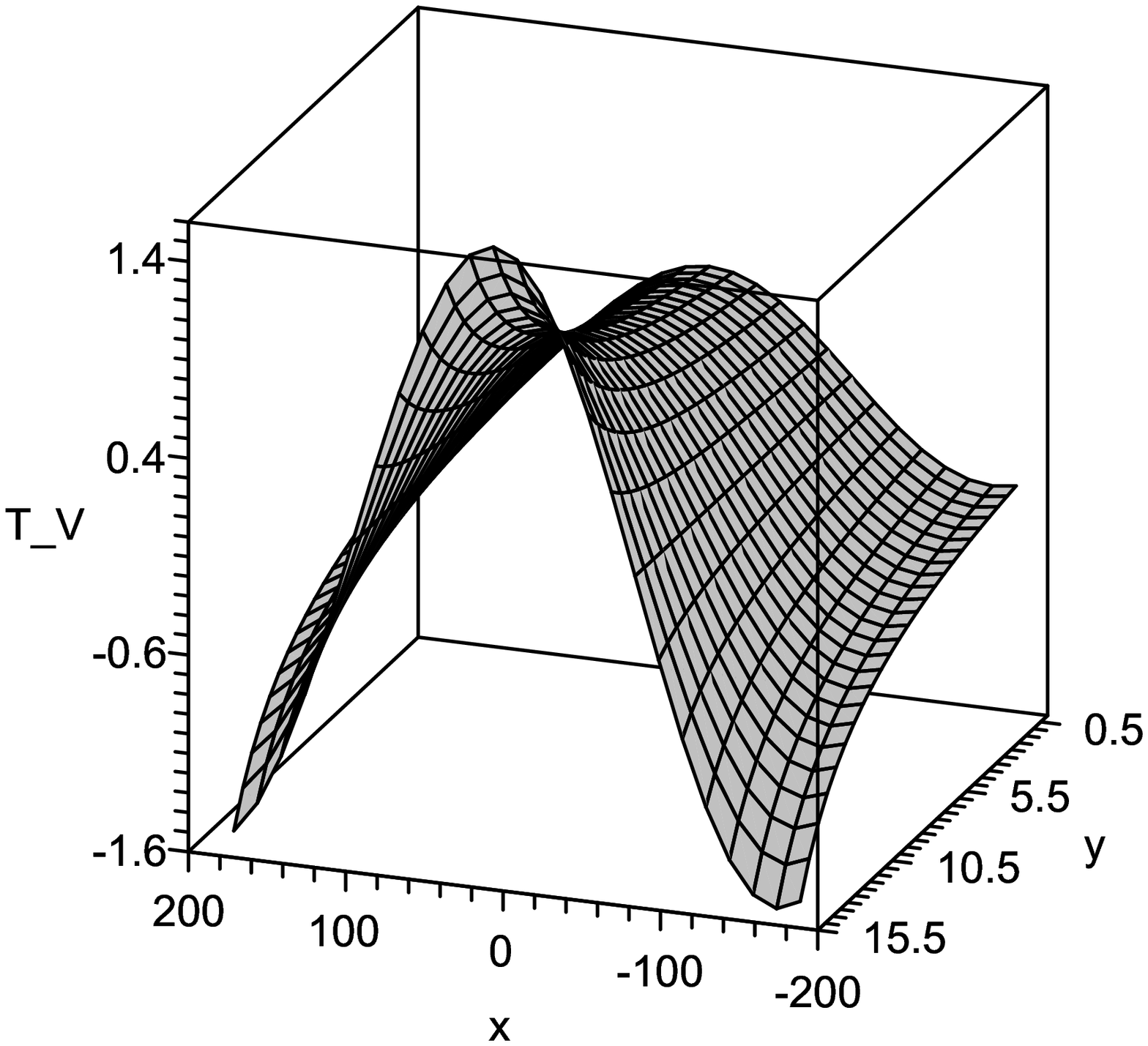}}
\end{picture}
\end{center}
\vspace*{17cm}\caption{The dependence of the transition form
factors on $q^2$ and $\theta_{s}$ for $B_c\to D_{s1}(2460)$
transition. In these figures, $x=\theta_s$ and $y=q^2$.}\label{F5}
\end{figure}
\normalsize
\begin{figure}[th]
\begin{center}
\begin{picture}(0,0)
\put(-110,-40){ \epsfxsize=7cm \epsfbox{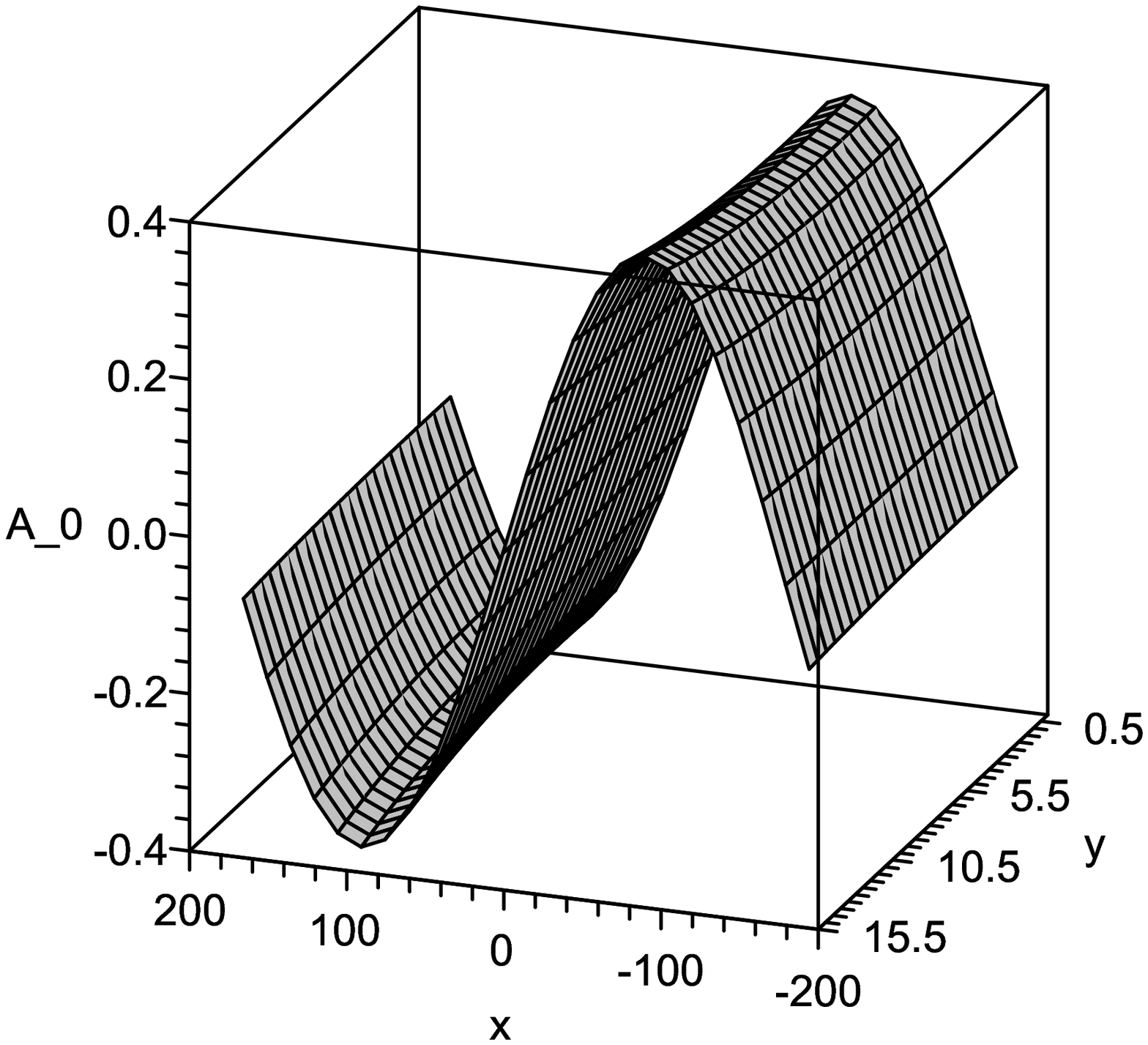}
\epsfxsize=7cm \epsfbox{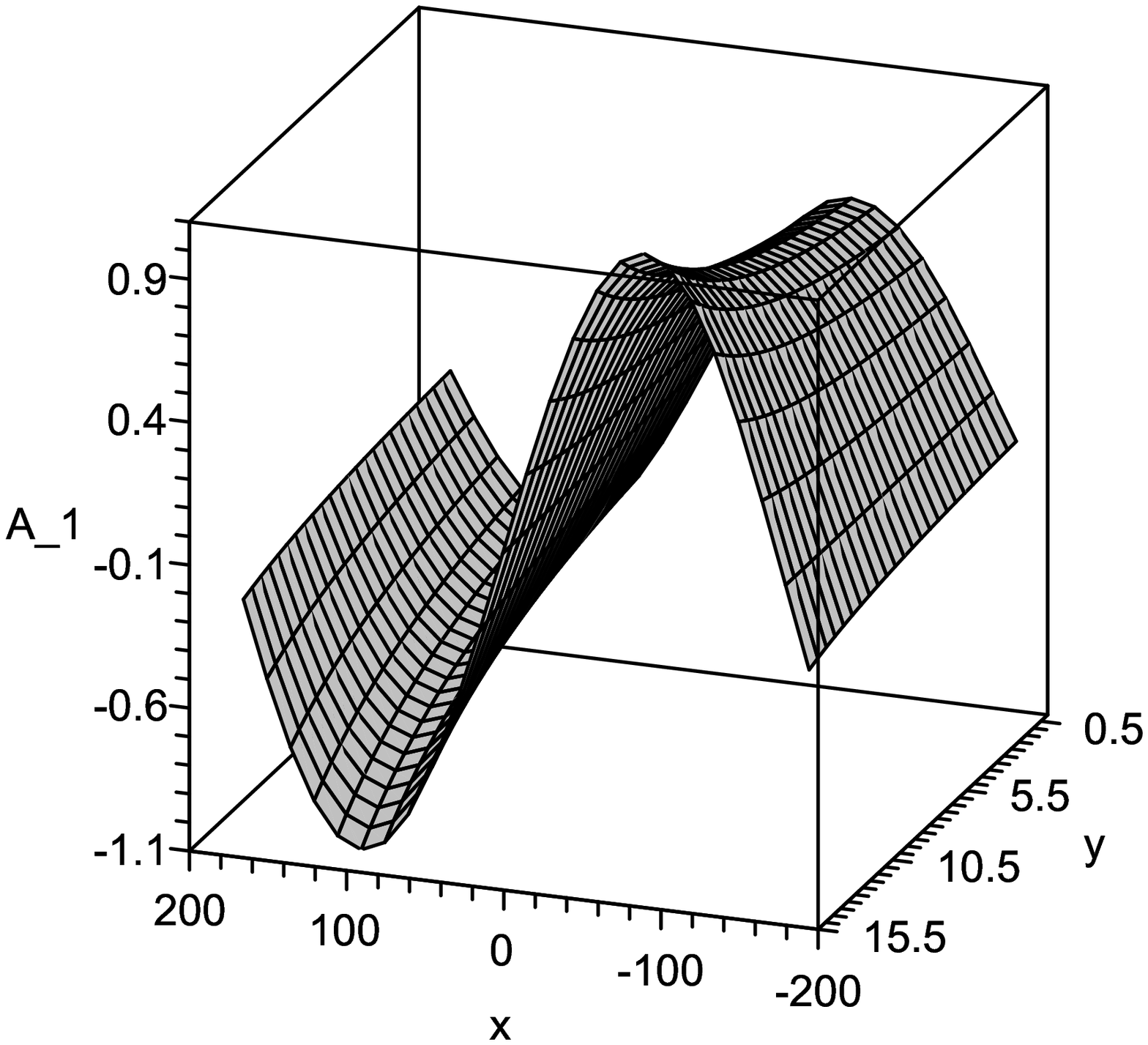} \epsfxsize=7cm
\epsfbox{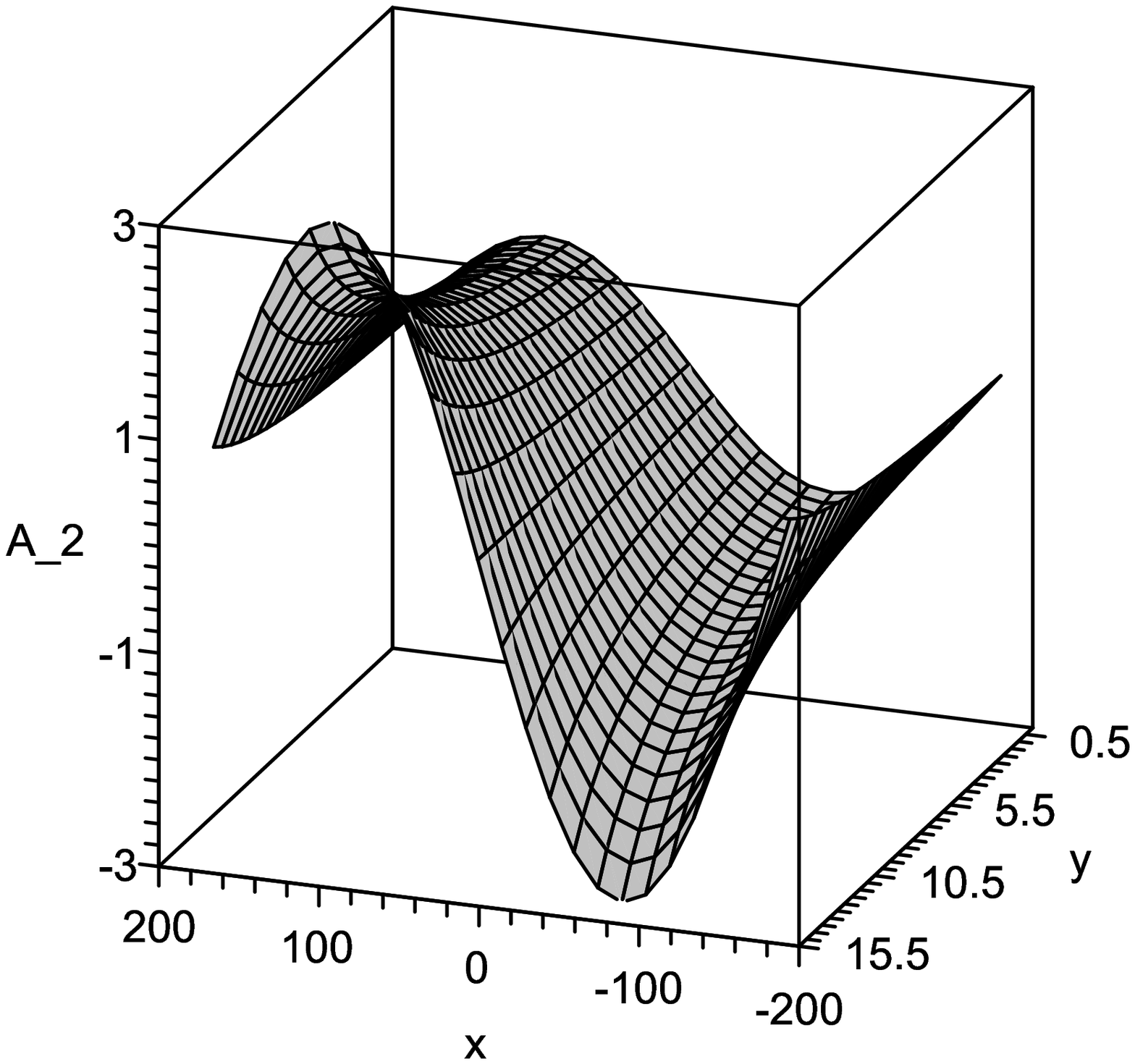}} \put(-110,-105){ \epsfxsize=7cm
\epsfbox{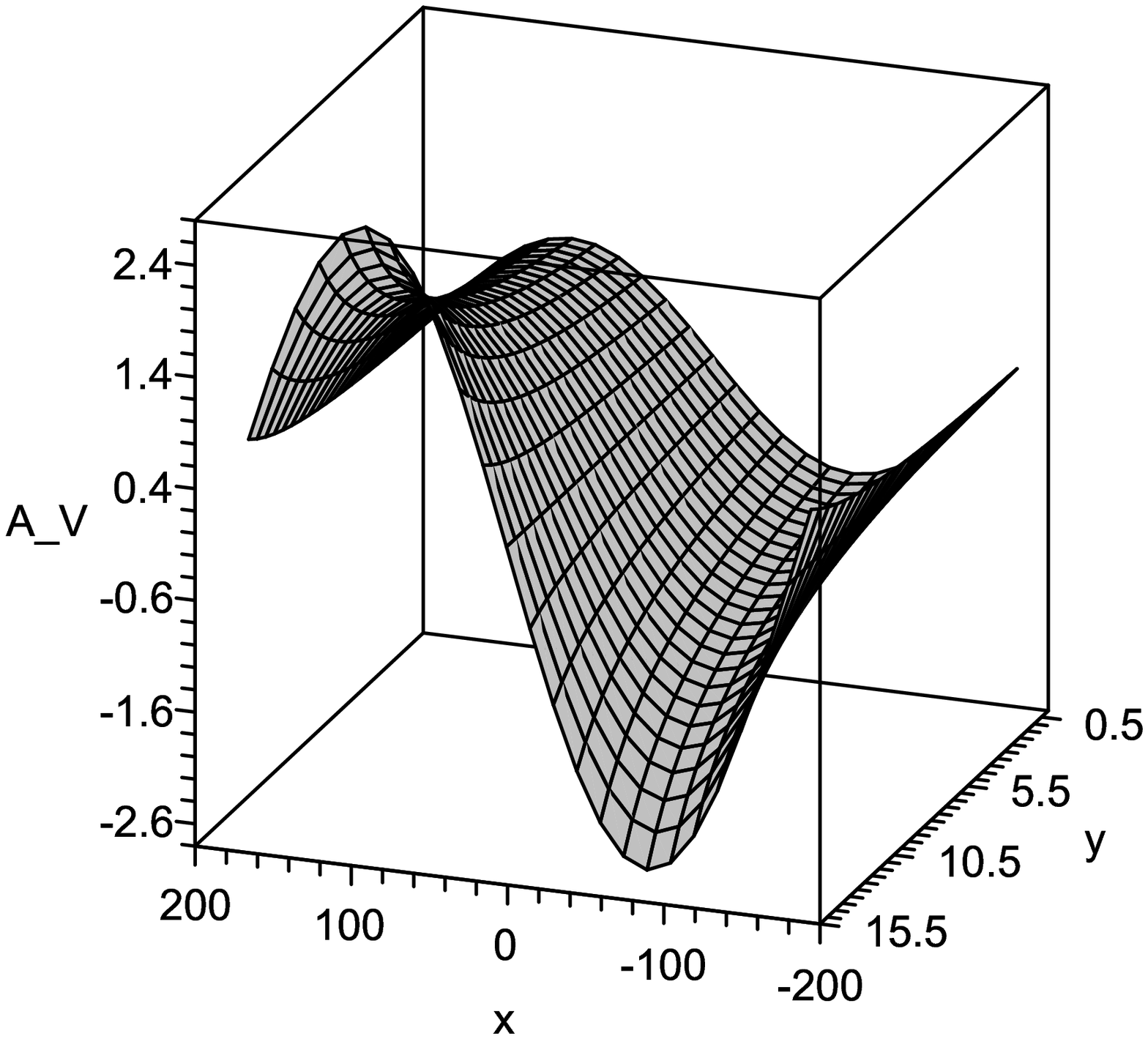} \epsfxsize=7cm
\epsfbox{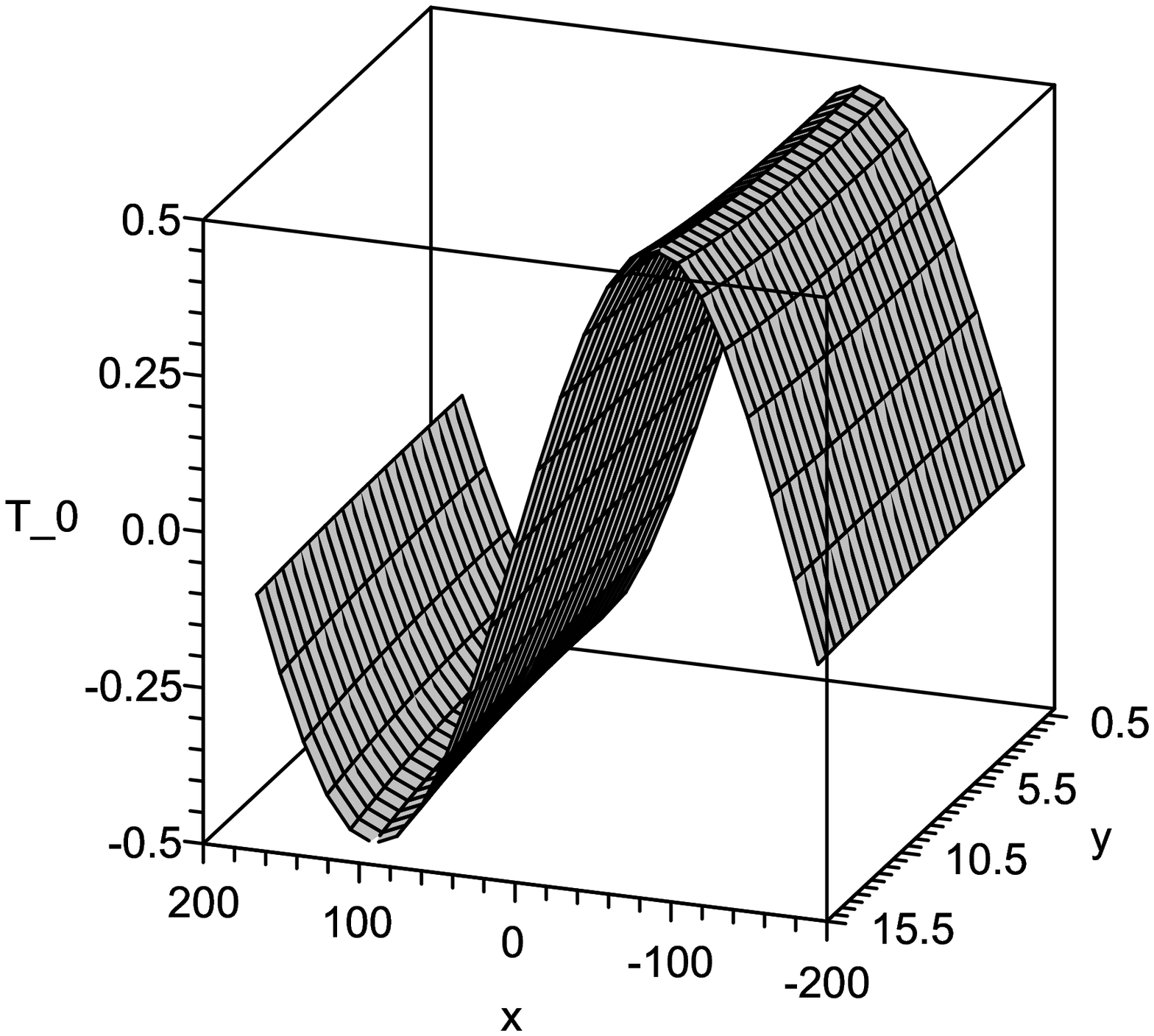}\epsfxsize=7cm \epsfbox{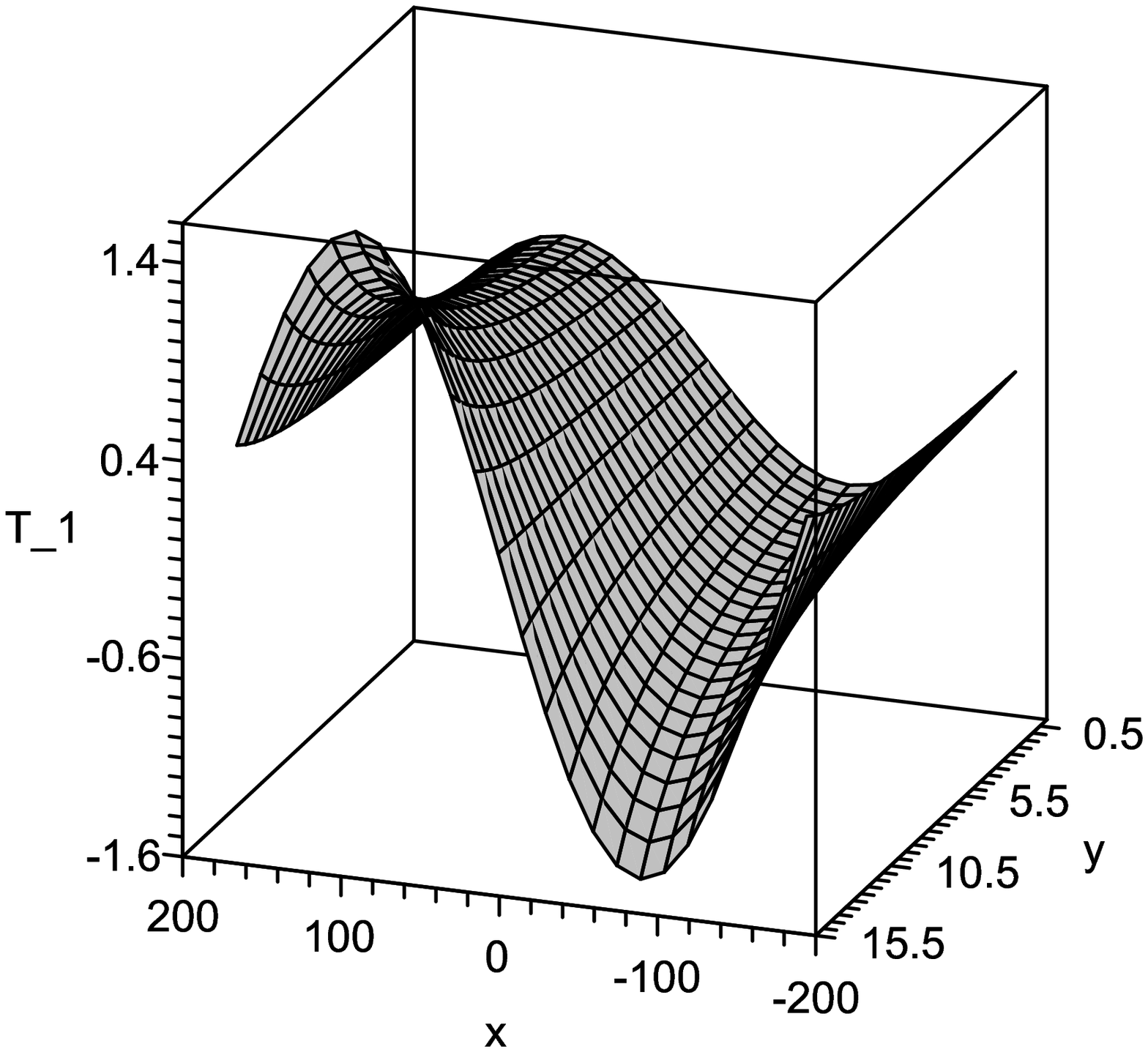}}
\put(-40,-170){ \epsfxsize=7cm \epsfbox{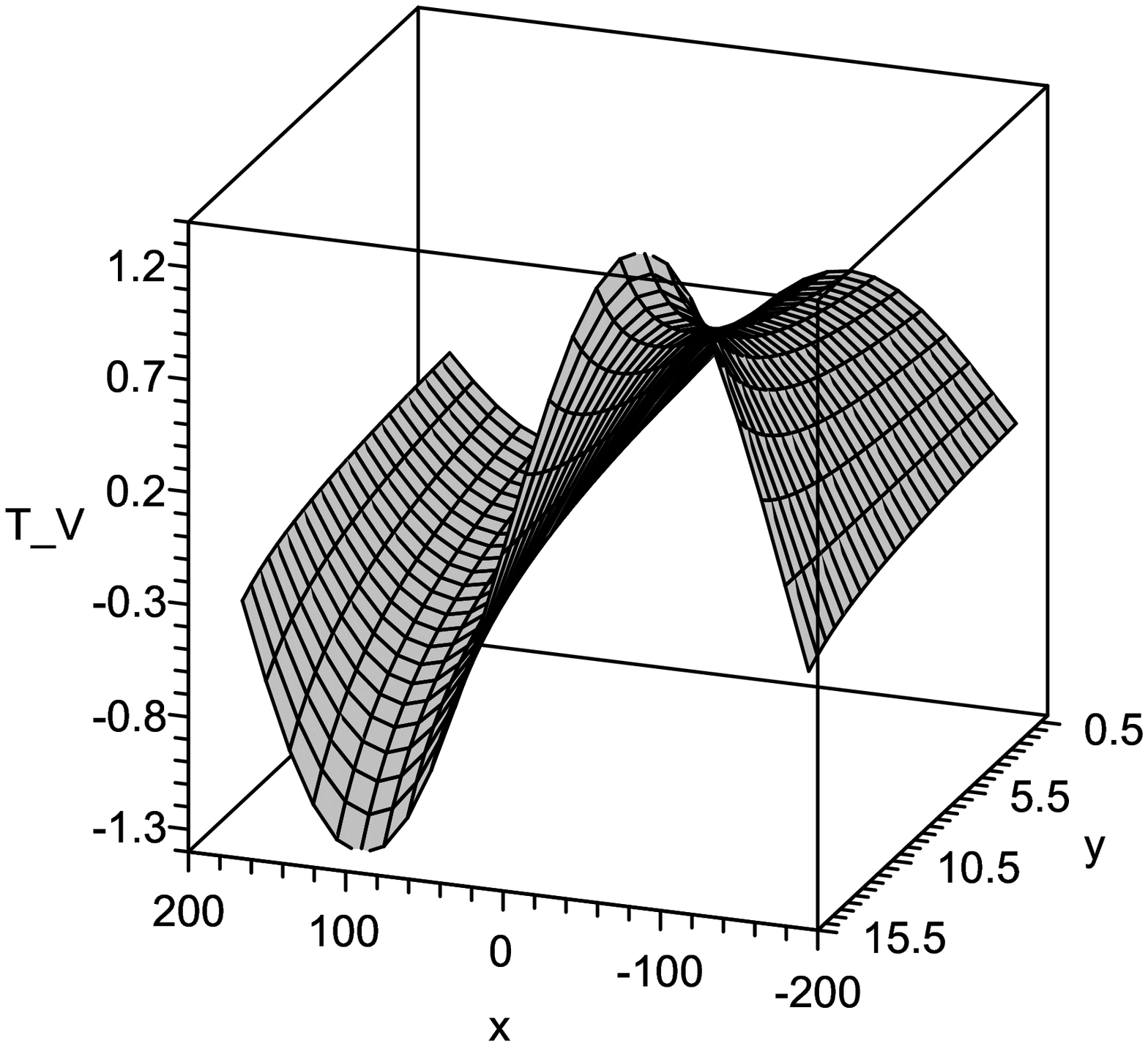}}
\end{picture}
\end{center}
\vspace*{17cm}\caption{The dependence of the transition form
factors on $q^2$ and $\theta_{s}$ for $B_c\to D_{s1}(2536)$
transition. In these figures, $x=\theta_s$ and $y=q^2$.}\label{F6}
\end{figure}
\normalsize
\newpage
\begin{figure}[th]
\begin{center}
\begin{picture}(160,100)
\put(-25,20){ \epsfxsize=10cm \epsfbox{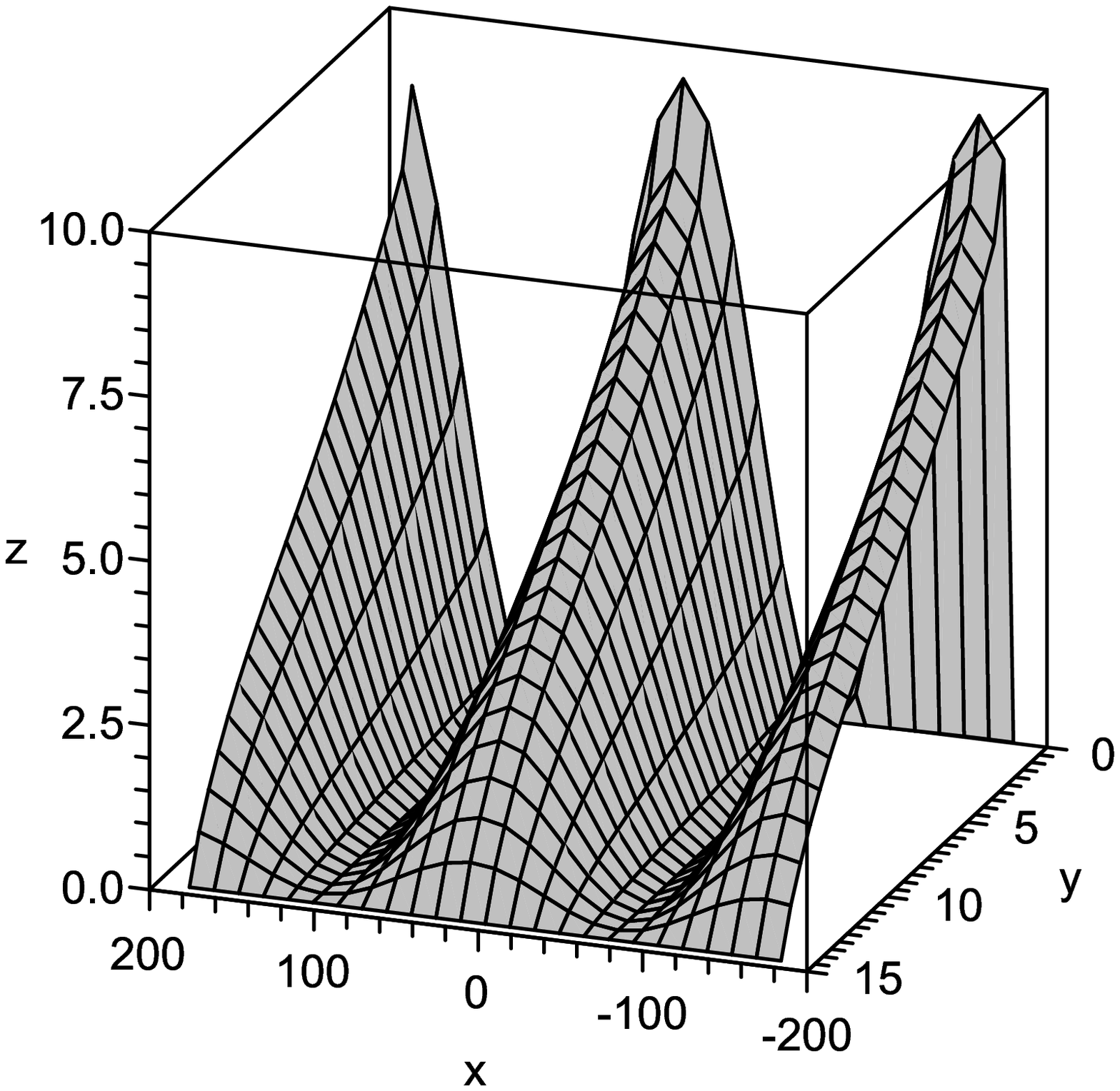}
\epsfxsize=10cm \epsfbox{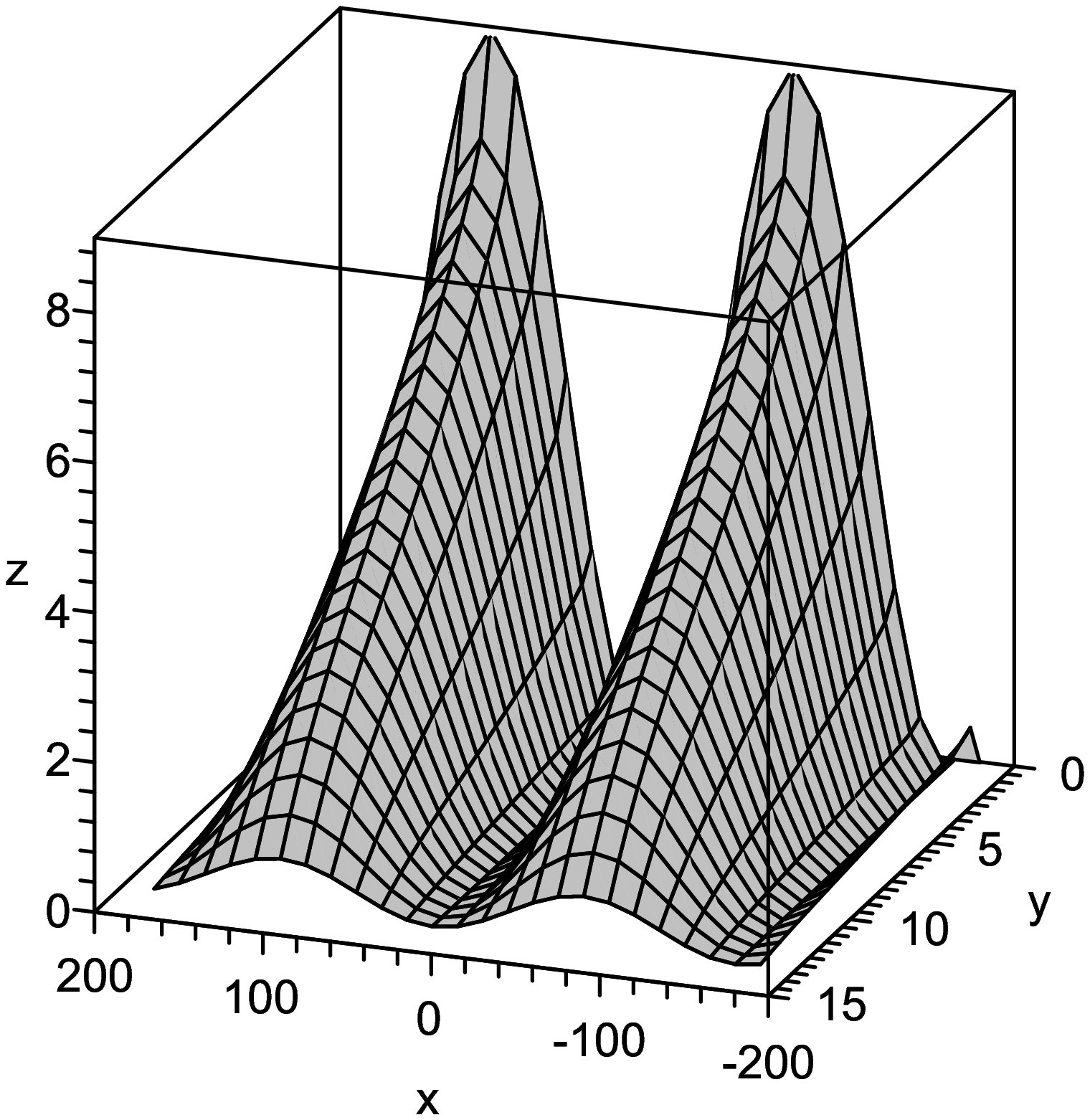}}
\end{picture}
\end{center}
\vspace*{-3cm}\caption{The decay width for $B_c\to
D_{s1}\mu^+\mu^-$ with respect to $\theta_s$ and $q^2$. The left figure
shows decay width of $B_c\to D_{s1}(2460)\mu^+\mu^-$ and right figure  belongs to
$B_c\to D_{s1}(2536)\mu^+\mu^-$. In these figures, $x=\theta_s$,
$y=q^2$ and $z=\Gamma(B_c\to D_{s1}\mu^+\mu^-)\times 10^{-20}$.
}\label{F7}
\end{figure}
\normalsize
\newpage
\begin{figure}[th]
\begin{center}
\begin{picture}(160,100)
\put(-25,0){ \epsfxsize=10cm \epsfbox{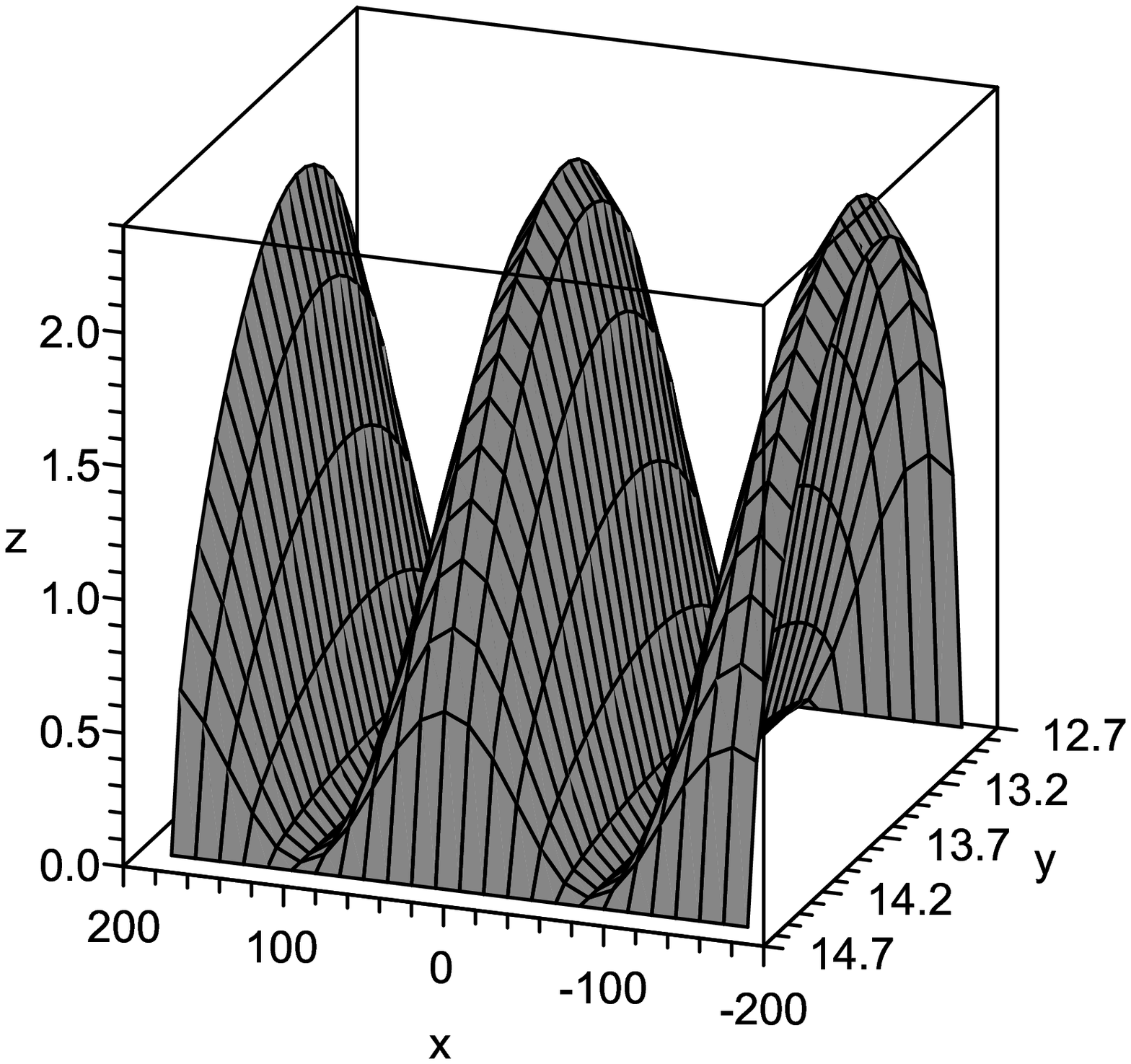}
\epsfxsize=10cm \epsfbox{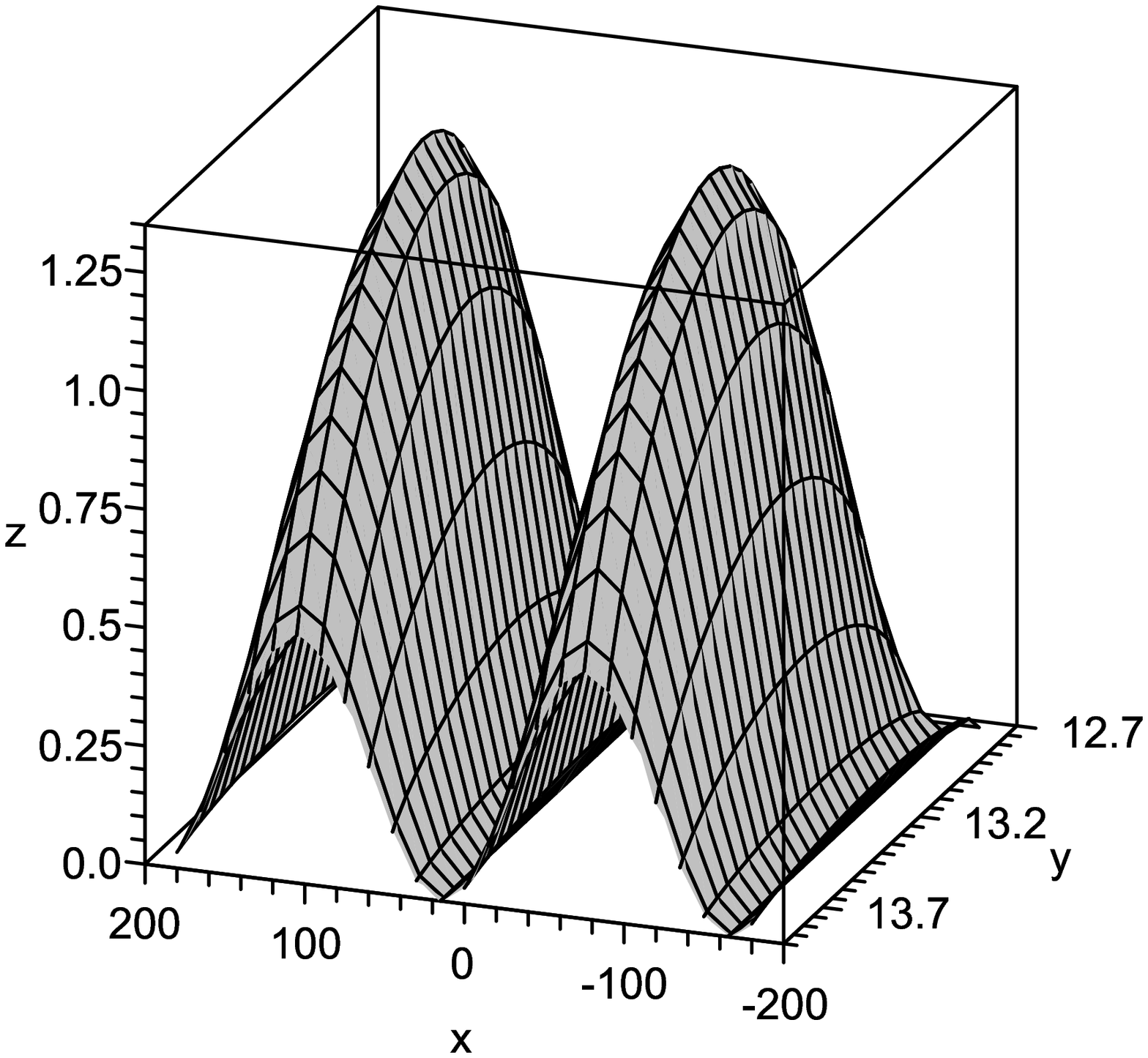}}
\end{picture}
\end{center}
\vspace*{0cm}\caption{The decay width for $B_c\to
D_{s1}\tau^+\tau^-$ with respect to $\theta_s$ and $q^2$. The left figure
shows decay width of $B_c\to D_{s1}(2460)\tau^+\tau^-$ and right figure  belongs to
$B_c\to D_{s1}(2536)\tau^+\tau^-$. In these figures, $x=\theta_s$,
$y=q^2$ and $z=\Gamma(B_c\to D_{s1}\tau^+\tau^-)\times 10^{-21}$.
}\label{F8}
\end{figure}
\normalsize
\newpage
\begin{figure}[th]
\begin{center}
\begin{picture}(160,100)
\put(-25,20){ \epsfxsize=10cm \epsfbox{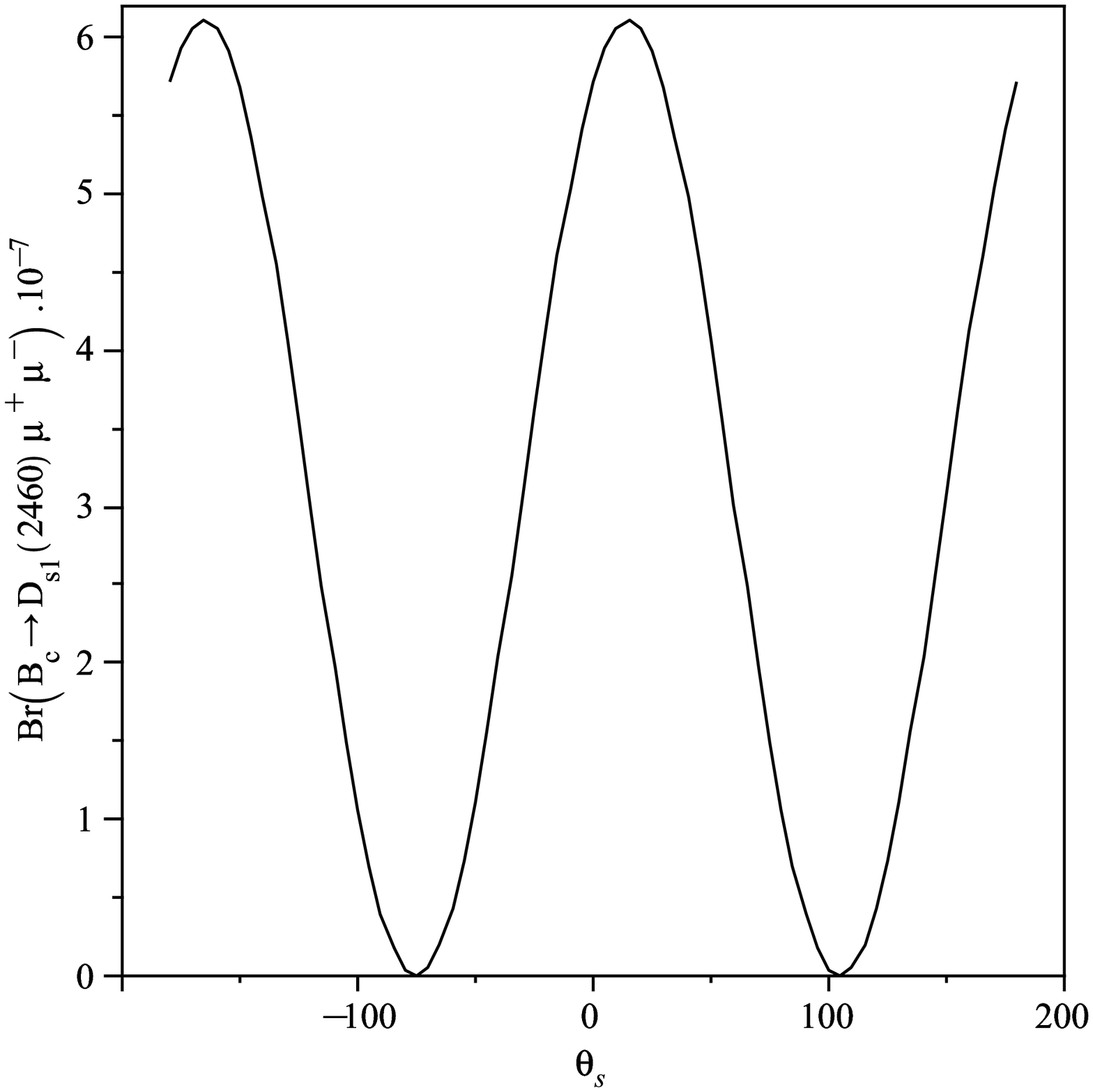}
\epsfxsize=10cm \epsfbox{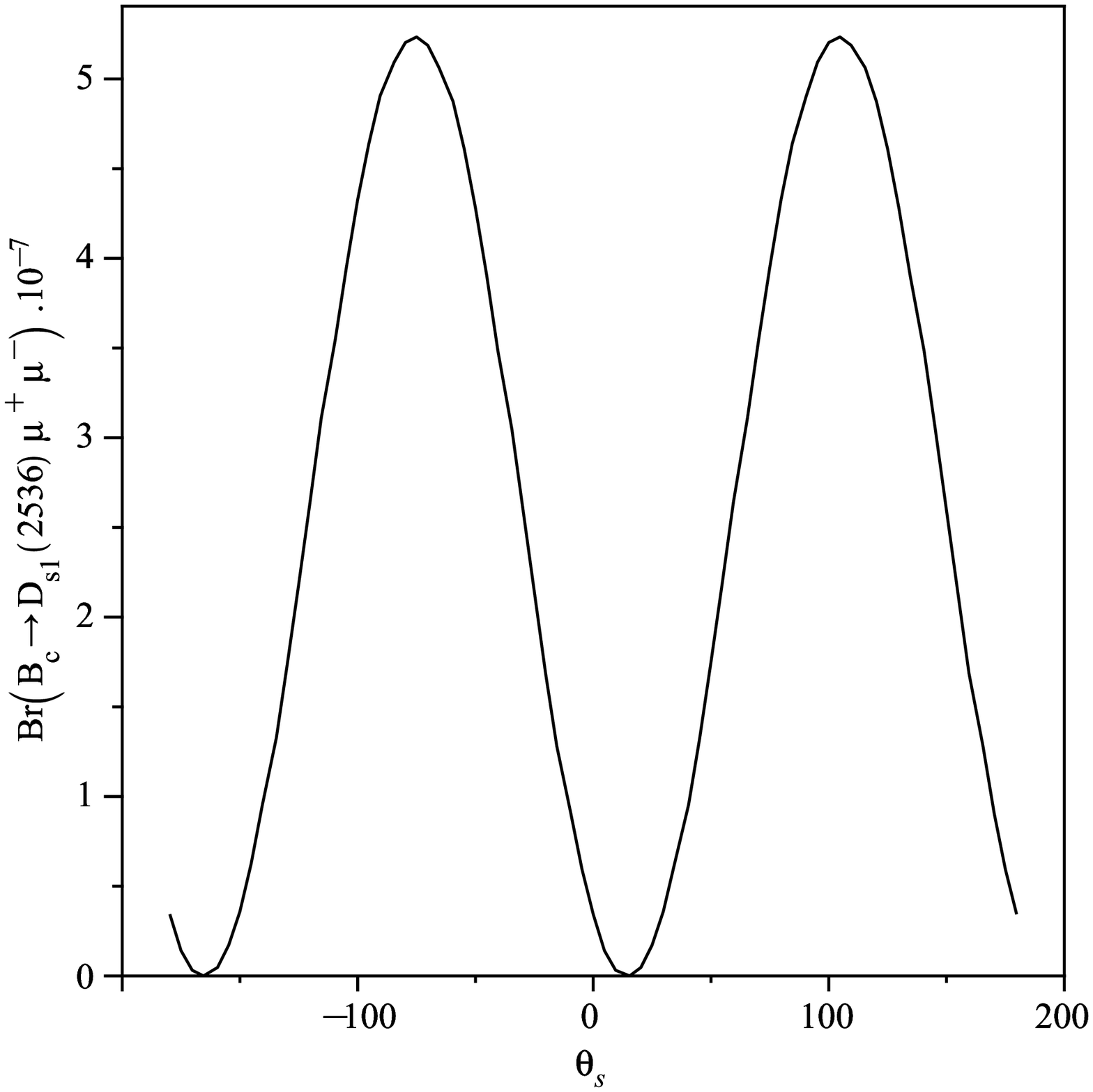}}
\end{picture}
\end{center}
\vspace*{-3cm}\caption{The branching ratios  of $B_c\to
D_{s1}\mu^+\mu^-$ with respect to $\theta_s$. }\label{F9}
\end{figure}
\normalsize
\newpage
\begin{figure}[th]
\begin{center}
\begin{picture}(160,100)
\put(-25,0){ \epsfxsize=10cm \epsfbox{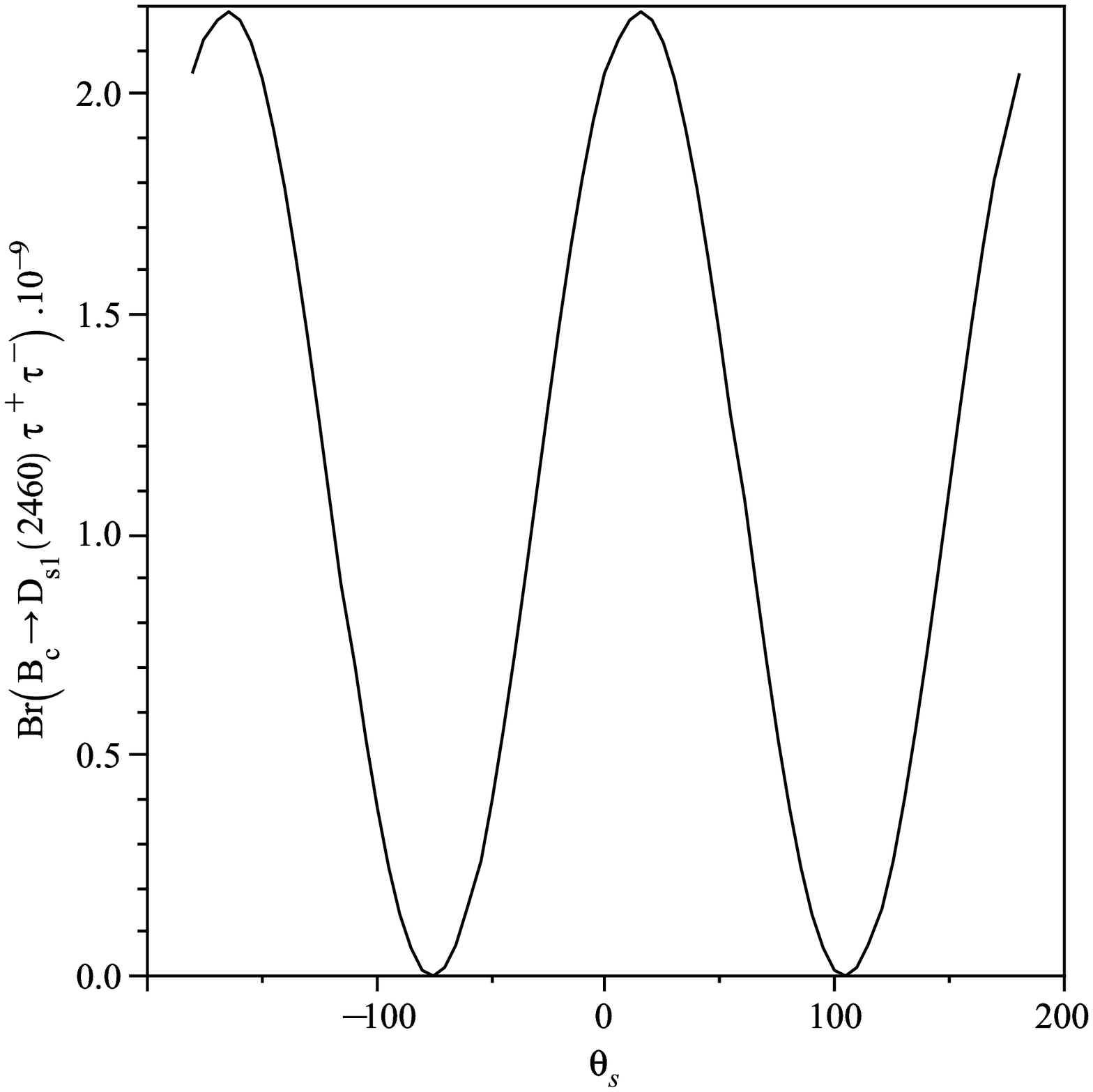}
\epsfxsize=10cm \epsfbox{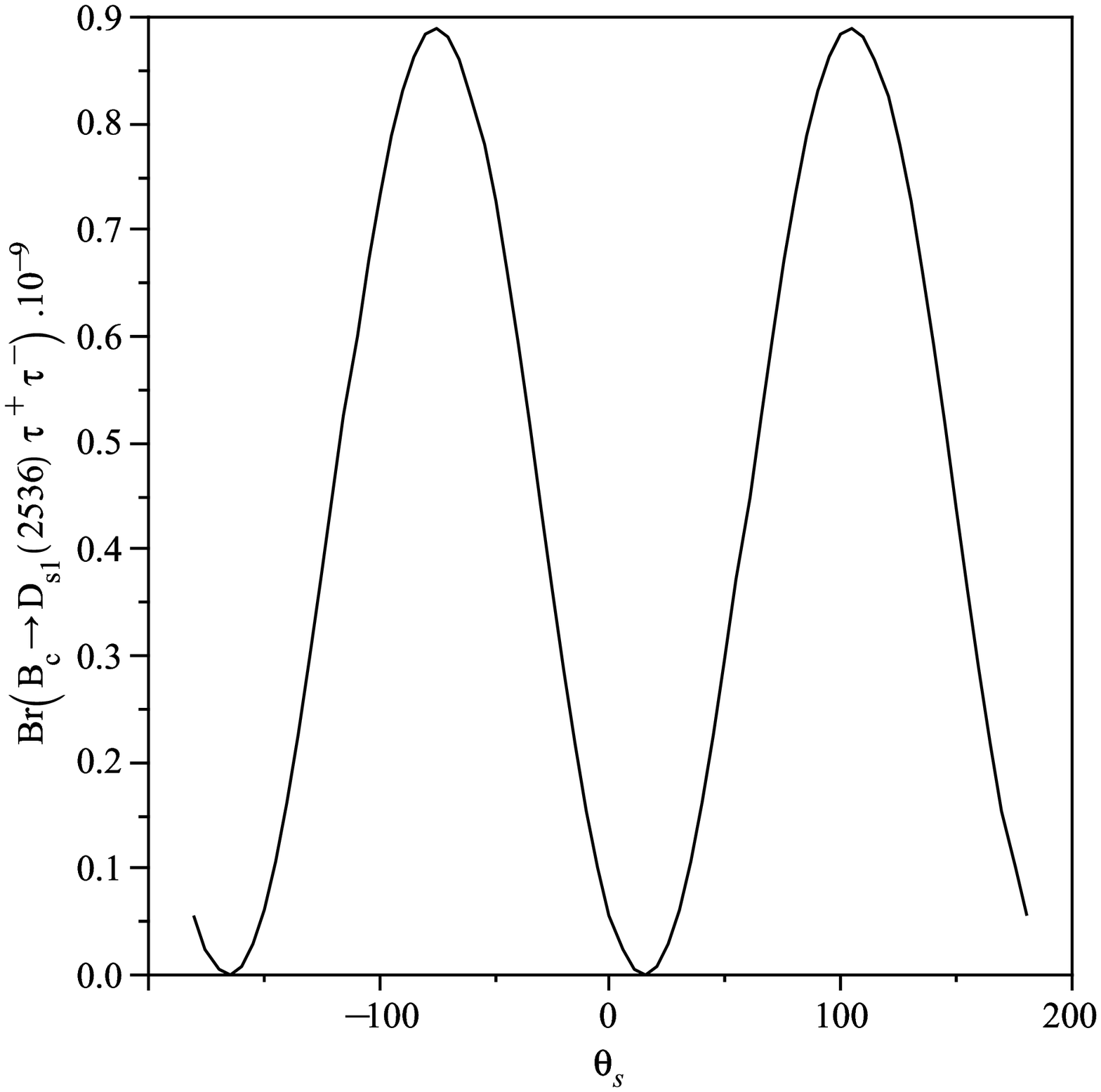}}
\end{picture}
\end{center}
\vspace*{0cm}\caption{The branching ratios  of $B_c\to
D_{s1}\tau^+\tau^-$ with respect to $\theta_s$. }\label{F10}
\end{figure}
\normalsize
\newpage
\begin{figure}[th]
\begin{center}
\begin{picture}(50,50)
\put(-25,0){ \epsfxsize=10cm \epsfbox{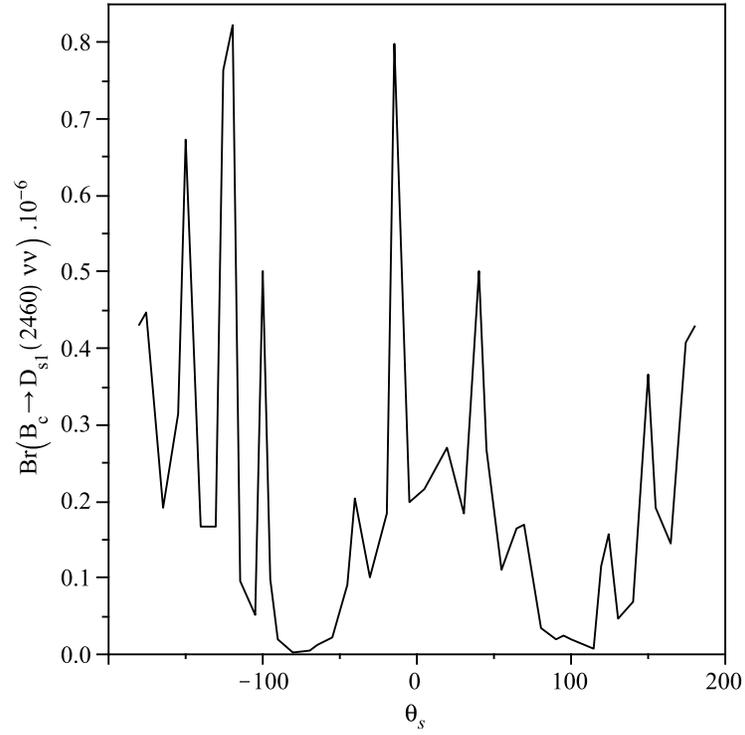} }
\end{picture}
\end{center}
\vspace*{0cm}\caption{The same as Figs. \ref{F9},
but for $B_c\to D_{s1}\nu \bar{\nu}$.}\label{F12}
\end{figure}
\normalsize

\end{document}